\documentclass{article}

\usepackage[T1]{fontenc}
\usepackage[utf8]{inputenc}
\usepackage{geometry}
\usepackage{amsmath}
\usepackage{amsthm}
\usepackage{graphicx}
\usepackage{ae,lmodern}
\usepackage{caption}
\usepackage{subcaption}
\usepackage[authoryear]{natbib}
\usepackage{amsmath}
\usepackage{amssymb}
\usepackage{authblk}
\usepackage{tikz}
\usepackage{diagbox}
\usepackage{float}
\usepackage{caption}
\usepackage{verbatim}
\usepackage[colorinlistoftodos]{todonotes}
\usepackage[colorlinks=true, allcolors=blue]{hyperref}

\usepackage{lineno}
\usepackage{setspace}
\usetikzlibrary{matrix,shapes,shapes.arrows,arrows.meta,positioning}

\usetikzlibrary{shapes,shadows,arrows}
\usetikzlibrary{positioning}

\usepackage[normalem]{ulem}

\newcommand\freddel{\bgroup\markoverwith{\textcolor{blue}{\rule[0.5ex]{2pt}{0.4pt}}}\ULon}

\newcommand\ogdel{\bgroup\markoverwith{\textcolor{blue}{\rule[0.5ex]{2pt}{0.4pt}}}\ULon}

\title{Fitting stochastic predator-prey models using both population density and kill rate data}

\author[1,2,*]{Frédéric Barraquand} 
\author[3]{Olivier Gimenez}
\affil[1]{\normalsize Institute of Mathematics of Bordeaux, CNRS, Talence, France}
\affil[2]{\normalsize Integrative and Theoretical Ecology, LabEx COTE, University of Bordeaux, Pessac, France}
\affil[3]{\normalsize Center for Evolutionary and Functional Ecology, University of Montpellier, CNRS, EPHE, IRD, University Paul Valéry Montpellier 3, Montpellier, France}
\date{}

\begin{document}
\maketitle
\thispagestyle{empty}

\begin{abstract}
Most mechanistic predator-prey modelling has involved either parameterization from process rate data or inverse modelling. Here, we take a median road: we aim at identifying the potential benefits of combining datasets, when both population growth and predation processes are viewed as stochastic. We fit a discrete-time, stochastic predator-prey model of the Leslie type to simulated time series of densities and kill rate data. Our model has both environmental stochasticity in the growth rates and interaction stochasticity, i.e., a stochastic functional response. We examine what the kill rate data brings to the quality of the estimates, and whether estimation is possible (for various time series lengths) solely with time series of population counts or biomass data. Both Bayesian and frequentist estimation are performed, providing multiple ways to check model identifiability. The Fisher Information Matrix suggests that models with and without kill rate data are all identifiable, although correlations remain between parameters that belong to the same functional form. However, our results show that if the attractor is a fixed point in the absence of stochasticity, identifying parameters in practice requires kill rate data as a complement to the time series of population densities, due to the relatively flat likelihood. Only noisy limit cycle attractors can be identified directly from population count data (as in inverse modelling), although even in this case, adding kill rate data -- including in small amounts -- can make the estimates much more precise. Overall, we show that under process stochasticity in interaction rates, interaction data might be essential to obtain identifiable dynamical models for multiple species. These results may extend to other biotic interactions than predation, for which similar models combining interaction rates and population counts could be developed. 
\end{abstract} 
* Corresponding author \url{frederic.barraquand@u-bordeaux.fr}\\
Published in \textit{Theoretical Population Biology} with \verb|doi:10.1016/j.tpb.2021.01.003|
\clearpage


\section{Introduction}

The parameterization of dynamical systems for interacting species usually involves either independent data on process rates \citep[e.g., reproduction and kill rates,][]{turchin1997ebm} or some form of inverse modelling (\citealp{turchin2000living,jost2000testing,jost2001pattern} or more recently, \citealp{delong2018life,rosenbaum2019estimating,curtsdotter2019ecosystem}). In inverse modelling approaches, a dynamical system (often deterministic) is fitted to time series of population counts, densities or biomasses \citep{stouffer2019all}. However, even in a perfect deterministic world, inverse modelling from time series is challenging due to numerous identifiability issues \citep{raue2009structural,eisenberg2014determining,kao2018practical}, that is, several sets of parameters can produce identical time series. For some models, even with an unlimited amount of data, two parameters cannot always be separated from each other if they contain redundant information \citep{cole2010determining,little2010parameter}, a problem that has long been known to statisticians and ecologists working on capture-recapture models \citep{catchpole1997detecting}. 

An additional concern when modelling ecological communities over time is that environmental stochasticity usually has a pervading influence on their dynamics \citep{lande2003stochastic,ives2003estimating,mutshinda2009drives,mac2010analysis}. It therefore makes sense to fit community-level models that allow for such environmental stochasticity ($\mathbf{e}_{t}$) to perturb the deterministic dynamics, which creates a stochastic mapping from state variables $\mathbf{x}_{t}$ at $t$ to those at $t+1$, such that $\mathbf{x}_{t+1} = \mathbf{f}(\mathbf{x}_{t},\mathbf{e}_{t})$. However, this increases rather than decreases the complexity of the inverse problem, since environmental stochasticity can then combine with nonlinearity in non-intuitive ways \citep{greenman2005impact}. When stochasticity arises from measurements or the observational setup  \citep[e.g.,][]{rosenbaum2019estimating}, the effects of stochasticity are not transferred to the next time step. However, with environmental stochasticity (also called process noise), perturbations to the growth rates of species within the community accumulate over time \citep{jost2001pattern,turchin2000living}, creating even more parameter configurations leading to similar time series. Even in simple, phenomenological statistical models for two species in interaction, there can therefore remain considerable uncertainty about the model structure and parameters, especially for short ecological time series of length $T\approx 30 - 50$ time steps \citep{barraquand2018predator}. 

How could the uncertainty on parameter estimates of stochastic community models be decreased? One strategy that has enjoyed a great success in population ecology is to use a combination of datasets \citep{besbeas2002integrating,maunder2004population,schaub2011integrated}. Combining multiple types of data increases the precision of parameters that could be estimated using a single data type (but with a large uncertainty), and often allows to estimate parameters that were not identifiable when the models were fit to a single data type \citep{besbeas2002integrating}. In the case of mechanistic predator-prey modelling, such strategy would be equivalent to combining process rate data with population densities, which is what we attempt here. Ours is not, of course, the first study to consider the relative benefits of forward modelling (process rate modelling) vs inverse modelling \citep[e.g.,][]{turchin2000living}. The novel perspective here is on merging different sources of data on process rates (i.e., kill rate or biomass flow) and densities (or biomasses) into a fully stochastic framework. 

Indeed, even in trophic interaction models that include environmental stochasticity on the population growth rates of predators and prey \citep[e.g.,][]{karban2010population,ives2008high}, the functional response \citep{solomon1949nca,holling1959cpr} describing the relationship between predator kill rate and population densities is almost always viewed as a deterministic object (outside of corrections for sampling variation, e.g., \citealp{fenlon2006modelling}). Exceptions to this rule are to be found in continuous-time ODE models where stochasticity has been added to \emph{all} parameters \citep{turchin1997ebm,turchin2000living} and in stochastic individual-based predator-prey models \citep[e.g.,][]{deroos1991mvd,wilson1998rdb,hosseini2003lcs,murrell2005lss}, in which interaction stochasticity naturally arises from the encounter process.
By contrast, in most models using stochastic difference or differential equations, kill rates of individual predators are deterministically determined by prey population densities. However, in addition to the vagaries of the interaction process (interaction stochasticity \emph{per se}, due to the randomness of encounters of pairs of individuals), the average predator kill rate typically depends on multiple quantities that vary over time (and space, though we will not discuss this here): the densities of several species and not only the main prey \citep{abrams1982functional,abrams2010quantitative}, abiotic factors such as temperature \citep{rall2012universal}, traits such as body size \citep{vonesh2005compensatory} which vary with population composition --- and the list goes on. 

With such an imperfect knowledge of the factors affecting the kill rate of the average individual predator, and measures of population densities restricted to two species (the predator and its prey), it can therefore make more sense to view the functional response as a stochastic object. There are also empirical arguments for this view: point clouds surrounding the functional response curve usually display considerable variation (e.g., Fig. 3 of \citealp{kalinkat2013body}, Fig. 3 of \citealp{pritchard2017frair}, Fig. 5 in \citealp{aldebert2018community}, Fig. 5 of \citealp{rosenbaum2018fitting}) that can hardly be ascribed solely to sampling and observation error. Yet, probably for historical reasons, reinforced by the prominence of ODE modelling, such random variation in kill rates is mostly viewed as observational noise rather than process noise. 

Considering the functional response as a stochastic rather than deterministic object, randomly varying over time even in the absence of changes in the focal population densities, makes therefore sense in light of both theory and empirical data. In fact, some researchers have chosen to move away from the classical food web modelling paradigm precisely because of its reliance on deterministic functional relationships \citep{planque2014non,subbey2016exploring}. In a more classic predator-prey parametric inference context, \citet{gilioli2008bayesian,gilioli2012nonlinear} embraced the stochasticity of the functional response using SDEs. However, they used only population densities as data for their stochastic model, which renders the estimation challenging, while we will consider additionally kill rate data. In this article, two key features are therefore that (1) the functional response comprises process rather than observational noise, and (2) kill rate data is added to population densities to help estimating functional response and other predator-prey parameters. 

To evaluate our ability to infer our stochastic predator-prey system parameters, we use simulations of parametric models. We consider a ground truth dynamical model for predator-prey systems with process noise on both growth rates and kill rates, in discrete time. We first simulate datasets using the model, and then fit this model and a variant to the simulated data. We use both frequentist and Bayesian estimation paradigms, to check that our results are congruent between the two frameworks and therefore robust. Identifiability is examined in both frameworks, using the Fisher Information Matrix in a frequentist setting and the posterior-prior overlap in a Bayesian setting, with or without kill rate data. We consider both long time series by ecological standards ($T = 100$) and then time series of realistic ecological length ($T = 50$ and $25$). Finally, we vary the percentage of data points for which kill rate data are available, which can contribute to a more optimal allocation of time and effort in the field, when designing surveys of community dynamics. 

\section{Models and statistical methods}

\subsection{Predator-prey model in discrete time}

We chose a model with a numerical response of the Leslie type, but similar analyses are performed for Rosenzweig MacArthur models in Supplement B1, to ensure generality. A Beverton-Holt function for density-dependence in the prey growth rate was chosen to avoid cycles in the prey in absence of the predator, so that the model behaviour is more reminiscent of its continuous-time counterpart (see \citealp{weide2018hydra} on connecting discrete-time to continuous-time predator-prey models). The functional response takes the form $g(N,\epsilon)$ where $N$ is prey density and $\epsilon$ a noise term. Our population-level model can be written as:

\begin{align}
N_{t+1} & = N_{t}\frac{e^{r+\epsilon_{1t}}}{1+\gamma N_{t}}\exp\left(-g(N_{t},\epsilon_{3t})\frac{P_{t}}{N_{t}}\right),\,\epsilon_{1t}\sim\mathcal{N}(0,\sigma_{1}^{2})\label{eq:prey_discreteLeslieMay}
\end{align}

\begin{align}
P_{t+1} & = P_{t}\frac{e^{s+\epsilon_{2t}}}{1+qP_{t}/N_{t}},\,\epsilon_{2t}\sim\mathcal{N}(0,\sigma_{2}^{2})\label{eq:predator_discreteLeslieMay}.
\end{align}

The roots of this discrete-time formulation can be traced back to \citet{leslie1948some,leslie1960properties} who included a Beverton-Holt regulation for the prey and predator to make discrete-time models more similar to their continuous-time counterparts. We added environmental stochasticity through log-normal noise. In addition, we consider a saturating functional response, which makes our model resemble the continuous time models of \citet{tanner1975stability} or \citet{may1973stability} and later \citet{turchin1997ebm}, whose notations we have kept. Except that here, the functional response is actually $g(N_t,\epsilon_{3t})$, with a stochastic term $\epsilon_{3t}$ included. 

One of the advantages of the Leslie-type parameterization over the Rosenzweig-MacArthur linear numerical response is that the former can be parameterized using the observed predator maximal (intrinsic) growth rate as \citet{tanner1975stability} did. Parameter values were inspired by small mammals \citep{turchin1997ebm}, which we modified slightly to be able to get both stable and cyclic dynamics. The division by $N_t$ in $\exp(-g(N_t,\epsilon_{3t})P_t/N_t)$ expresses the fact that all terms within the exponential are on the prey fitness scale (per capita mortality), a very classic representation \citep[e.g.,][]{ives1995predicting}. This exponential term is reminiscent of the Nicholson-Bailey model and its spin-offs \citep{weide2018hydra}.

Until now, we have not specified a model for the functional response $g(N_{t},\epsilon_{3t})$. With a deterministic functional response, we have a classic stochastic predator-prey model with log-normal environmental noise but an otherwise deterministic skeleton. A stability analysis of the deterministic skeleton of the model is performed in Appendix~\ref{sec:stability}, and was used to determine which parameters led ---  in the stochastic model --- to a perturbed fixed focus or node vs a noisy limit cycle. We use the following equation for a stochastic Holling type II functional response: 

\begin{equation}\label{eq:FR}
G_{t} = \frac{CN_{t}}{D+N_{t}} + \epsilon_{3t}, \;  \epsilon_{3t} \sim \mathcal{N}\left(0,\sigma_{3}^2\right)
\end{equation}

Equation \eqref{eq:FR} corresponds to mild Gaussian fluctuations around the functional response (we cannot model strong fluctuations with a Gaussian model, as they may create negative values). Because there can be substantial process noise in real functional response data, we also considered much stronger fluctuations in the kill rates, by allowing the parameters $C$ and $D$ to themselves vary over time: these models and results are presented in Supplement B2. 

\subsection{Statistical methodology}

We consider a discrete-time stochastic dynamical system (stochastic nonlinear difference equation).  The main state variables that we consider ($\mathbf{x}_{t}$) are log-transformed population counts (i.e., abundances) or densities (abundances divided by area). Although we apply here data integration to a predator-prey case, the methodology is more general and can in principle be applied to any addition of auxiliary information to such a discrete-time system. Our auxiliary variables can be any component of process rates (e.g., an interaction rate such as the predator kill rate, a demographic rate such as predator fertility) which is recorded over time and not simply a deterministic function of the state variables. In that sense, our definition of auxiliary variables is included within that of \citet{benaim2019persistence}, who define a larger theoretical framework for multispecies population persistence.

Our approach share similarities with Integrated Population Modelling \citep{besbeas2002integrating,maunder2004population,peron2012integrated}, in that it fits a population or community model to multiple data types, but differs slightly due to the kind of data considered: population densities and interaction rate data, similar to stocks and flows in ecosystem ecology. Logically, our approach also share similarities with data fusion/assimilation approaches of ecosystem ecology \citep{keenan2013rate,niu2014role}, but these tend to be applied to more complex models without a closed form likelihood.  Within population ecology, in a predator-prey context, we can relate to the seal predation model of \citet{cook2015grey} or the recent endeavour of \citet{ferguson2018integrating} to combine population count and isotopic (diet) data: their models differ from ours, however, in that they do not view predation as a fully stochastic process. 
 
We illustrate our methodology with the abovementioned predator-prey model. Log-densities for both predator and prey are gathered in a vector $\mathbf{x}_{t}=(\ln(N_{t}),\ln(P_{t}))^{T}$. The time unit $t$ can be thought of as a year for vertebrate species. The auxiliary variable is, in this case, the observed kill rate per predator which is denoted $G_t$. $G_t$ is considered here to be a positive real number, representing the average kill rate across individual predators for year $t$, computed from field observations \citep{gilg2003cyclic} or reconstructed from dietary requirements and observed diet \citep{nielsen1999gyrfalcon}. To this can be added other demographic (vital) rates $R_{t}$, such as fertilities, that are stacked in a vector as well, $\mathbf{a}_{t}=(G_{t}, R_{t})^{T}$. Currently
we only use $\mathbf{a}_{t}=G_{t}$ but it may be useful to add more variables in other applications; hence the derivation is kept general.

The population dynamics part of the model gives us the probability distribution of $\mathbf{x}_{t+1}|(\mathbf{x}_{t},\mathbf{a}_{t})$, since the population counts at time $t+1$ depend both on past abundances and the interaction or demographic data based on the chosen mechanistic model. We also know the probability distribution of $\mathbf{a}_{t}|\mathbf{x}_{t}$ (in our simple predator-prey case, the functional response). We can therefore write down easily the joint likelihood for both data sources in quite general terms, denoting $\mathbf{X}=(\mathbf{x}_{1},...,\mathbf{x}_{t_{m}})$ and $\mathbf{A}=(\mathbf{a}_{1},...,\mathbf{a}_{t_{m}})$: 

\begin{equation}
\mathcal{L}(\mathbf{X},\mathbf{A})=p(\mathbf{x}_{1},\mathbf{a}_{1})\prod_{t=1}^{t_{m}-1}p_{C}(\mathbf{x}_{t+1},\mathbf{a}_{t+1}|\mathbf{x}_{t},\mathbf{a}_{t})
\end{equation}

where $p(\mathbf{y})$ and $p_{C}(\mathbf{y}|\mathbf{y}')$ are continuous probability densities
for the matrix $\mathbf{y}=(\mathbf{x},\mathbf{a})$ and its conditional pdf,
respectively. The conditional pdf can be further decomposed using
the chain rule of probability (i.e., relying on conditional independence)

\begin{align}
p_C(\mathbf{x}_{t+1},\mathbf{a}_{t+1}|\mathbf{x}_{t},\mathbf{a}_{t})=p_{2}(\mathbf{a}_{t+1}|\mathbf{x}_{t+1},\mathbf{x}_{t},\mathbf{a}_{t})\times p_{1}(\mathbf{x}_{t+1}|\mathbf{a}_{t},\mathbf{x}_{t})=p_{2}(\mathbf{a}_{t+1}|\mathbf{x}_{t+1})\times p_{1}(\mathbf{x}_{t+1}|\mathbf{a}_{t},\mathbf{x}_{t})
\end{align}

where $p_{1}(\mathbf{x}_{t+1}|\mathbf{a}_{t},\mathbf{x}_{t})$ represents the dynamical system, given auxiliary information $\mathbf{a}_{t}$ on process rates (here, the functional response), and $p_{2}(\mathbf{a}_{t+1}|\mathbf{x}_{t+1})$
is the functional response model (or a demographic model), which is conditional only to the current set of abundances. We therefore end up with a closed form expression for the likelihood

\begin{equation}\label{eq:full-likelihood}
\mathcal{L}(\mathbf{X},\mathbf{A})=p_{1}(\mathbf{a}_{1}|\mathbf{x}_{1})p(\mathbf{x}_{1})\prod_{t=1}^{t_{m}-1}\underbrace{p_{1}(\mathbf{x}_{t+1}|\mathbf{a}_{t},\mathbf{x}_{t})}_{\text{dynamical system}}\times\underbrace{p_{2}(\mathbf{a}_{t+1}|\mathbf{x}_{t+1})}_{\text{auxiliary information model}}
\end{equation}

where we swapped $p_{1}$ and $p_{2}$ to get the dynamical system model first. 
Finally, it would be possible to wrap the above community dynamics model within a state-space framework \citep[as done in][]{ives2003estimating,karban2010population}, adding an observation submodel for population counts, e.g., with log-abundance
$\mathbf{y}_t \sim \mathcal{N}(\mathbf{x}_t,\Sigma_y)$ if measurement error is lognormal, or abundance $\mathbf{y}'_t \sim \mathcal{P}(\exp\left(\mathbf{x}_t)\right)$ for small counts. However, state-space models have important identifiability issues \citep{knape2008estimability,auger-methe_state-space_2016,auger2020introduction} when no information or data is provided for the measurement error; adding a measurement error is best done whenever error magnitude can be specified, or estimated through replicated samples at each point in time \citep{dennis2010replicated}. Distinguishing the best observation model is often impossible unless there are replicated samples \citep{knape2011observation}, which introduces another layer of complexity: we have therefore left the issue of measurement error for further work.

\subsection{Application to the stochastic predator-prey model}

In the simplest case highlighted by our two-species discrete-time stochastic dynamical system, $p_{1}(\mathbf{x}_{t+1}|\mathbf{a}_{t},\mathbf{x}_t)=p_{11}(x_{1,t+1}|\mathbf{a}_t,\mathbf{x}_t)p_{12}(x_{2,t+1}|\mathbf{a}_t,\mathbf{x}_t)$ is the product of the two Gaussian pdf for log-densities conditional on past densities ($\mathbf{x}_{t}$) and auxiliary information ($\mathbf{a}_{t}$). For simplicity we denote $\mathbf{x}_{t}=(\ln(N_{t}),\ln(P_{t}))^{T} = (n_{t},p_{t})^{T}$. Using the equations \eqref{eq:prey_discreteLeslieMay}  and \eqref{eq:predator_discreteLeslieMay}, we obtain the following transition probabilities for log-scale densities:

\begin{equation}
n_{t+1}|(\mathbf{a}_{t},\mathbf{x}_{t})\sim\mathcal{N}(\mu_{1t},\sigma_{1}^{2})\,\,,\mu_{1t}=n_{t}+r-G_{t}\frac{P_{t}}{N_{t}}-\ln(1+\gamma N_{t})\label{eq:prey_discreteLeslieMay_logscale}
\end{equation}

and 

\begin{equation}
p_{t+1}|(\mathbf{a}_{t},\mathbf{x}_{t})\sim\mathcal{N}(\mu_{2t},\sigma_{2}^{2})\,\,,\mu_{2t}=p_{t}+s-\ln(1+qP_{t}/N_{t})\label{eq:predator_discreteLeslieMay_logscale}. 
\end{equation}
The functional response model ($p_{2}$ in eq.~\eqref{eq:full-likelihood}) is also given by a Gaussian pdf, although it should be noted that our results are robust to this approximation (see Discussion and Supplement B2): 

\begin{equation}
G_{t}|N_t \sim\mathcal{N}(\mu_{3t},\sigma_{3}^{2}),\;\mu_{3t}=\frac{CN_{t}}{D+N_{t}}. 
\end{equation}

We considered two parameter sets for the model, given in Table~\ref{tab:parameter_table}.    
\begin{table}[]
    \centering
    \begin{tabular}{c|c|c|c|c}
    Parameter & Ecological meaning & Unit & Perturbed FP & Noisy LC\\
    \hline
    r & Low-density prey pgr & $\text{year}^{-1}$ & 2 & 2\\
    s & Low-density predator pgr & $\text{year}^{-1}$ & 0.5 & 0.5\\
    $\gamma$ & Prey density-dependence & NA & 1 & 1\\
    Q & Prey-to-predator ratio & NA & 10 & 10\\
    C & Max prey consumption per predator & [prey units $\times \text{year}^{-1}$] & 2.5 & 15  \\
    D & Half-saturation constant & [prey units] & 1.0  & 0.25
    \end{tabular}
    \caption{Parameter values for both types of attractors considered (Perturbed FP: Perturbed fixed point; noisy LC: limit cycle perturbed by noise).
    pgr: per capita growth rate. 
    The rest of the parameters are the variances on the error terms, set to $\sigma_1^2=\sigma_2^2=\sigma_3^2=\sigma^2=0.05$.} 
    \label{tab:parameter_table}
\end{table}
The first parameter set (FP) corresponds to a forced fixed point (focus), and although the eigenvalues are imaginary (Appendix~\ref{sec:stability}) in this case there are no discernible cyclic oscillations (i.e., no quasi-cycles \emph{sensu} \citealp{nisbet1982modelling}). Parameter set FP is a crucial example because not all fluctuating predator-prey systems give rise to limit cycle oscillations. We also consider a noisy limit cycle (LC), i.e., parameters that give rise to a limit cycle without the noise, so that the cycle has a broad amplitude and well-defined attractor even as noise is added, but an irregular periodicity due to the random perturbations \citep{louca2014distinguishing}. We use here the wording `limit cycle' by analogy to the continuous time theory, although quite formally, this parameter set gives rise to an invariant loop rather than a discrete-time limit cycle \citep[see, e.g.,][chapter 16]{caswell2001matrix}. 
Representative time series of the model for both parameter sets are provided in Fig. 1.

\begin{figure}[h]
    \centering
    \includegraphics[width=16cm]{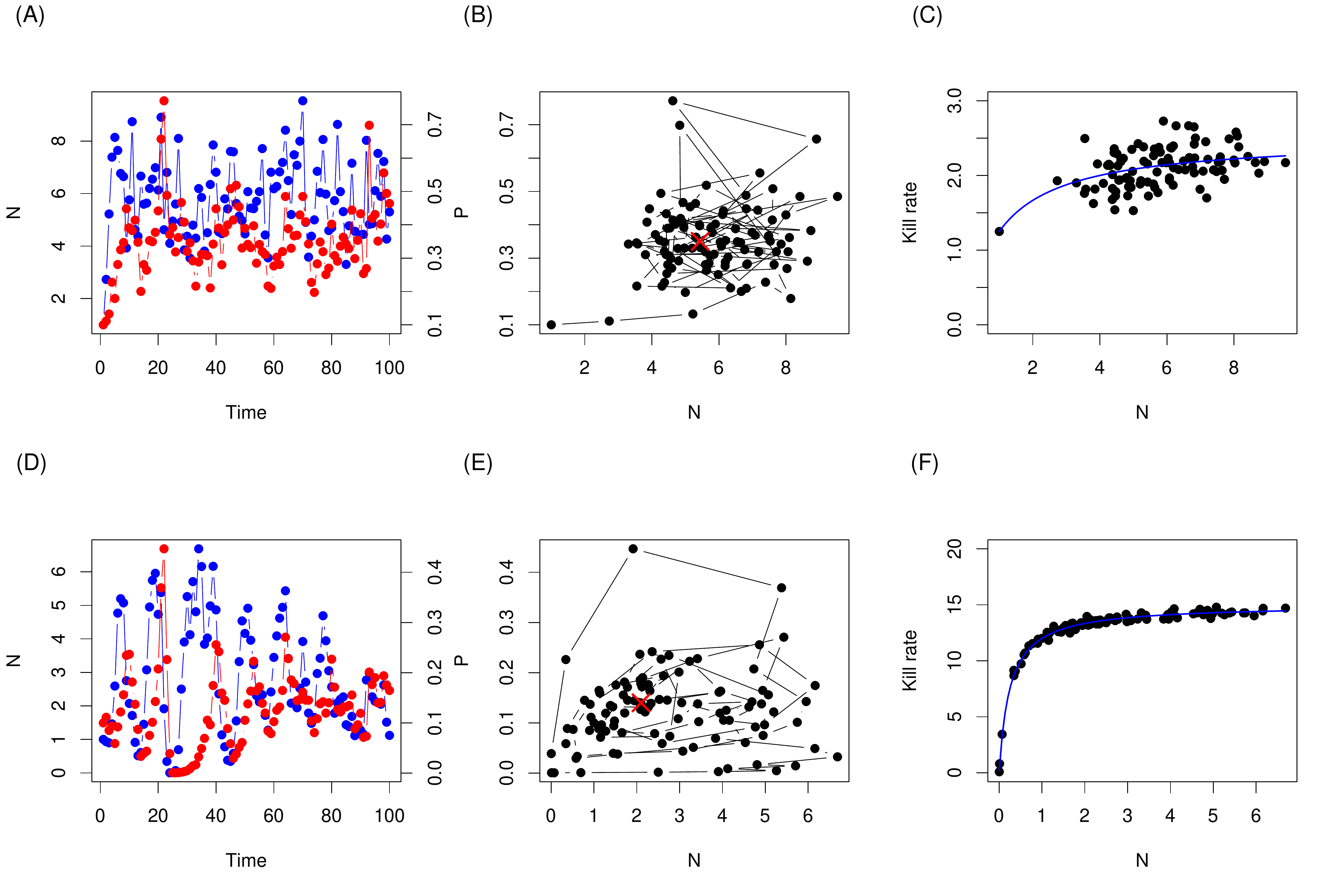}
    \caption{System dynamics for T=100 (1 simulation). In panel (A), we show densities of prey $N(t)$ in blue and predator in red $P(t)$, for the perturbed FP case, (B) corresponding trajectories in phase plane (the red cross materializes the fixed point) and (C) functional response: kill rate of individual predator as a function of prey density. In the second row, identical panels (D-E-F) for the noisy LC simulation. }
    \label{fig:system_dynamics}
\end{figure}

These two sets of parameters are crossed with several scenarios of data availability:
\begin{itemize}
    \item time series of length $T=100$, $50$ and $25$
    \item we consider that kill rate (KR) data is available along a fraction $p_{KR}=100\%$, $25\%$, or $0\%$ of the time series data. This is meant to emulate common scenarios in which the kill rates are not monitored over the whole time series, and to quantify the benefits of adding even small amounts of kill rate data. 
\end{itemize}
Formal identifiability results involving the Fisher Information Matrix (FIM) or prior-posterior overlap are reported for $T=100$ and $p_{KR} = 0$ or $1$, and will be presented first. Throughout the manuscript, kill rate data will refer only to the predator intake rates $(G_t)$ whilst functional response data will refer to the pair $(G_t,N_t)$.

Note that in the case without kill rate data ($p_{KR} = 0$), we fit the model without noise on the functional response, since there is no data to fuel the statistical model of the functional response and we wanted to have a meaningful comparison to the rest of the literature. Having a latent, completely unobserved stochastic state variable for the kill rate was considered in early simulations but typically leads to even worse estimates, and is more complicated to fit. 

For each parameter $\times$ data availability scenario, we fit the models in a Bayesian framework. A Bayesian framework is here more convenient as our scenarios with $p_{KR} = 0.25$ require to estimate a latent state for partially unobserved kill rate data (it would be doable in a frequentist setting using an EM algorithm, with more effort). Because we consider 100 simulations for each parameter x data availability scenarios, we have distributions even for the point estimates (means over the posteriors), whose upper and lower bounds reveal the precision of the estimator in a frequentist sense. 
In cases where $p_{KR} = 0$ or $1$, we also fit models through optimization in Appendix~\ref{sec:optim}, as detailed below. 

\subsection{Model fitting}

Most results rely on fits of the model by Markov Chain Monte Carlo (MCMC) in JAGS \citep{plummer2003jags} which uses the Gibbs sampler, using 3 chains and vague priors. For instance, parameters $C$ and $D$ that are defined on the half-line have prior $\text{Gamma}(0.01,0.01)$, which is near-exponential and therefore the maximum entropy prior \citep{mcelreath2020statistical}. The full code is provided in \citet{barraquand_gimenez_2020} \footnote{Github version at \texttt{https://github.com/fbarraquand/predatorpreyDynamics\_wFRdata}} with all the details of the implementation. We also derived mathematically the full model likelihood based on eq.~\eqref{eq:full-likelihood}, which we used to plot likelihood surfaces and ascertain identifiability from a frequentist viewpoint. Having a well-defined likelihood function then allows us to find point estimates of the parameters through quasi-Newton methods (hill-climbing algorithms), which we have done using the L-BFGS-B algorithm provided by \verb|optim()| in R (Appendix~\ref{sec:optim}). These matched qualitatively the results obtained by MCMC. 

\subsection{Identifiability diagnostics}

Identifiability can be defined as the fact that $\mathcal{M}(\theta_1)  = \mathcal{M}(\theta_2) \Rightarrow \theta_1 = \theta_2$ \citep{rothenberg1971identification,cole2010determining,cole2016parameter,auger2020introduction,cole2020parameter} where $\mathcal{M}$ is a model fit descriptor (in a frequentist setting, $\mathcal{M}$ will be the likelihood $\mathcal{L}$ or some sufficient statistic).
We classically distinguish \textit{intrinsic} or \textit{structural} identifiability \citep{eisenberg2014determining,cole2020parameter}, that is not dependent upon a particular dataset, from \textit{extrinsic} or \textit{practical} identifiability, that is pertains to a given dataset. In both cases, identifiability refers to our ability to find a unique maximum of the likelihood. Unfortunately, the literature on identifiability is fraught with ambiguities; for instance \citet{raue2009structural} referred to \textit{practical identifiability} as being able to find finite confidence intervals, our more classic definition focuses on the identifiability of \textit{point estimates} rather than intervals. 

Identifiability was inspected in different ways depending on the statistical framework chosen. In a frequentist setting, in order to ascertain structural identifiability, we computed the Fisher Information Matrix (FIM) $\mathcal{I}_T(\theta_{\text{true}})$ from the likelihood, defined as 
\begin{align}
\mathcal{I}_T(\theta_{\text{true}})_{ij} = \mathbb{E} \left[ \left(\frac{\partial \ln \mathcal{L} (\theta;\mathbf{Y}_T)}{\partial \theta_i} \right) \left(\frac{\partial \ln \mathcal{L} (\theta;\mathbf{Y}_T)}{\partial \theta_j} \right) \right] = - \mathbb{E} \left[ \left(\frac{\partial^2 \ln \mathcal{L} (\theta;\mathbf{Y}_T)}{\partial \theta_i \partial \theta_j} \right) \right]
\end{align}

for a dataset of length $T$, $\mathbf{Y}_T = (\mathbf{X}_T,\mathbf{A}_T)$. The FIM is critical to evaluating \textit{structural} identifiability \citep{rothenberg1971identification,eisenberg2014determining,cole2020parameter}, as it does not depend on a particular dataset. It is obtained here by taking expectations over many time series of length $T$. 
In practice, we use $T=100$ and $k_r=100$ simulations; we previously computed the Hessian matrix for very large $T$, with similar results (the stochastic process is likely ergodic). A parameter set $\theta$ is structurally identifiable whenever the FIM is non-singular (i.e., no zero eigenvalue) because this condition is equivalent to the matrix being positive definite, the FIM being real and symmetric. Positive definiteness, in turn, is essential to having a unique minimum of the negative log-likelihood \citep{rothenberg1971identification}. It is therefore possible to quantify identifiability by assessing whether the FIM has some zero eigenvalues. Geometrically, the FIM is a curvature as it is obtained as the expected Hessian of $-\ln(\mathcal{L})$, i.e., the expected second derivative of the negative log-likelihood. If the vector of parameters $\theta \in \Theta_0$, where $\Theta_0$ is some subset of parameter space where all $\theta$ maximize the likelihood (e.g., there is a true ridge on the likelihood surface), then both the slope and the curvature of $\ln \mathcal{L} (\theta;\mathbf{Y}_T)$ must be zero for $\theta \in \Theta_0$ along some direction, which is equivalent to having a zero FIM eigenvalue. 

Moreover, as classical statistical theory dictates, asymptotically (as the time series length $T$ gets large or as we average of over many datasets), the vector of parameters $\hat{\theta} \rightarrow \mathcal{N}(\theta_{\text{true}},\mathcal{I}_T(\theta_{\text{true}})^{-1})$. We therefore also computed the expected pairwise correlation between the parameters: the expected variance-covariance matrix of the parameters is defined by $\Sigma = \mathcal{I}^{-1}$ so that we get easily the expected pairwise parameter correlation matrix $\mathbf{\rho} = (\rho_{ij})$, with each element defined as $\rho_{ij} = \frac{\Sigma_{ij}}{\sqrt{\Sigma_{ii}\Sigma_{jj}}}$. This analysis was done here for the FIM, or expected Fisher Information, but as a check we have also performed similar inspections of the variance-covariance matrix derived from the Hessian at the maximum likelihood estimate (which is the observed FIM for one simulation, \citealp{viallefont1998parameter}) with similar results.  
Still in a frequentist setting, we computed extrinsic/practical identifiability by looking at parameter estimator distributions for 100 datasets of length $T=100$; these results are presented in Appendix~\ref{sec:optim}. 

In a Bayesian setting, we examined whether the prior and posterior distributions overlapped \citep{gimenez2009weak,cole2020parameter}, which is a classical Bayesian way to check that the model is identifiable \emph{in practice}, in the sense that the likelihood brings enough information to the posteriors. 
We also inspected the correlations in the Markov chains for pairs of parameters, which translates into pair posterior distributions of parameters. Parameters whose chains were too positively or negatively correlated were considered to be unidentifiable separately, even when the means of the posterior distributions were close to the true values. 

\section{Results}

\subsection{Identifiability diagnostics}

\subsubsection{Frequentist analysis}

\paragraph{Fisher Information Matrices}
We consider here two contrasted cases, for $T=100$: $p_{KR}=1$, we have data on kill rates for the whole time series vs $p_{KR}=0$, no data on kill rates for the whole time series. 

\begin{figure}[h]
    \centering
    \includegraphics[width=15cm]{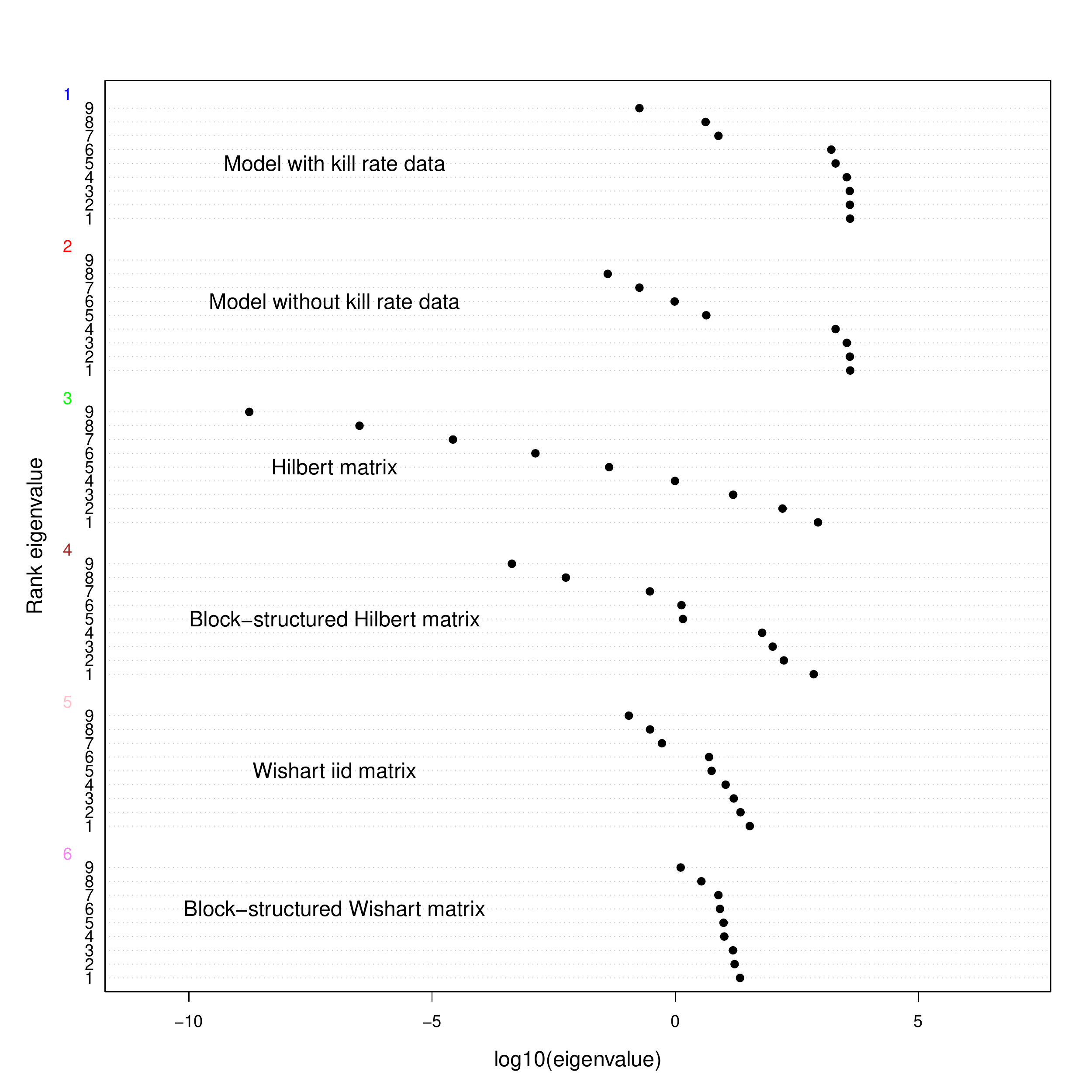}
    \caption{Distribution of eigenvalues of the FIM, perturbed fixed point case. First row, eigenvalues of the FIM for the model with kill rate data, second row, eigenvalues for the model without kill rate data. Third and fourth rows, eigenvalues for Hilbert matrices of similar size for comparison; fifth and sixth rows, eigenvalues for Wishart matrices of similar size for comparison.}
    \label{fig:FIM_eigenvalues_FP}
\end{figure}

We have reported the eigenvalues of the FIM with and without kill rate data for the fixed point in Fig~\ref{fig:FIM_eigenvalues_FP} (first and second row, respectively). It is challenging to interpret whether eigenvalues are truly zero \citep{viallefont1998parameter,gimenez2004methods}, thus the condition number, the ratio between the largest and smallest eigenvalue, is usually given. As eigenvalues are reported in log-scale in Fig~\ref{fig:FIM_eigenvalues_FP}, the distance between the smallest and largest eigenvalue on the graph represents the condition number. To aid the interpretation, we have plotted in Fig~\ref{fig:FIM_eigenvalues_FP} the eigenvalue spectrum of two types of matrices of similar size: Hilbert matrices (with $(i,j)$ elements $\frac{1}{i+j-1}$), notorious for being on the border of invertibility, and random Wishart matrices, generated as $XX'$ with $X$ a matrix with i.i.d. $\mathcal{N}(0,1)$ entries, the simplest random matrices meeting the necessary requirements to mimic an ideal-case FIM (positive definiteness, symmetry). The structure of the FIM being block-diagonal, which helps invertibility, we also constructed two equivalent block-diagonal Hilbert and Wishart matrices, whose elements are set to zero whenever FIM elements are zero. The comparison of the FIMs to these matrices show that while the FIMs with and without kill rate data have lower condition number than random Wishart matrices (which should come close to what the FIM can look like under ideal circumstances), their condition numbers are also lower than those of ill-behaved Hilbert matrices. Model FIMs should therefore be considered to have non-zero eigenvalues both with and without kill rate data. Similar results are presented in Appendix~\ref{sec:noisyLC-results} for the noisy limit cycle parameter set. 

\paragraph{Variance-Covariance Matrices}

\begin{figure}[htbp]
\centering
\begin{subfigure}[b]{0.45\textwidth}
    \centering
    \includegraphics[width=\textwidth]{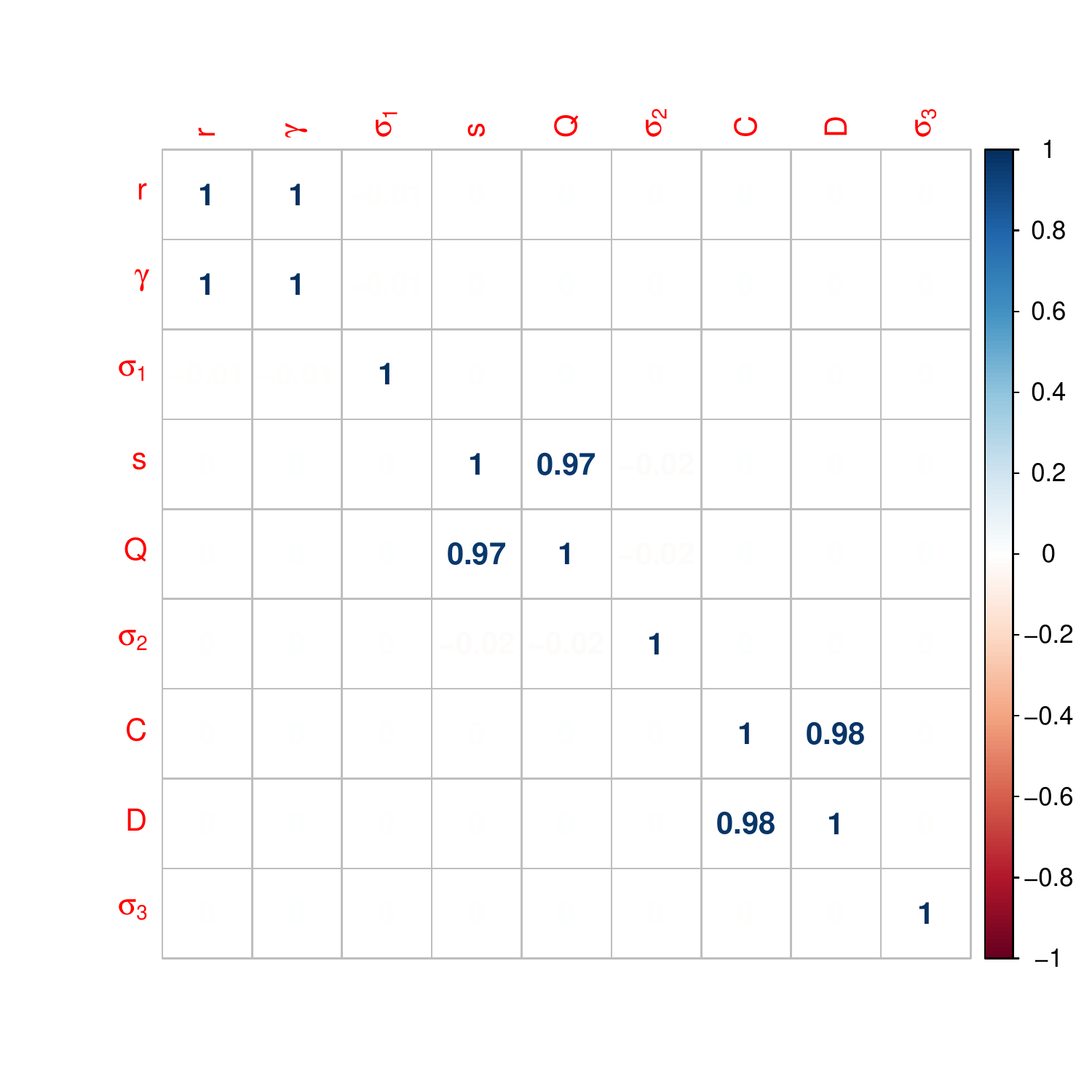}
    \caption{With kill rate data, FP}
    \label{fig:inverseFIM1_FP}
\end{subfigure}
~ 
\begin{subfigure}[b]{0.45\textwidth}
    \centering
    \includegraphics[width=\textwidth]{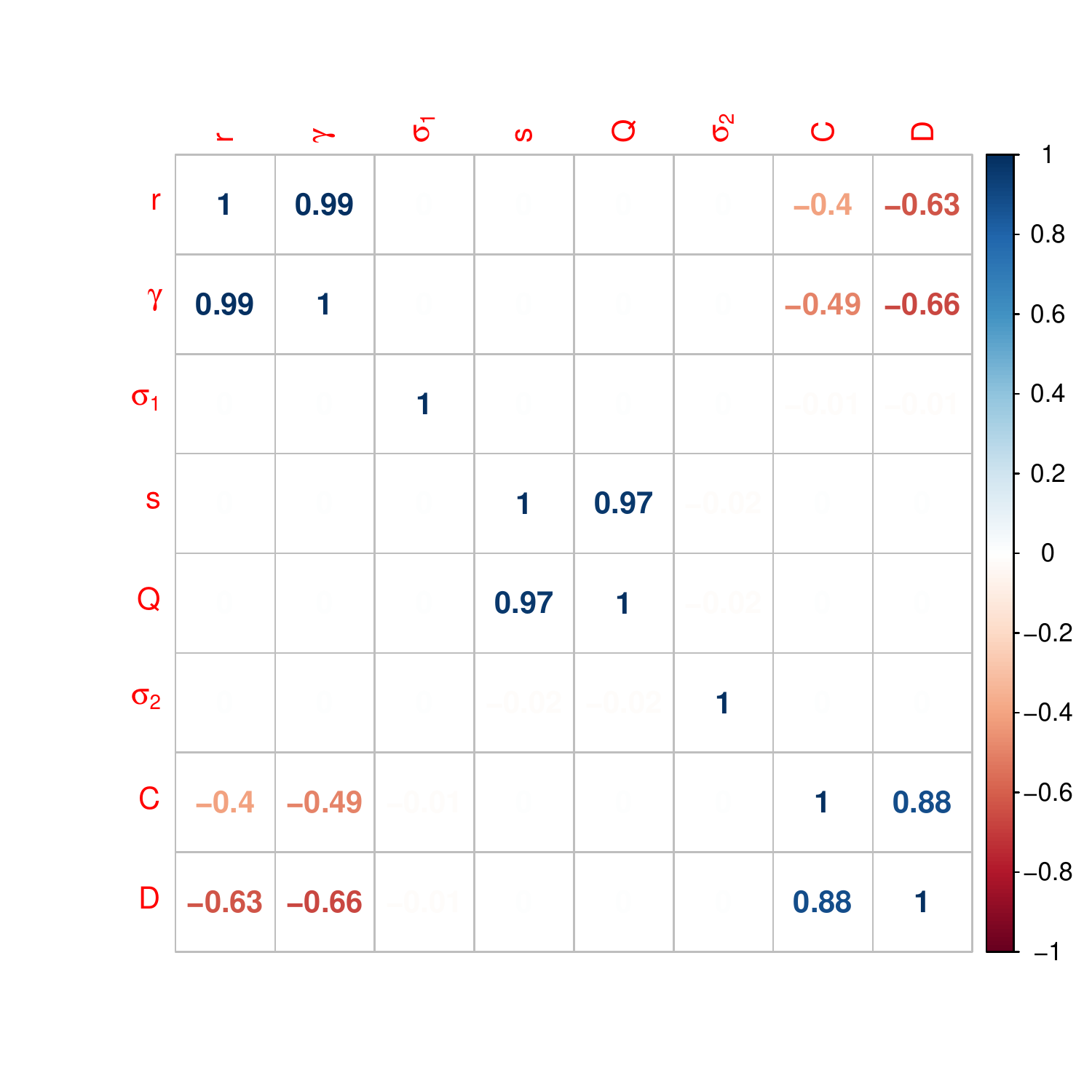}
    \caption{Without kill rate data, FP}
    \label{fig:inverseFIM2_FP}
\end{subfigure}

\begin{subfigure}[b]{0.45\textwidth}
    \centering
    \includegraphics[width=\textwidth]{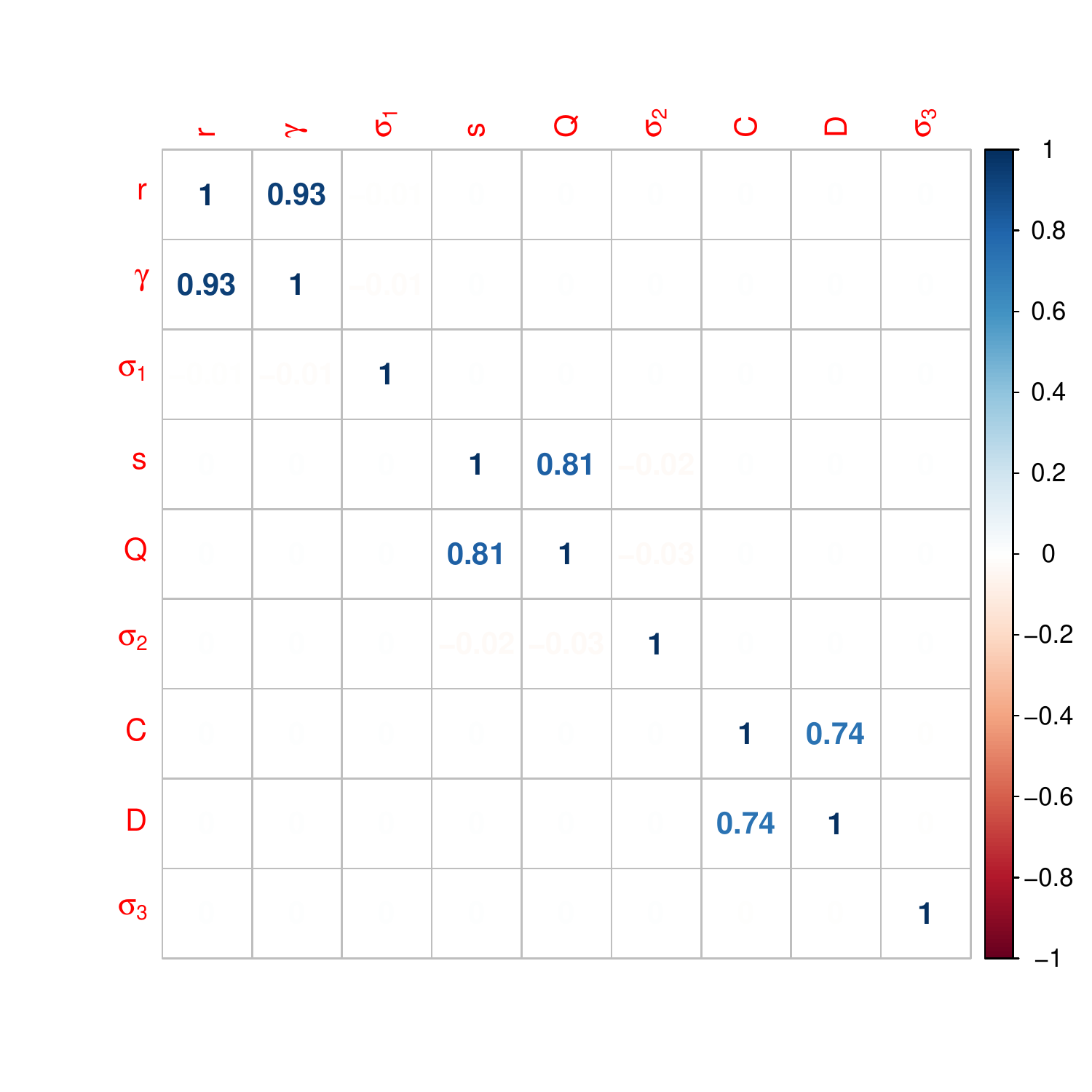}
    \caption{With kill rate data, LC}
    \label{fig:inverseFIM1_LC}
\end{subfigure}
~ 
\begin{subfigure}[b]{0.45\textwidth}
    \centering
    \includegraphics[width=\textwidth]{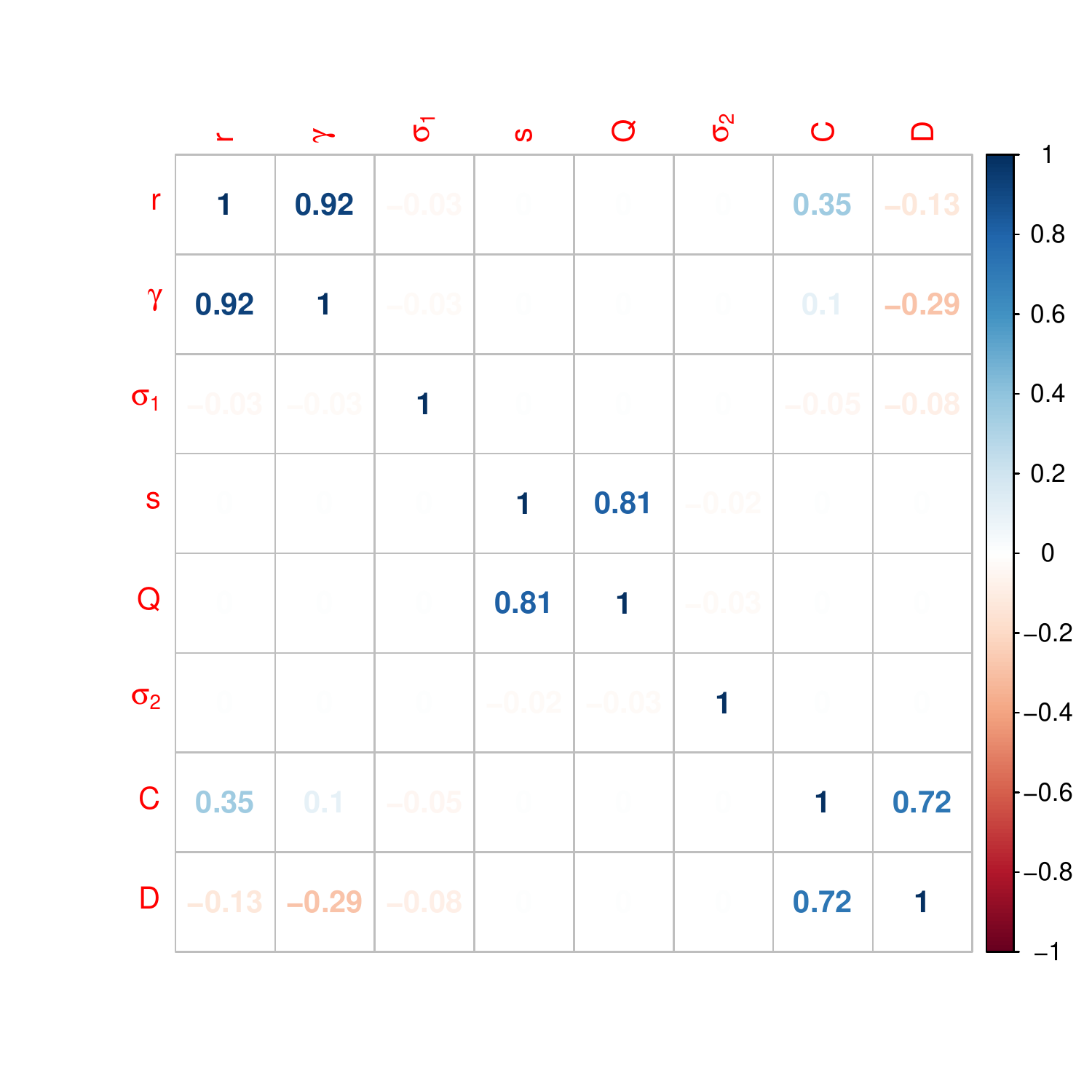}
    \caption{Without kill rate data, LC}
    \label{fig:inverseFIM2_LC}
\end{subfigure}
  \caption{Correlation matrix between parameters at the true parameter value (FP: perturbed fixed point; LC: noisy limit cycle)}
  \label{fig:correlation-matrix}
\end{figure}

The variance-covariance matrices display considerable correlation between pairs of parameters that belong to the same functional form of the model ($r$ and $\gamma$, $s$ and $Q$, $C$ and $D$), and null to weak correlation between parameters that belong to different functional forms. These within-functional forms correlations are very strong, suggesting identifiability issues since parameter estimates covary, but as we show below, these are less problematic than could be thought. Fig.~\ref{fig:correlation-matrix} shows that the within-functions correlations emerge with and without kill rate data, and for both parameter sets, the perturbed fixed point as well as the limit cycle. However, without the kill rate data, and for both parameter sets, additional between-functions correlations between the functional response $(C,D)$ and prey growth parameters ($r,\gamma$) start to emerge. 

\renewcommand{\thesubfigure}{\Alph{subfigure}}

\paragraph{Likelihood surfaces}

\begin{figure}[H]
    \centering
    \begin{subfigure}[b]{0.8\textwidth}
        \centering
        \includegraphics[width=\textwidth]{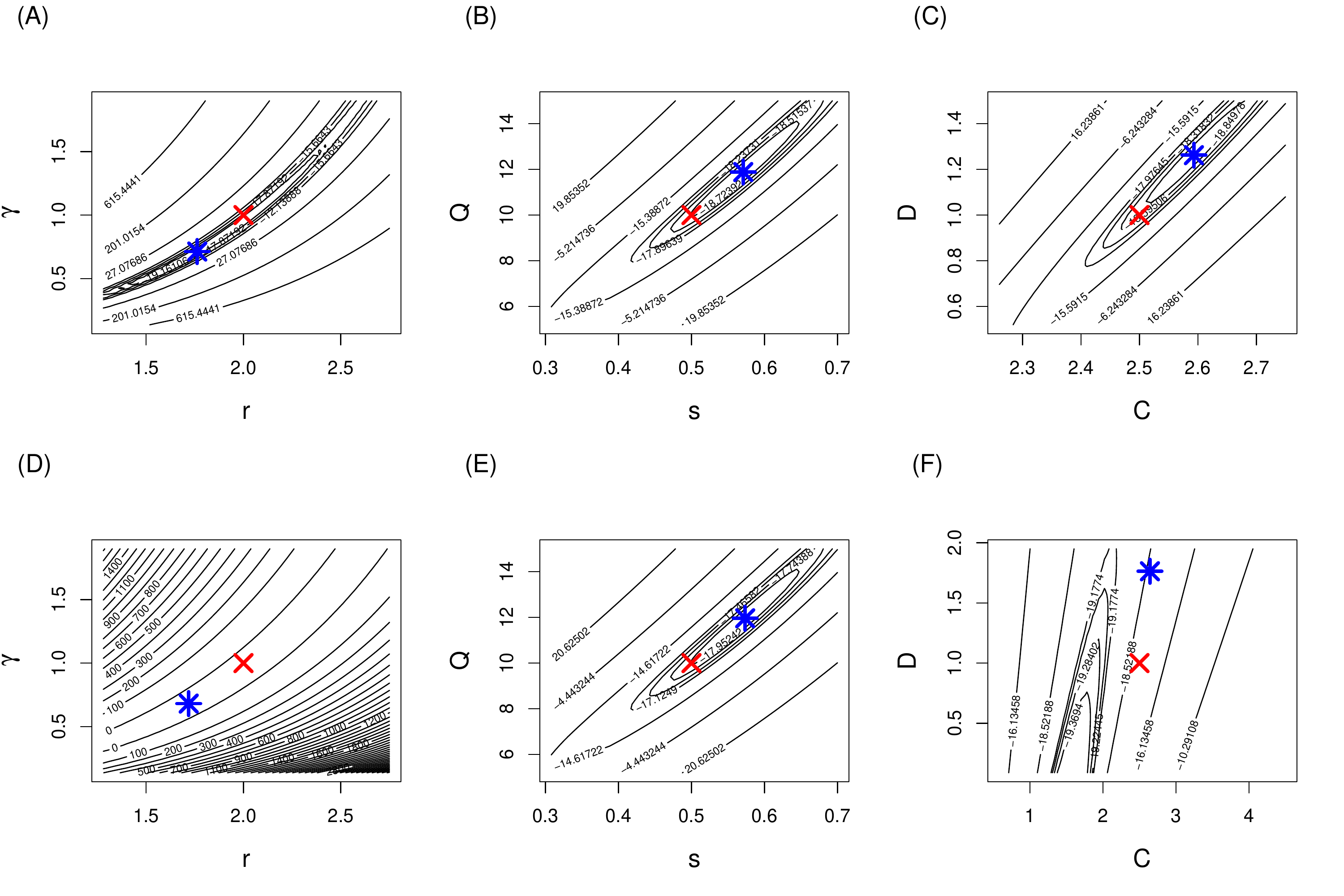}
        \caption{Perturbed fixed point parameter set}
        \label{fig:likelihood_surfaces_FP}
    \end{subfigure}

~

    \begin{subfigure}[b]{0.8\textwidth}
        \centering
        \includegraphics[width=\textwidth]{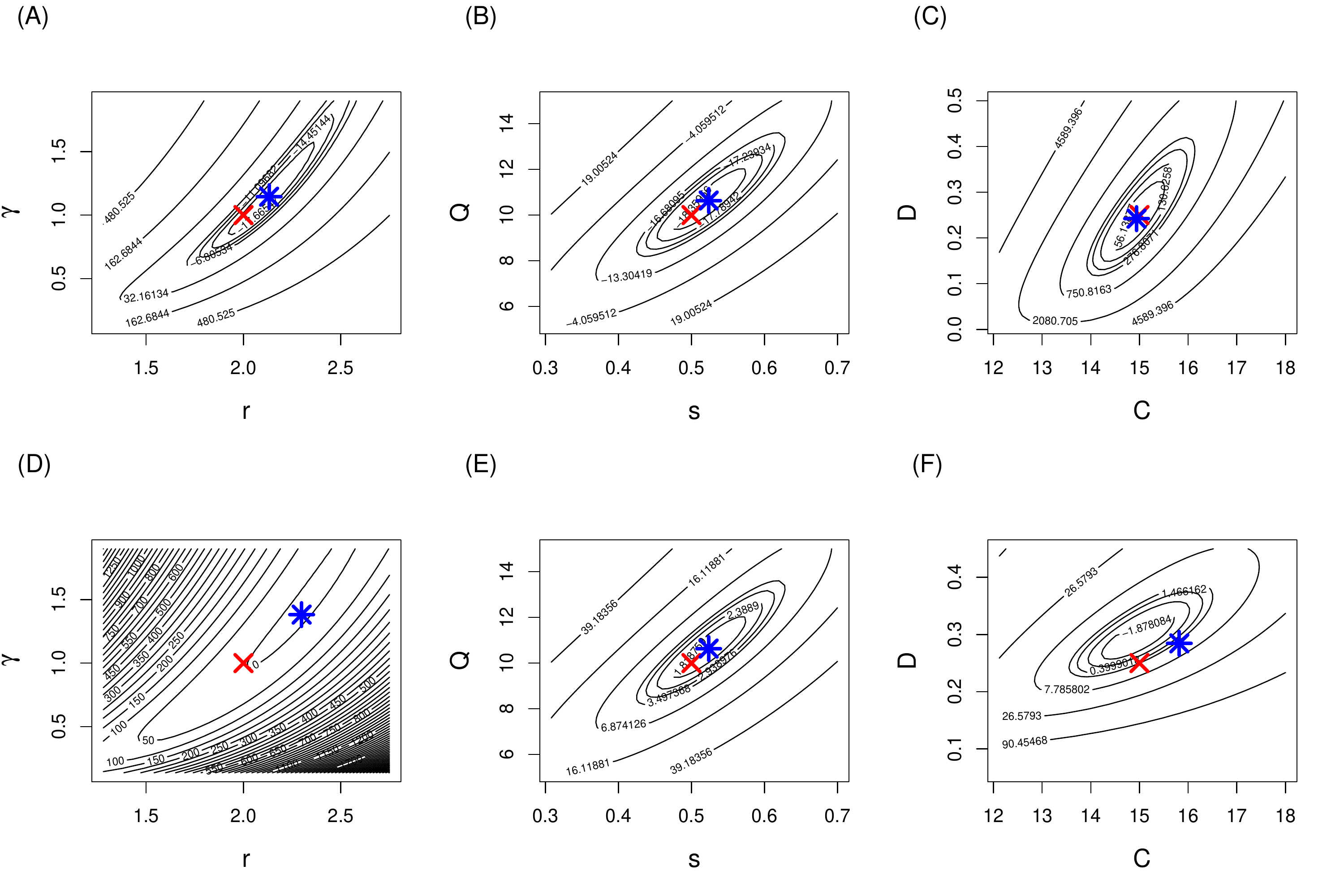}
        \caption{Noisy limit cycle parameter set}
        \label{fig:likelihood_surfaces_LC}
    \end{subfigure}
\caption{Negative log-likelihood surfaces for pairs of parameters. In columns, pairs of parameters $(r,\gamma)$, $(s,Q)$ and $(C,D)$; upper row: with kill rate data (panels A-B-C); bottom row: without kill rate data (panels D-E-F). Both parameter sets are considered. True parameter values are shown as red crosses and point estimates as blue stars.}
\label{fig:likelihood_surfaces}
\end{figure}

Based on the variance-covariance matrix results, that are local to the point in parameter space considered (here, the true parameter set), we further explore the relationships between parameters by examining the (negative) log-likelihood surfaces for the pairs of parameters that were correlated $(r,\gamma)$, $(s,Q)$ and $(C,D)$. Very similar results can be obtained using directly the residual sum of squares, without taking into account the residual variances $\sigma_i^2$ \citep{barraquand_gimenez_2020}; we present only the negative log-likelihood in Fig.~\ref{fig:likelihood_surfaces} for consistency and simplicity. 

\renewcommand{\thesubfigure}{\roman{subfigure}}

In spite of the correlation between parameters, we see in Fig.~\ref{fig:likelihood_surfaces_FP} that a maximum can be found on the likelihood when kill rate data is provided, as the eigenvalues of the FIM suggested. However, when the kill rate data is not included (D-E-F in Fig.~\ref{fig:likelihood_surfaces_FP}), the prey growth rate and functional response parameter values do not correspond clearly to minima of the negative log-likelihood. Fig.~\ref{fig:likelihood_surfaces_LC} shows the same plots for the  noisy limit cycles parameter set, where with or without the kill rate data, the minima are very well-defined. 




\subsubsection{Bayesian analysis}

\begin{figure}
    \centering
    \begin{subfigure}[b]{0.8\textwidth}
    \centering
    \includegraphics[width=\textwidth]{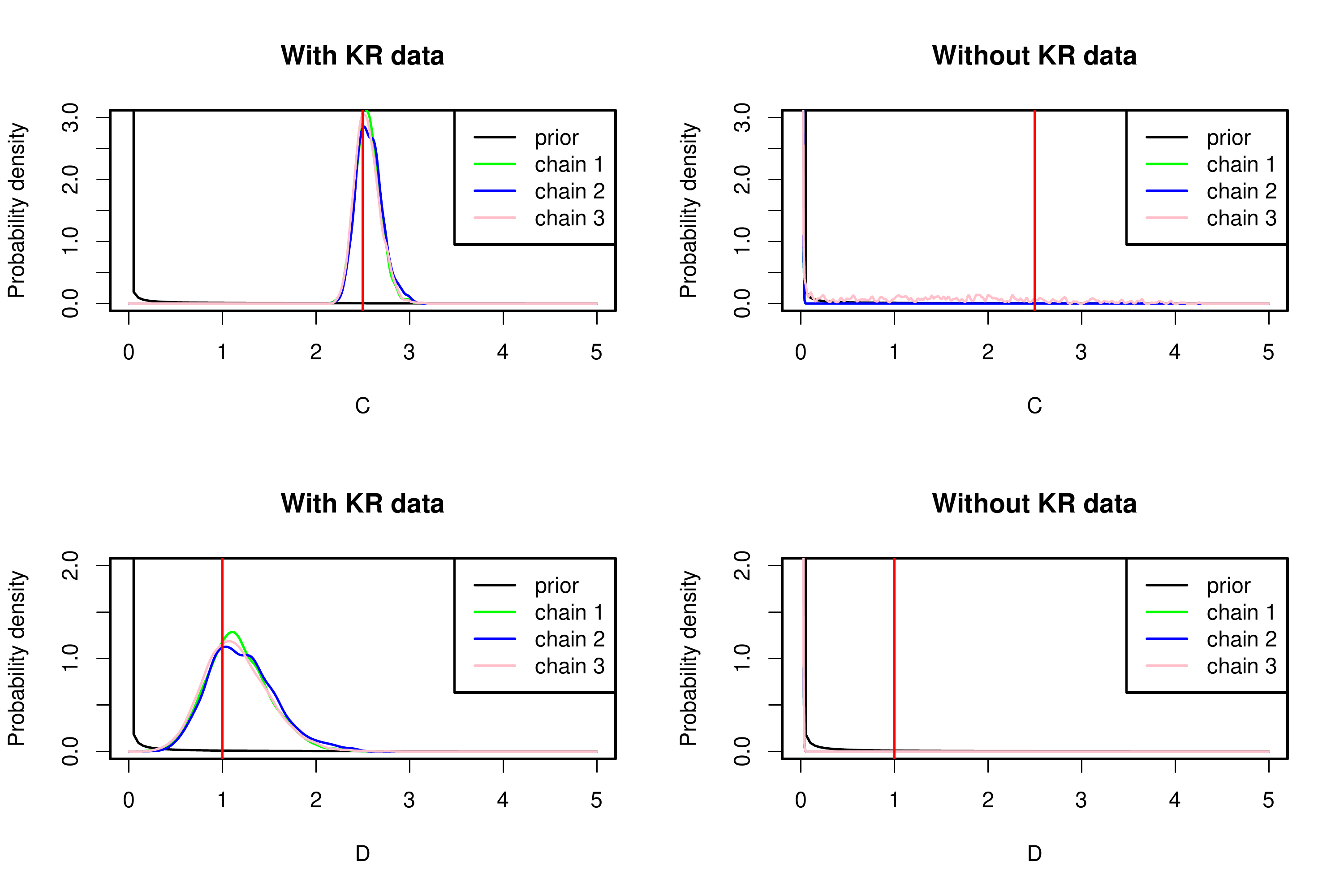}
    \caption{Perturbed fixed point dataset}
    \label{fig:PPO_FP}
     \end{subfigure}
     
    \begin{subfigure}[b]{0.8\textwidth}
    \centering
    \includegraphics[width=\textwidth]{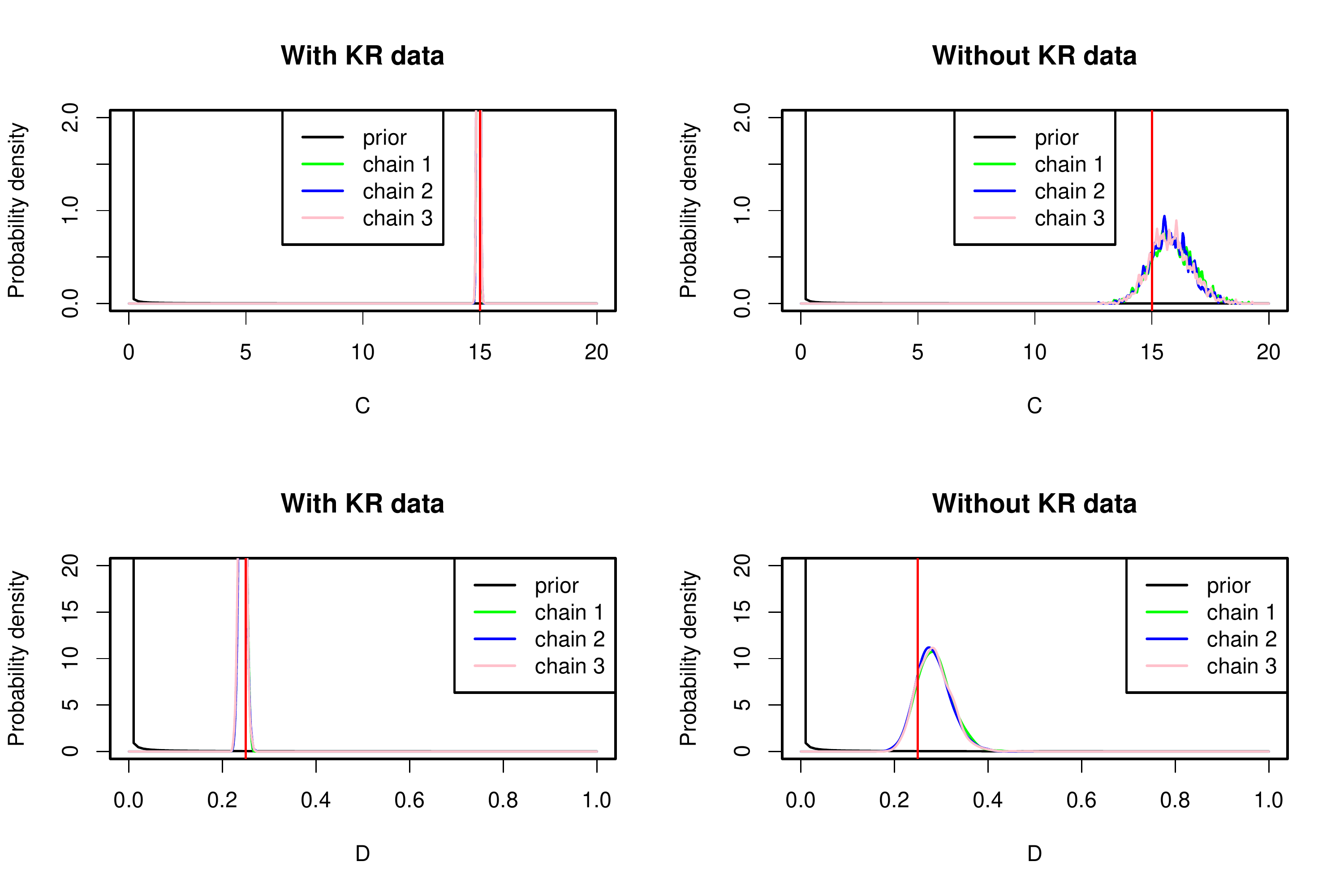}
    \caption{Noisy limit cycle dataset}
    \label{fig:PPO_LC}
    \end{subfigure}
    \caption{Prior-posterior overlap with kill rate (KR) data (left) and without kill rate data (right). Top row, parameter $C$; bottom row, parameter $D$. True parameter values are red vertical lines. 
    }
    \label{fig:PPO}
\end{figure}

\paragraph{Prior-posterior overlap} For one simulation, we now plot the overlap between prior and posterior distributions in Fig.~\ref{fig:PPO_FP}. For parameters $C$ and $D$, the near-absence of overlap with kill rate data and near-perfect prior-posterior overlap in the case without kill rate data demonstrates the practical unidentifiability of the functional response parameters without kill rate (KR) data, for the perturbed fixed point parameter set.

These results hold for all simulations, as exemplified by the reported $(C,D) = (0,0)$ estimates for the perturbed fixed point case without kill rate data: the posterior mass concentrates in this case at the mode of the prior (see section Data availability scenarios). Only in the noisy LC case can we identify correctly in practice $C$ and $D$ (Fig.~\ref{fig:PPO_LC}) without the use of kill rate data. And, even though all panels of Fig.~\ref{fig:PPO_LC} demonstrate low prior-posterior overlap, the posterior mass for both $C$ and $D$ parameters is much more narrowly distributed with kill rate data, so that the final precision of the estimates is much improved with kill rate data, no matter what parameter set is considered. 

\paragraph{Pairwise correlations in the joint posterior distribution}

\renewcommand{\thesubfigure}{\Alph{subfigure}}

\begin{figure}[h]
\centering
\begin{subfigure}[b]{0.4\textwidth}
    \centering
    \includegraphics[width=\textwidth]{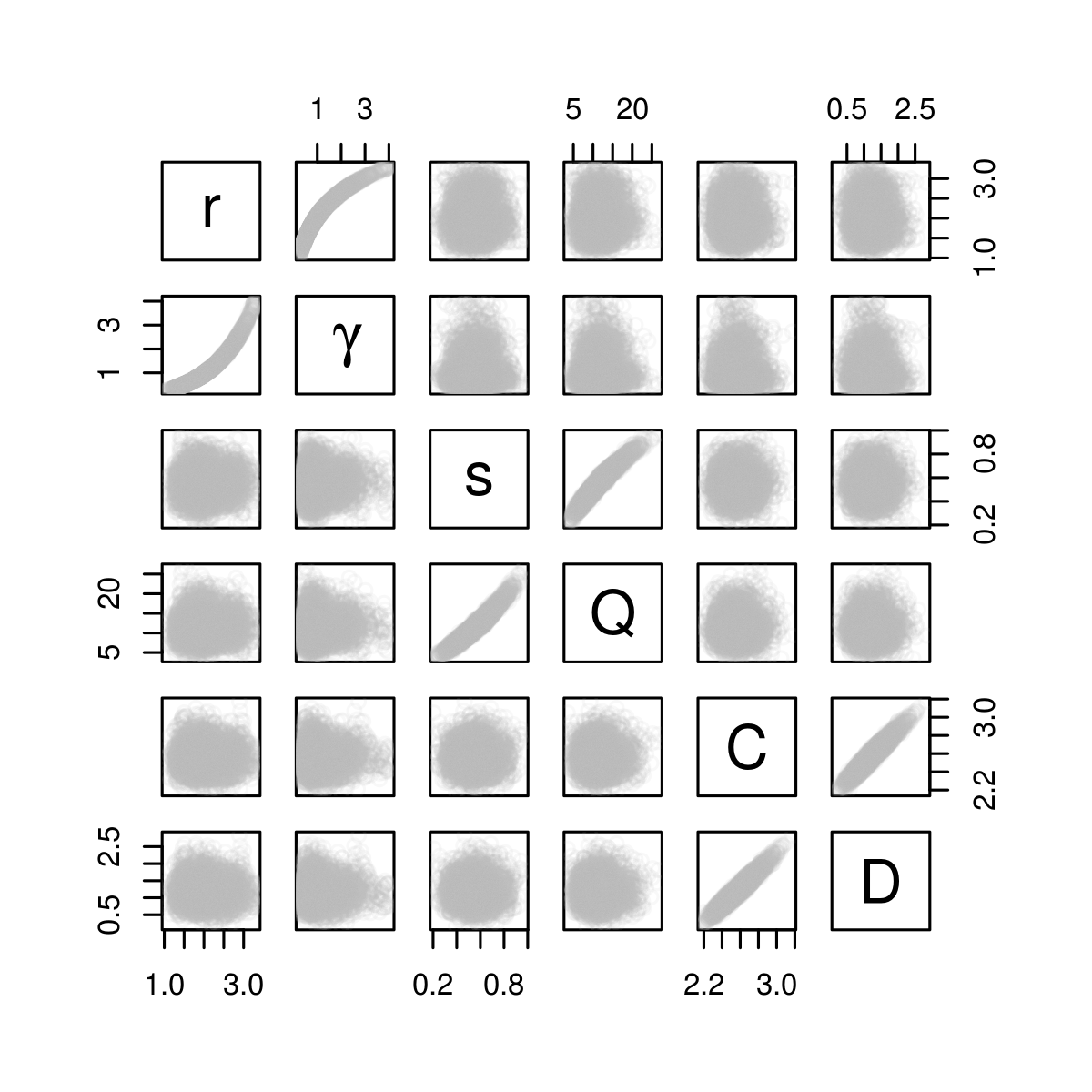}
    \caption{With kill rate data}
    \label{fig:correlation_posteriors_wKR}
\end{subfigure}
~
\begin{subfigure}[b]{0.4\textwidth}
    \centering
    \includegraphics[width=\textwidth]{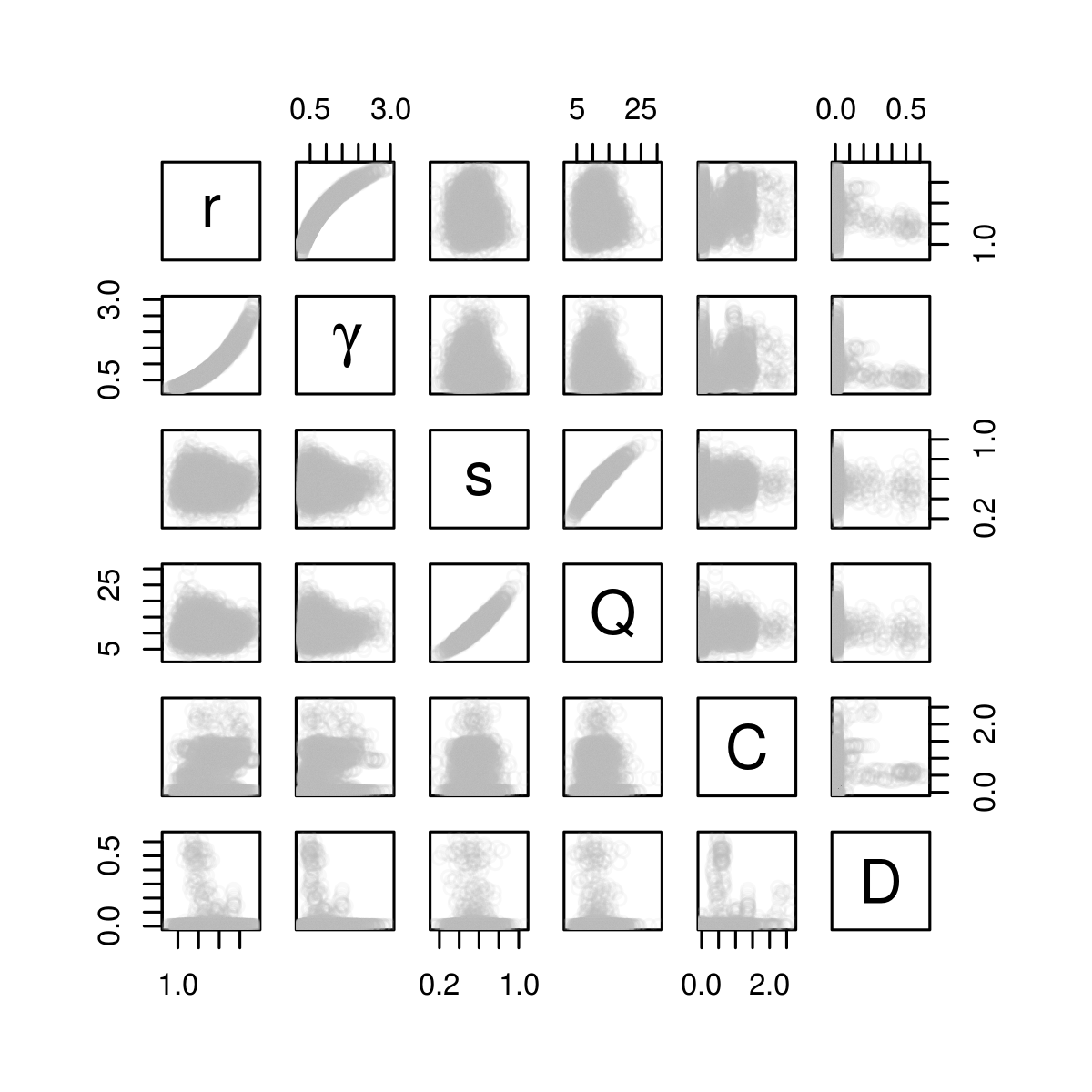}
    \caption{Without kill rate data}
    \label{fig:correlation_posteriors_woutKR}
\end{subfigure}
 \caption{Biplots of MCMC samples per pair of parameters, perturbed fixed point (FP) parameter set.}
\label{fig:correlation_posteriors_FP}
\end{figure}

The correlation in the posteriors mirrors the frequentist results obtained on the variance-covariance matrix, and is shown in Fig. \ref{fig:correlation_posteriors_FP} for the fixed point parameter set. Pairs of parameters belonging to the same functional form are not estimated as independent. The correlation between parameters typically becomes worse without the kill rate data (Fig.~\ref{fig:correlation_posteriors_woutKR}). Very similar results are presented in Appendix~\ref{sec:noisyLC-results} for the noisy limit cycle parameter set.

\paragraph{Estimation of curves rather than single parameters}


\begin{figure}[h]
    \centering
    \includegraphics[width=0.8\textwidth]{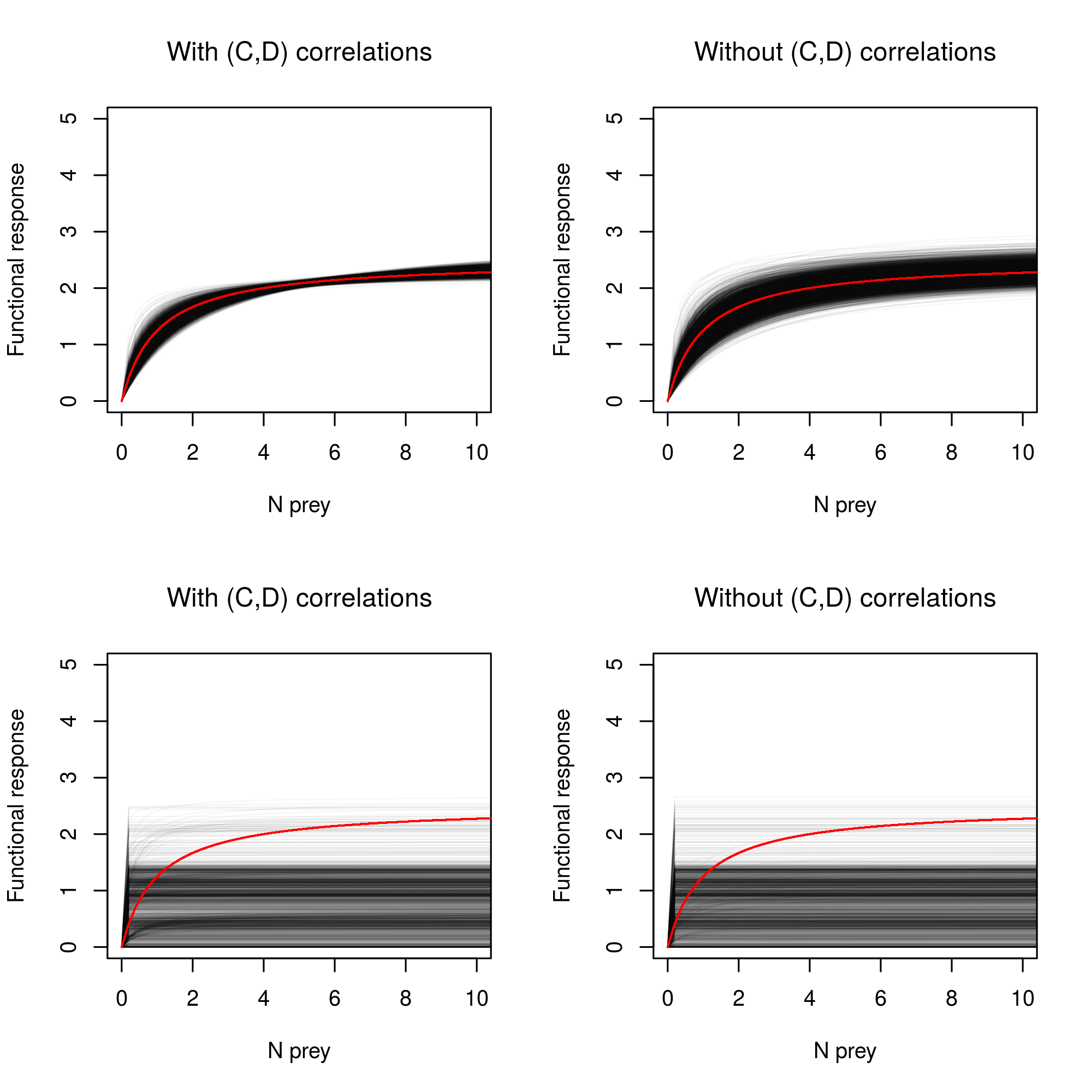}
    \caption{Estimated average functional response with vs without correlation between parameters, with (top row) and without (bottom row) kill rate data, for the FP parameter set. Thin grey lines each represent one iteration of the MCMC chains. These are overlayed on top of each other, so that the black envelope represents likely functional response values. The thick red line is the true, simulated average functional response (note that Gaussian noise is added to that curve in the stochastic predator-prey model, eq.~\eqref{eq:FR}).}
    \label{fig:estimation_curves_FR_FP}
\end{figure}

\begin{figure}[h]
    \centering
    \includegraphics[width=0.8\textwidth]{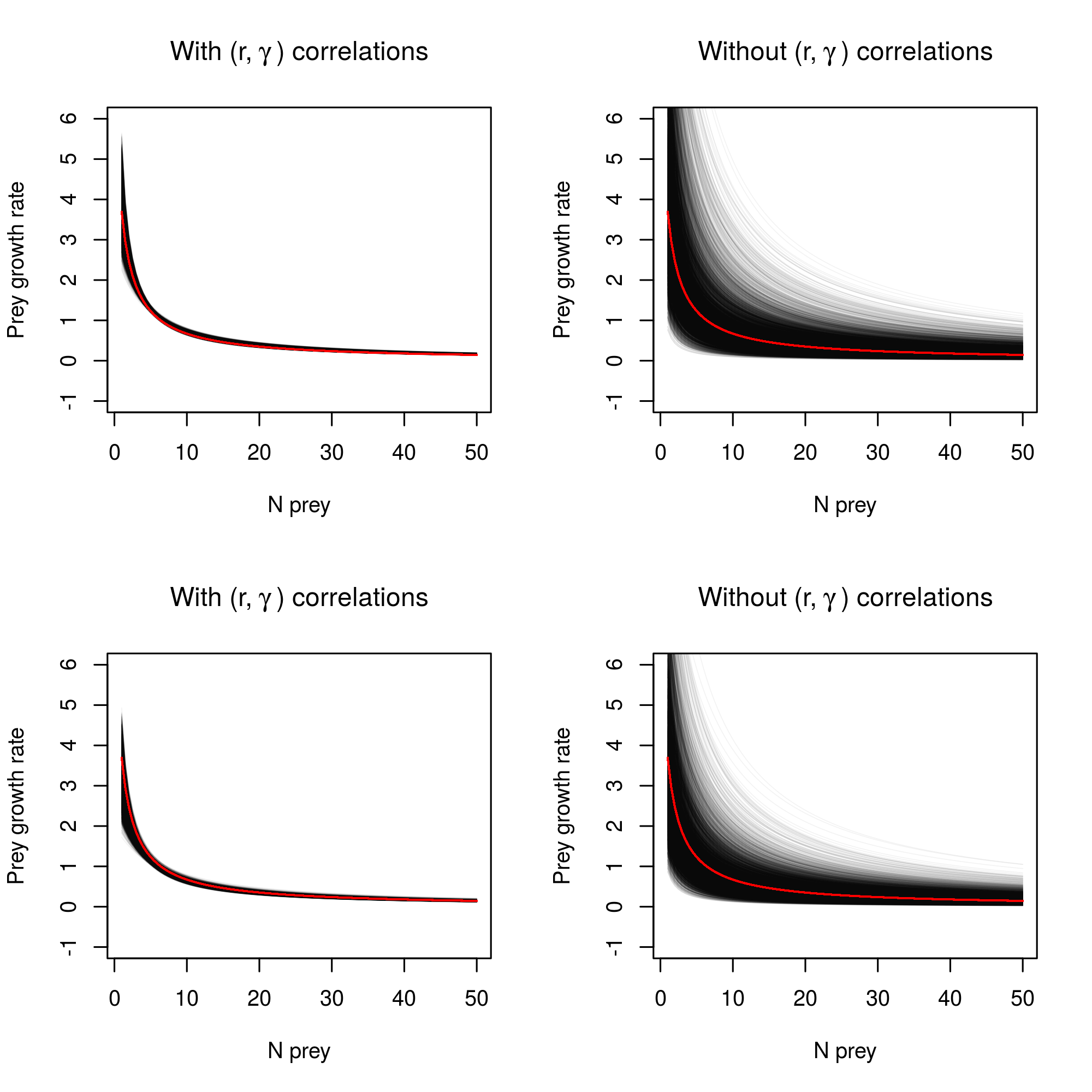}
    \caption{Estimated average prey growth rate-density curve with vs without correlation between parameters, with (top row) and without (bottom row) kill rate data, for the FP parameter set. The true curve is drawn in red.}
    \label{fig:estimation_curves_preydd_FP}
\end{figure}

Examination of the shape of the growth rate-density curves for both predator and prey, as well as that of the average functional response, reveals interesting model properties. Namely, the correlation between parameters belonging to the same function serves as to maximize the precision of the estimation of the function as a whole. In other words, we estimate well the function and the parameter pair $(C,D)$ but not the $C$ and $D$ parameters separately. This can be seen by overlay of the functions for each iteration in the Markov chain, with vs without the correlations between parameters (Fig.~\ref{fig:estimation_curves_FR_FP}). Correlations are removed by permutating independently the samples from posterior distributions for $C$ and $D$. We see that the cloud of lines that corresponds to the posterior distribution of the average functional response itself is much thinner when the correlations are included (when the functional response can be estimated, that is, with kill rate data in Fig.~\ref{fig:estimation_curves_FR_FP}). The same holds for the prey growth rate-density curve in Fig.~\ref{fig:estimation_curves_preydd_FP}, which can be well estimated with or without the kill rate data. Therefore, even though correlations between parameters make individual parameters more difficult to identify, the opposite is actually true for functional forms: correlation between parameters are instrumental in making the functional forms of the model more precise. A reparameterization proposed in the next section allows to get rid of most of those correlations, but does not markedly improve the estimation of the functional curves, when compared to the present parameterization with posterior correlations included. 

We show here plots for the prey density-dependence curve but similar results can be obtained for the predator density-dependence curve. Similar plots are also provided in Appendix~\ref{sec:noisyLC-results} regarding the noisy LC parameter set. 

\paragraph{Re-parameterization of the model}

In this section, we consider a re-parameterized form of the model given by a prey carrying capacity $K=(e^r-1)/\gamma$
as in the classical Beverton-Holt model, and a transformation of $q$ to $q(e^s-1)$ following the same logic. This is known to help identifiability \citep{lele2019consequences}. The functional response has for its part being changed from $\frac{CN}{D+N}$ (classic Monod or Michaelis-Menten formulation) to the more mechanistic Holling parameterization $\frac{aN}{1+ahN}$. We then arrive at the equations:

\begin{align}
N_{t+1} & = N_{t}\frac{e^{r+\epsilon_{1t}}}{1+(e^r-1) N_{t}/K}\exp\left( \left(-\frac{a N_t}{1+a h N_t} +\epsilon_{3t}\right)\frac{P_{t}}{N_{t}}\right),\,\epsilon_{1t}\sim\mathcal{N}(0,\sigma_{1}^{2})\label{eq:prey_discreteLeslieMay_reparam}
\end{align}

\begin{align}
P_{t+1} & = P_{t}\frac{e^{s+\epsilon_{2t}}}{1+(e^s-1)qP_{t}/N_{t}},\,\epsilon_{2t}\sim\mathcal{N}(0,\sigma_{2}^{2})\label{eq:predator_discreteLeslieMay_reparam}
\end{align}

\begin{figure}[htbp]
\centering
\begin{subfigure}[b]{0.45\textwidth}
    \centering
    \includegraphics[width=\textwidth]{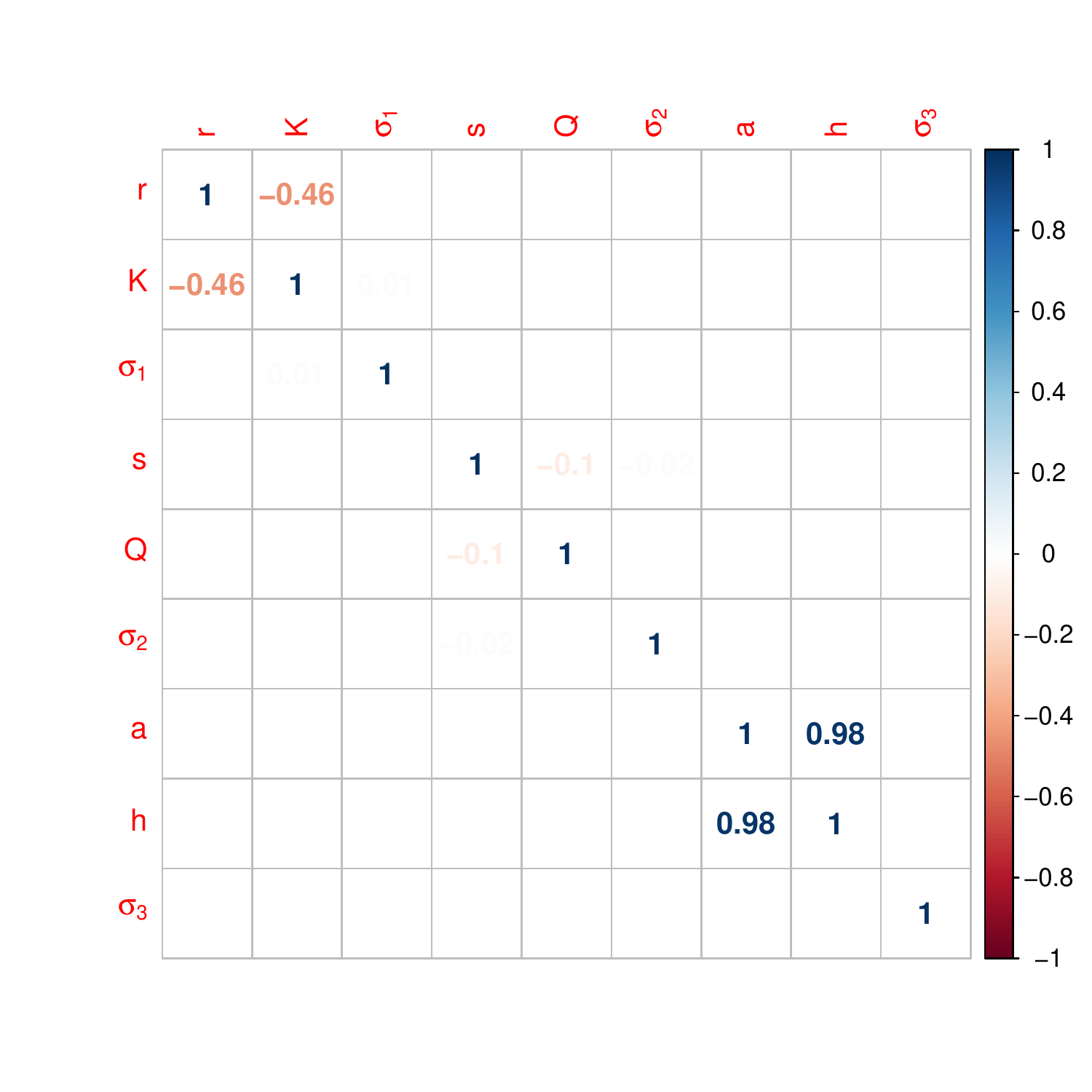}
    \caption{With kill rate data, FP}
    \label{fig:inverseFIM1_FP-reparam}
\end{subfigure}
~ 
\begin{subfigure}[b]{0.45\textwidth}
    \centering
    \includegraphics[width=\textwidth]{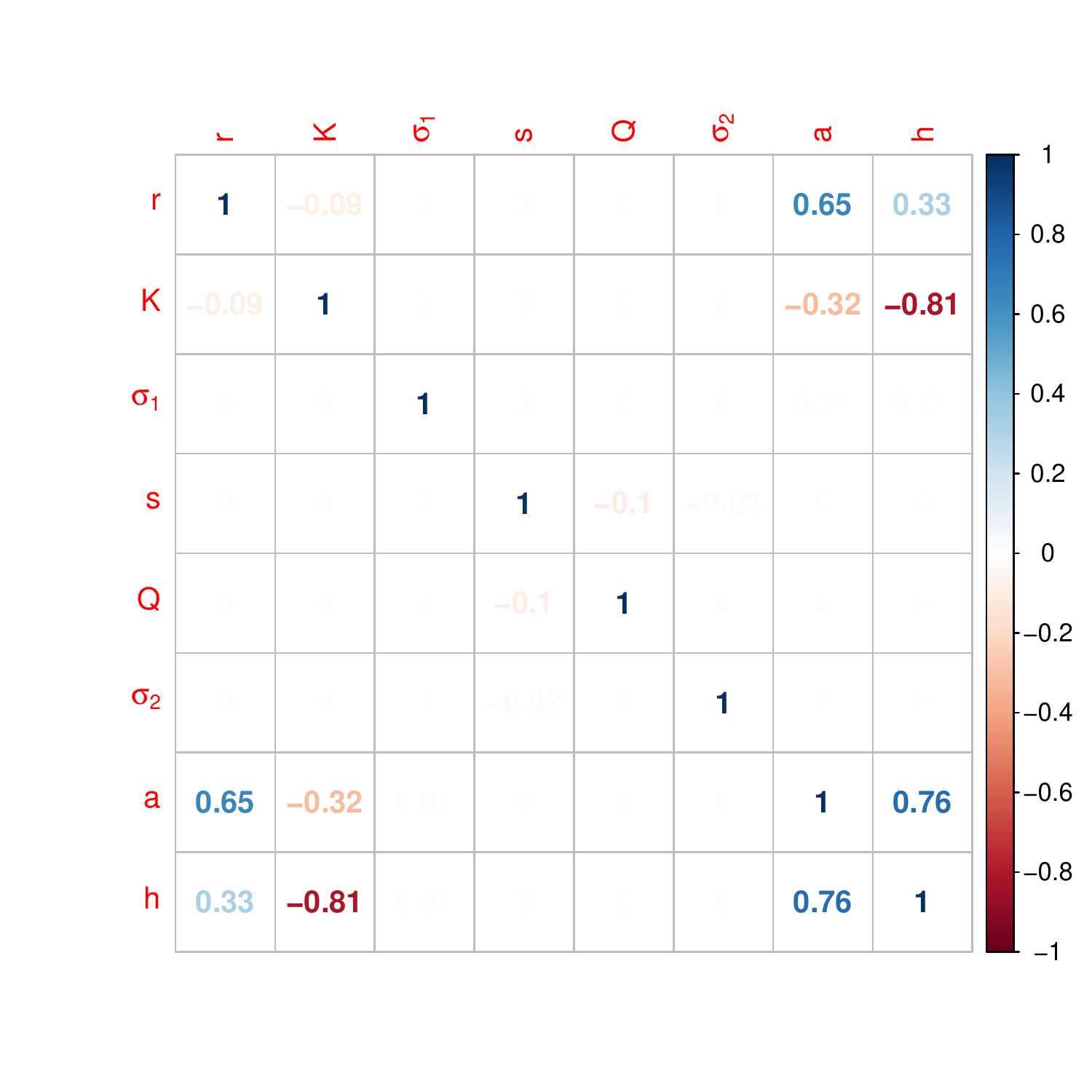}
    \caption{Without kill rate data, FP}
    \label{fig:inverseFIM2_FP-reparam}
\end{subfigure}

\begin{subfigure}[b]{0.45\textwidth}
    \centering
    \includegraphics[width=\textwidth]{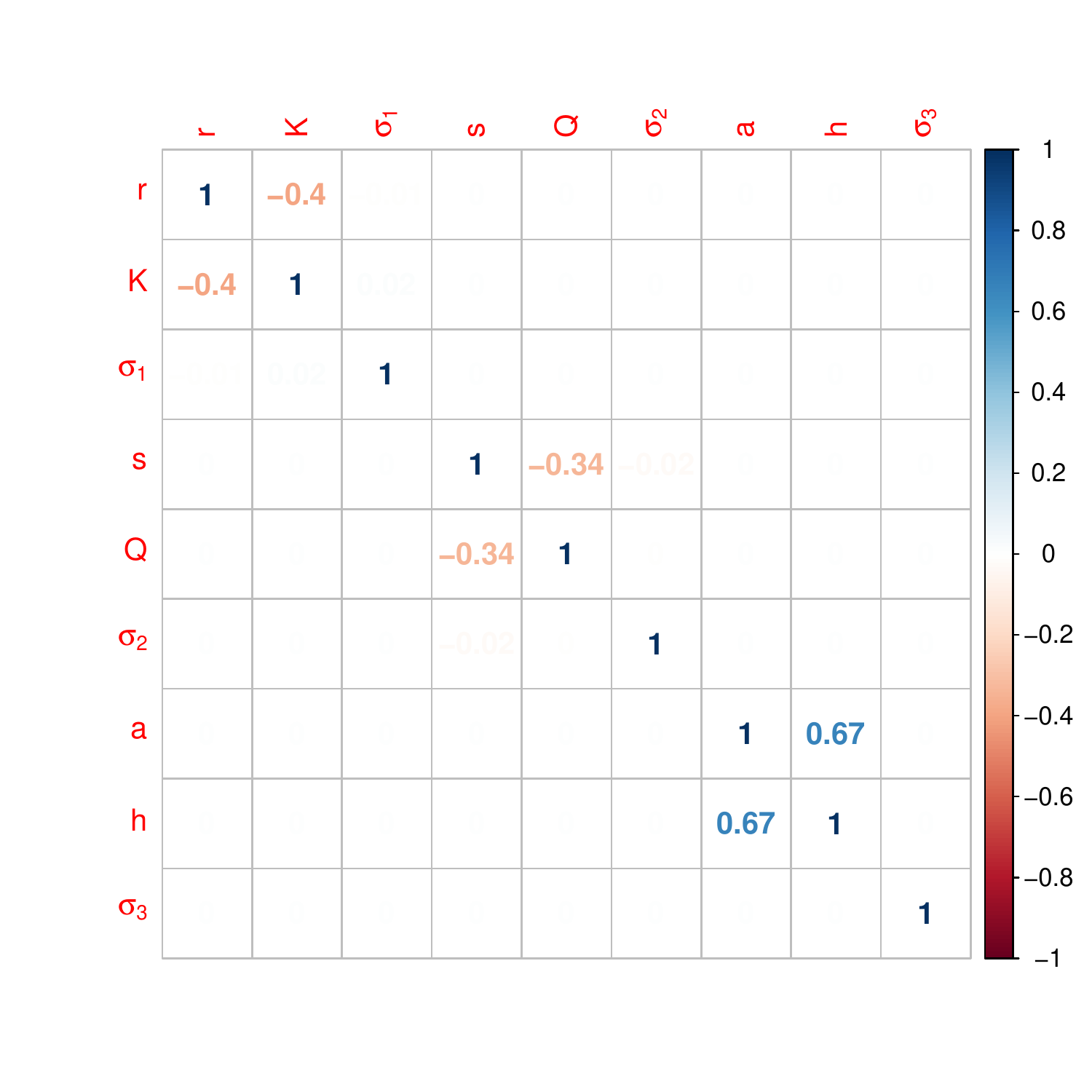}
    \caption{With kill rate data, LC}
    \label{fig:inverseFIM1_LC-reparam}
\end{subfigure}
~ 
\begin{subfigure}[b]{0.45\textwidth}
    \centering
    \includegraphics[width=\textwidth]{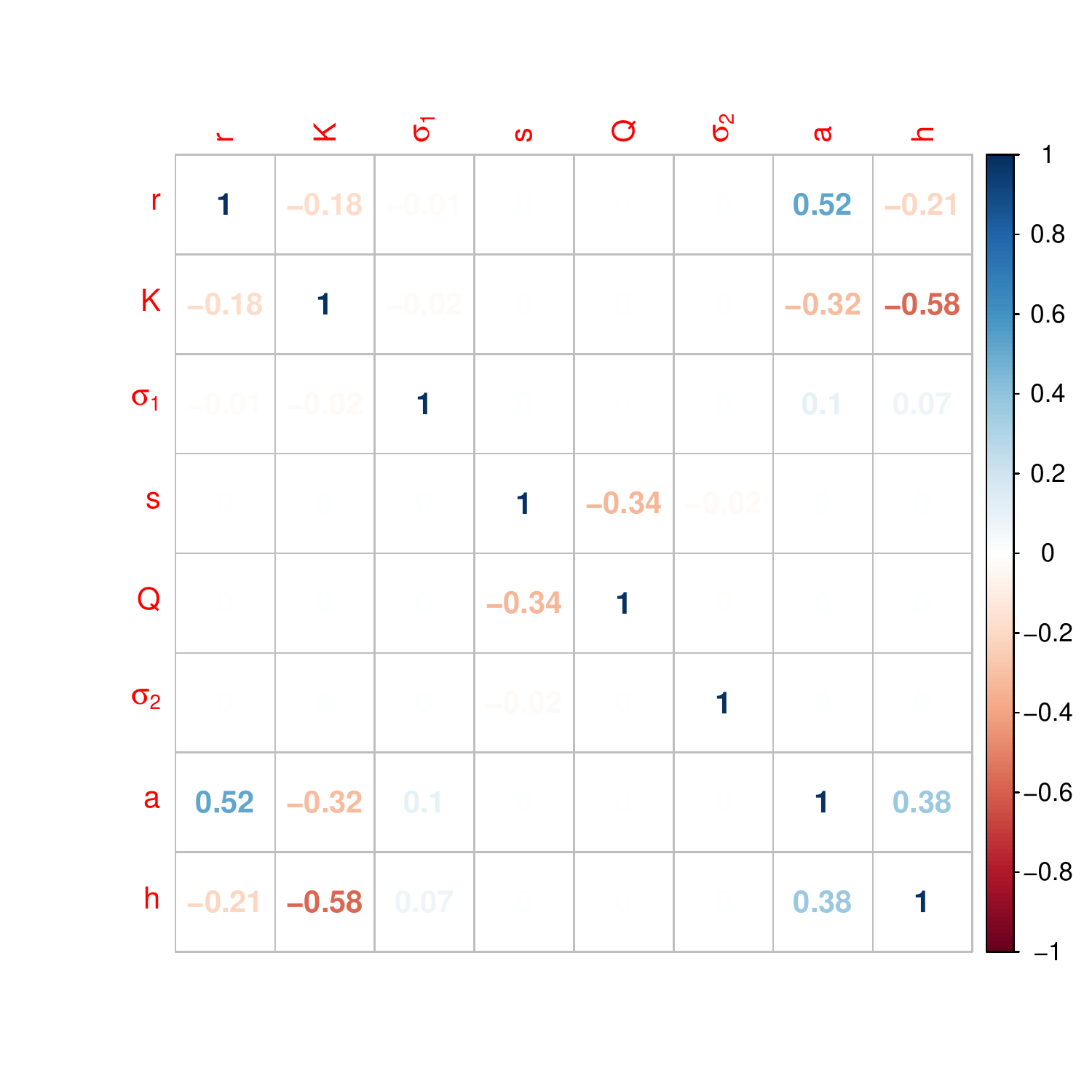}
    \caption{Without kill rate data, LC}
    \label{fig:inverseFIM2_LC-reparam}
\end{subfigure}
  \caption{Correlation matrix between parameters at the true parameter value (FP: perturbed fixed point; LC: noisy limit cycle), for the re-parameterized Leslie predator-prey model.}
  \label{fig:correlation-matrix-reparam}
\end{figure}

This re-parameterization conserves the exact same number of parameters. The correlations between pairs of parameters have decreased with the reparameterized model, as exemplified by the correlation matrices derived from the new Fisher Information Matrix at the true parameter value shown in Fig.~\ref{fig:correlation-matrix-reparam} (similar results have been obtained in a Bayesian setting, Appendix~\ref{sec:reparam-results}).

In this case, the reparameterization in terms of carrying capacities worked well, especially for the predator, but less so for the Holling-type parameterization of the functional response. However, reparameterization did not solve in any way the unidentifiability of the model without the kill rate data, in the perturbed fixed point case (Appendix~\ref{sec:reparam-results}). Moreover, it is important to realize that while improving the identifiability of \emph{individual parameters}, such reparameterization does not markedly improve the quality of the estimation of the functional response and growth rate curve \emph{per se} (see Appendix~\ref{sec:reparam-results}). We therefore have no reason to strongly favour the reparameterized form over the original form of the model given by eqs.~\eqref{eq:prey_discreteLeslieMay}-\eqref{eq:predator_discreteLeslieMay}.

\subsubsection{Data availability scenarios}

\begin{figure}[h]
    \centering
    \includegraphics[width=0.9\textwidth]{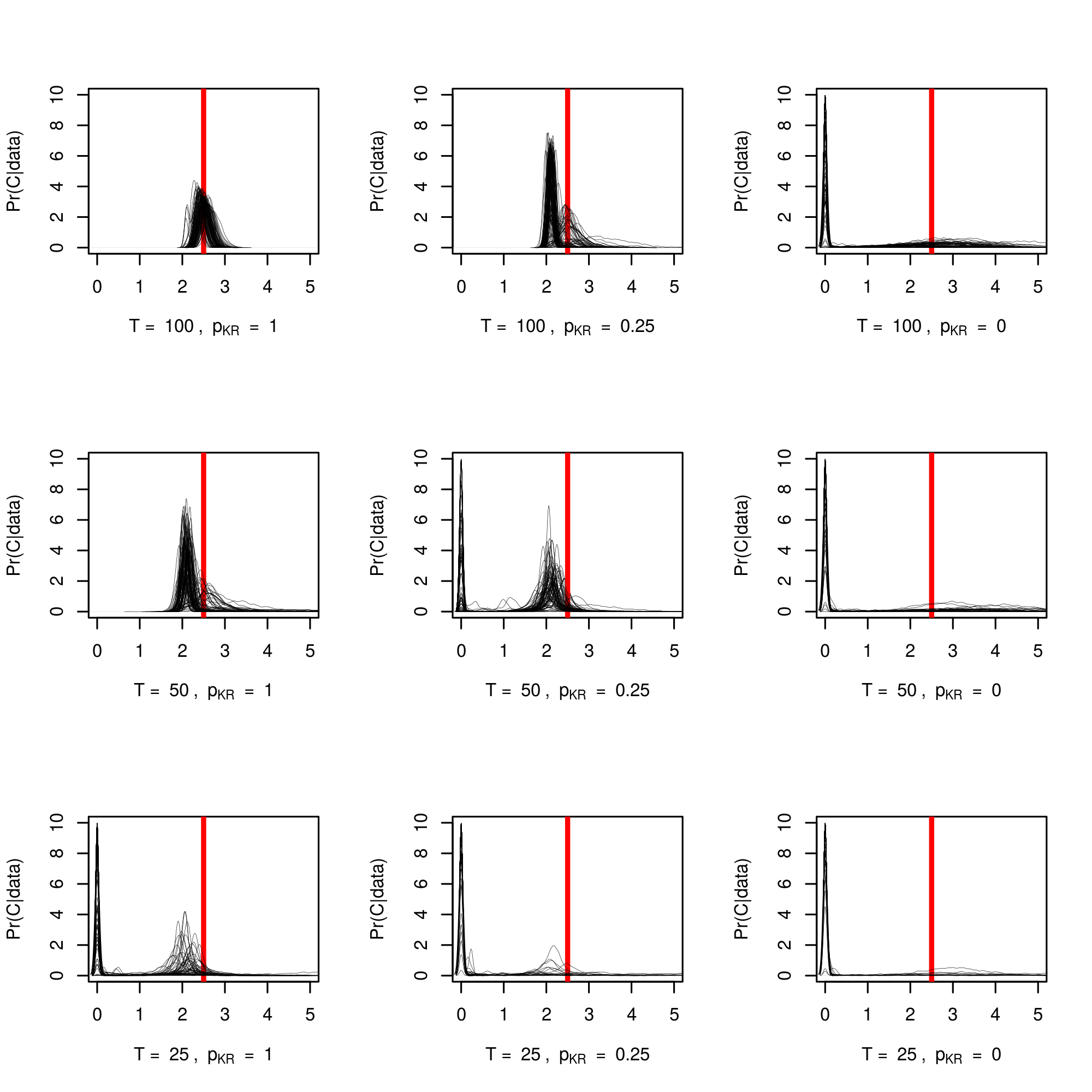}
    \caption{Posterior probability densities for the parameter $C$, for 100 simulations (each black line is one posterior probability density) using the FP parameter set. The vertical red line materializes the true parameter value.}
    \label{fig:PosteriorsC_FP}
\end{figure}

\begin{figure}[h]
    \centering
    \includegraphics[width=0.9\textwidth]{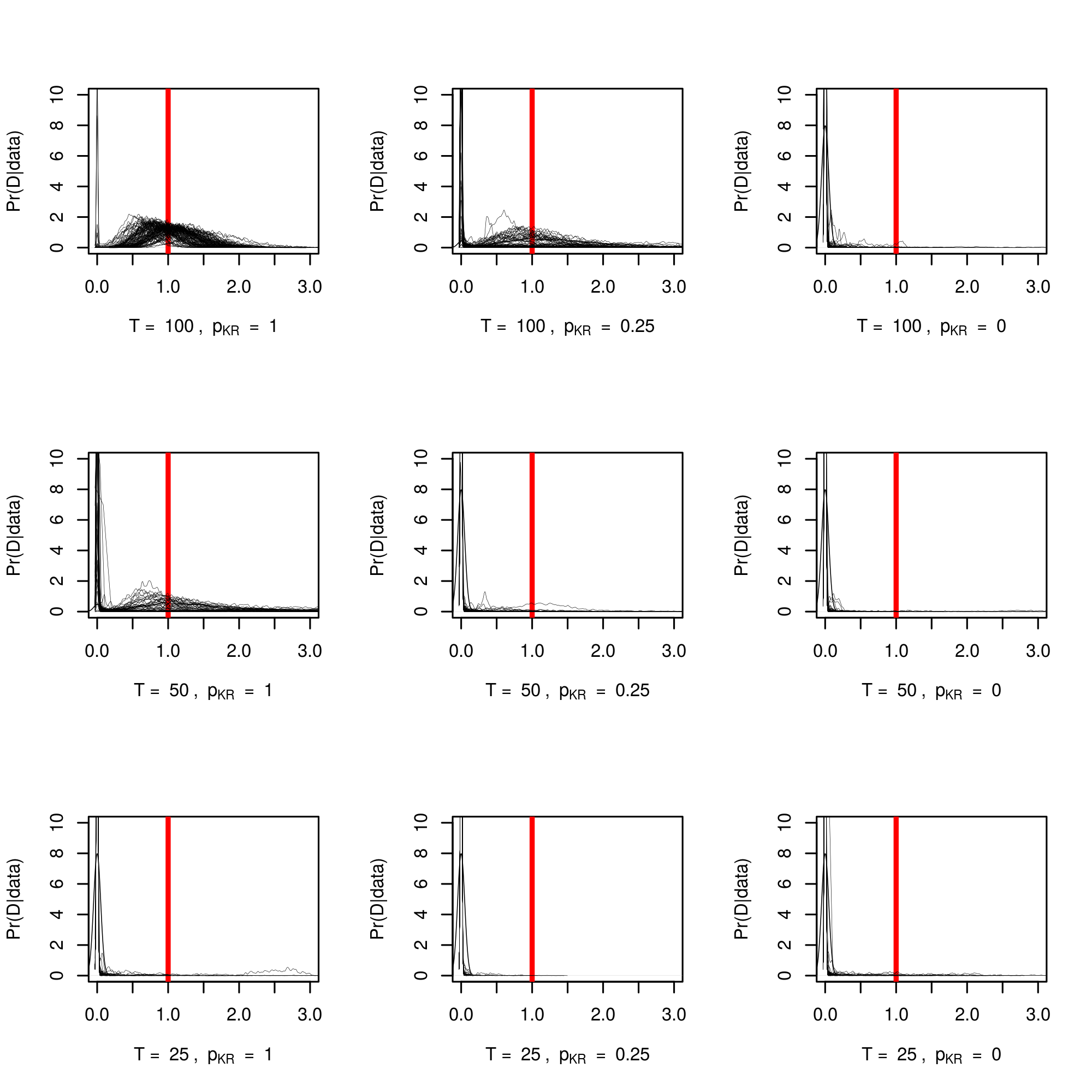}
    \caption{Posterior probability densities for the parameter $D$, for 100 simulations (each black line is one posterior probability density) using the FP parameter set. The vertical red line materializes the true parameter value.}
    \label{fig:PosteriorsD_FP}
\end{figure}

Our numerical experiment crosses 3 time series lengths and 3 data availability scenarios (fraction of time units for which the kill rate is available, 0\%, 25\% and 100\%). 
Figs.~\ref{fig:PosteriorsC_FP} and \ref{fig:PosteriorsD_FP} illustrate the bias and precision for the $C$ and $D$ parameters of the functional response, for the fixed point (FP) parameter set. The previous Bayesian results using a single simulated dataset are confirmed: total absence of kill rate data prevents proper estimation of parameters $C$ and $D$ (for all 100 simulated FP datasets). However, adding kill rate data for only one fourth of the duration of the time series ($p_{KR}=0.25$) gives estimates of $C$ that are almost as precise as with kill rate data for the whole of the time series. The half-saturation constant $D$ benefits as well from the addition of kill rate data, though less than $C$ when looking at the marginal posterior distribution. But it should be kept it mind that as these parameters co-vary strongly, if $C$ is better estimated, $D$ must be too. 

Results regarding $C$ and $D$ for the noisy LC parameter set are shown in Appendix~\ref{sec:noisyLC-results}. 
For the noisy limit cycle (LC) parameter set, estimation of $C$ and $D$ is possible in absence of kill rate data but adding kill rate data greatly decreases bias and increases precision. Especially, adding kill rate data for one fourth of the time series $(p_{KR}=0.25)$ in the case $T=100$ yields bias and precision that are barely different from $p_{KR}=1$. The benefits are still large for $T=50$. For $T=25$ even with both density and kill rate data, there remains considerable uncertainty about the parameter values, with absence of practical identifiability for some simulations.   

\section{Discussion}
We have simulated data from a stochastic predator-prey model with a stochastic functional response, for parameters where the deterministic skeleton of the model would predict either convergence towards a stable fixed point or a limit cycle. These dynamic models, when environmental and interaction stochasticity were included, exhibited considerable variation in both population densities and kill rates, as could be expected from theory \citep{nisbet1982modelling,greenman2005impact}. 

We then studied the identifiability of parameters in theory (structural identifiability, using the Fisher Information Matrix) and in practice (for a given dataset of finite time series length). We examined the relative contributions to identifiability of density time series and kill rate data. We have found that kill rate data is essential to identifiability in the case of a fixed point equilibrium perturbed by stochasticity, and that it greatly reduces bias and improves precision of functional response parameters in the case of a noisy limit cycle (formally, a noisy invariant loop). Finally, we have shown that a small amount of data on kill rates might go a long way towards improving practical identifiability for both parameter sets. 

\subsection{Identifiability of the model with kill rate data}

The model is always structurally identifiable, in the sense of having a nonsingular Fisher Information Matrix. Model parameters are also always identifiable in practice when kill rate data is included, in the sense of a low prior-posterior overlap when starting with vague priors.

However, for the initial parameterization of the model, pairs of parameters belonging to the same functional forms of the models are highly correlated. These pairs are $(r,K)$, $(s,Q)$, and $(C,D)$. Substantial correlations within those pairs occur in both the (joint) posterior distributions of Bayesian estimates and the variance-covariance matrix in a frequentist setting. 

Reparameterizing using carrying capacities substantially decreased parameter correlations within density-dependence functions. Regarding the functional response, whilst \citet[][chapter 6, p. 200]{bolker2008ecological}, who fitted a Holling-type functional response, suggested that a Michaelis-Menten formulation $CN/(D+N)$ may improve identifiability compared to the Holling $aN/(1+ahN)$ formulation (our original guess as well), we did not find this here: correlation seemed slightly lower for the $(a,h)$ pair. However, some correlation between the two parameters always seemed to remain \citep[as recently found by][]{uszko2020fitting}. 
But correlations between parameters can be either problematic or beneficial depending on which model components they affect; a key question to ask when noticing correlations between estimated parameters is whether they occur \emph{between} functional forms of interest -- which is a considerable hassle -- or \emph{within} functional forms, which can even be advantageous for the estimation of the function (so long as an MLE exists). Indeed, we have shown (Figs. \ref{fig:estimation_curves_FR_FP}--\ref{fig:estimation_curves_preydd_FP}) that correlations between functional response parameters are instrumental in making the estimated functional response closer to the ground truth. 

\citet{kao2018practical} found, in an epidemiological model fitted on short incidence time series, that in spite of the unidentifiability of individual transmission parameters (that were very strongly correlated), they could estimate the basic reproduction number successfully. Our setup here is less extreme as even with the correlations, the MLE is well-defined (for $T=100$ at least), thus it is doable to estimate the parameters individually with kill rate data. Still, their findings resonate with ours: while the values for the $C$ and $D$ parameters should be interpreted individually with caution due to the correlations in the estimates (unless the posterior distribution is very narrow), the functional response average value and shape can always be estimated with precision when population counts are complemented with kill rate data. In other words, $C$ and $D$ form a \emph{practically identifiable parameter combination}, \emph{sensu} \citet{eisenberg2014determining}. Similar results have been obtained by \citet{uszko2020fitting} in the context of functional response model fitting (without population dynamics). Estimation of functional forms, as in the present article, or aggregate properties, as in \citet{kao2018practical}, could therefore be given primacy over estimation of individual parameters, that are sometimes of more limited ecological significance. 

\subsection{Identifiability without kill rate data}

The absence of kill rate data can substantially compromise the practical identifiability of the functional response (and its component parameters), but structural identifiability (sensu FIM) is always maintained. Practical unidentifiability occurs for simulations with a perturbed fixed point, showing overdamped convergence to a point equilibrium in absence of noise, which results in fluctuations around the fixed point with noise (fluctuations are non-cyclic here). In the main text, we report simulations with length $T=100$ but additional simulations using $T=1000$ show the same results; hence this is not just a question of time series length, but of the likelihood being too flat.  

These results are true for the model and parameters that we consider in the main text, but also other model types such as a discrete-time version of the Rosenzweig-MacArthur (RMA) model (Supplement B1), which includes a more direct coupling between prey consumption and predator dynamics, as well as models with increased temporal variability in the functional response (Supplement B2). \citet{turchin2000living}, using nonlinear forecasting to fit a similar stochastic predator-prey model to ours (whose likelihood had no closed form, because of continuous time), also reported a lesser identification of parameters for stable fixed points. We are therefore confident that our main result has some degree of generality. However, we concede that quasi-cycles \citep{nisbet1982modelling} with more discernible cyclic fluctuations (e.g., closer to the discrete Hopf bifurcation, \citealp{Wiesenfeld1985Noisy,neiman1997coherence}) should logically be easier to identify without kill rate data. In fact, quasi-cycles might, in many cases, be difficult to distinguish from limit cycles (but see \citealp{louca2014distinguishing}). Surprisingly, our investigations of the RMA model in Supplement B1 do not quite appear to support that theory, since the functional response was poorly estimated without kill rate data for the QC parameter set. But this could be due to the specifics of the simulation or the discrete-time model, an investigation with more emphasis on quasi-cycles would be needed for a definite answer.

We also tried, when kill rate data was omitted, to fit models with a stochastic rather than deterministic functional response. These did not perform better (and typically performed worse) than the models with a deterministic functional response. This poor performance is likely associated to an absence of information available to dissociate functional response signal vs noise: the stochastic kill rate becomes an unobserved latent state  which is in this case poorly estimated --- a problem that is reminiscent of identifiability issues in state-space models without knowledge of observation error \citep{auger-methe_state-space_2016}. 



We used prior-posterior overlap to quantify practical identifiability, together with inspection of likelihood surfaces for pairs of parameters. Another useful tool is the profile likelihood \citep{bolker2008ecological,raue2009structural,eisenberg2014determining}, which would likely yield similar results. However, as explained in \citet{eisenberg2014determining}, likelihood profiles can sometimes be ambiguous when combinations of multiple parameters are at play \citep{raue2009structural}. They are also very computationally intensive, since for all values of one parameter, optimization has to be performed for all other parameters. We have therefore preferred to evaluate the practical identifiability with the prior-posterior overlap, which is available as a by-product of any Bayesian estimation.  

\subsection{Relative performances of the data combinations}

We found that adding kill rate data on just one fourth of the time series greatly increased model performance with respect to no kill rate data, in the sense of decreasing the bias and increasing the precision of the estimators. In some cases (noisy limit cycle), kill rate data availability for only one fourth of the time series was almost as good as having measured kill rates for the whole time series. This is remarkable, as this requires the algorithm to estimate the stochastic kill rate as a latent state for 75\% of the time series, which is usually a hard task. 

Time series in ecology are typically short, from a dozen to often at best a hundred of time points, especially for vertebrates (some fast-living species can have much longer time series), and functional response data are rarely available for the whole duration of the study \citep[e.g.,][]{gilg2003cyclic}. While very short time series of $T = 25$ time steps seemed beyond saving for our nonlinear predator-prey models --- at least without more informed priors --- the results for $T = 100$ and $T = 50$ with 25\% of functional response data are quite encouraging, especially for cyclic species. For non-cyclic species, $T = 100$ and $p_{KR}=25\%$ worked well but a time series length $T=50$ presented many unidentifiable simulations for max intake $C$ and especially the half-saturation constant $D$. It may be possible, in this case, to improve practical identifiability by complementing the kill rate data with other data types, typically survival and reproduction data \citep{peron2012integrated,barraquand2019integrating}, or more informative priors on the intrinsic growth rates. 


\subsection{Avenues for methodological development of predator-prey models}

The behaviour of deterministic predator-prey models is typically sensitive to the type of functional response that is used \citep{fussmann2005community,aldebert2018community}. Here, we used a type II functional response of the Holling type.  Different functional responses, e.g., an equivalent Ivlev-type $g(N) = C\left(1-\exp(-\ln(2)\frac{N}{D})\right)$ may yield slightly different results in a deterministic setting (e.g., a change in values making the model switch between from an equilibrium to a limit cycle), though it is known that stochasticity smoothes over the transition between different dynamical behaviours \citep{Wiesenfeld1985Noisy,neiman1997coherence,barraquand2017moving}, making the dynamical behaviour probably less sensitive to the exact functional response formulation in stochastic models. 

There would, however, be several ways to extend the present results exploiting different functional response formulations:
(i) simulating and fitting the model with the Ivlev functional response, to check the robustness of our results to the functional form used. Or, (ii) simulating with Holling and fitting with Ivlev, or vice versa, keeping in mind that exact parameter match is impossible because of slightly different shapes. Moreover, the functional response can depend on additional variables such as predator densities \citep{skalski2001functional,abrams2000npp} or multiple species, which could be added as well. A simple and elegant way to model implicitly many species is to consider a stochastic, temporally variable half-saturation constant: such model was successfully fitted to data in Supplement B2. 

Our models used discrete time. Continuous-time equivalents using stochastic differential equations could be considered \citep{gilioli2008bayesian,gilioli2012nonlinear}, though two difficulties arise. First, the likelihood of a nonlinear stochastic differential equation cannot usually be written in a closed form, which requires evaluation by more elaborate algorithms. These include nonlinear forecasting using attractor reconstruction methods \citep{turchin2000living}, particle filtering \citep{ionides2006inference} or other likelihood-free methods including Approximate Bayesian Computation (see \citealt{fasiolo2016comparison} for a review). In a stochastic differential equation context, \citet{gilioli2008bayesian,gilioli2012nonlinear} discretized their equations with an Euler scheme to simplify the inferential problem; in that case, it can be as simple and more transparent with regard to biological assumptions to work with a discrete-time predator-prey model from the start. Second, the model presented here can easily be extended to incorporate environmental variation that is correlated in time. That is typically more difficult in stochastic differential equations, whose noise terms are Brownian (Wiener) processes. At the moment, discrete time therefore seems a more convenient option in many respects, but continuous-time stochastic models also have clear advantages (e.g., no ordering of events and a good connection to most of the multi-species deterministic theory) that should make them an interesting target for future work. 

Finally, we glossed over some of the difficulties in actually measuring kill rates from the number of kills, while taking into account non-replacement of prey items \citep{vonesh2005compensatory,fenlon2006modelling,rosenbaum2018fitting}. While we believe that these observational issues deserve attention, this type of work may be best done as a follow-up on a case study with real data. Moreover, many if not most sources of kill rate data are not numbers of observed kills, but in fact reconstructed kill rates combining observed diet with some maximum or allometrically reconstructed consumption \citep{christensen1992ecopath,nielsen1999gyrfalcon,gilg2006functional}; our model framework is ideal for the latter data types and can be readily applied in this context. 

\subsection{Towards a better understanding of identifiability in general multispecies stochastic systems}

Here, we relate our results to more general community dynamics and formulate some conjectures. Most population dynamic models of interactions between species are formulated so that a state vector $\mathbf{x}_t$ has a dynamics 

\begin{equation}\label{eq:dynsyst}
    \mathbf{x}_{t+1} = \mathbf{f}(\mathbf{x}_t,\mathbf{g}_\alpha(\mathbf{x}_{t}),\mathbf{e}_t)
\end{equation}

where $\alpha \in \mathbb{R}^p$ is a given set of interaction parameters (possibly a matrix), $\mathbf{g}_\alpha(\mathbf{x}_{t})$ is a matrix of functions characterizing the effect of other species onto the focal one, and $\mathbf{e}_t$ the noise vector -- the latter being often a perturbation on the intrinsic growth rate.
These models might well be identifiable in many scenarios, but that is because all interaction rates contained within $\mathbf{g}_\alpha(\mathbf{x}_{t})$ are solely determined by the state vector $\mathbf{x}_{t}$. Assuming that noise affects interaction rates too (even in small amounts), i.e., $\mathbf{g}_\alpha(\mathbf{x}_{t},\mathbf{e}'_t)$, may well change that. What we have done in this study is to consider a ground truth model of the form  \begin{equation}\label{eq:dynsyst-stochint}
    \mathbf{x}_{t+1} = \mathbf{f}(\mathbf{x}_t,\mathbf{g}_{\alpha}(\mathbf{x}_t,\mathbf{e}'_t),\mathbf{e}_t)
\end{equation}
In eq. \eqref{eq:dynsyst-stochint}, we consider that interaction rates randomly vary from one time step to the next, even if the densities did not change. In the present paper, we have assumed that the predator kill rate (defining the functional response) was the unique stochastic interaction rate, and have shown that when the attractor is a fixed point in absence of stochasticity, the average interaction parameters $\alpha$ may not be identifiable using eq. \eqref{eq:dynsyst} as fitted model (or eq. \eqref{eq:dynsyst-stochint} without adding data on interaction rates). This is true even in seemingly harmless cases, for example,  $e'_{it} \sim \mathcal{N}(0,\sigma_{e'_{it}}^2)$ with $\sigma_{e'_{it}}^2$ small. This finding -- stochastic interaction rates can compromise identifiability unless data about them is provided -- could well generalize to other types of interactions and systems of larger dimensionality. One could imagine that stochastic competition or mutualism rates, for instance, would render competition/mutualism versions of eq.  \eqref{eq:dynsyst-stochint} very difficult to identify without adding additional interaction rate data. We conjecture that for competition/mutualistic systems as well, fixed point attractors will require more interaction data to be identifiable when compared to non-point attractors (e.g., cycles, loops, chaos). 

One could hypothesize as well that similar phenomena may occur in structured population models, where vital rates such as fertilities can be randomly varying over time; identifiability may be compromised without adding data on such randomly varying vital rates, so long as the attractor shape is not clear (i.e., no cycle or strange attractor). Our results therefore connect with identifiability issues in integrated population models \citep{besbeas2002integrating,abadi2010assessment}. 
Integrated community models \citep{peron2012integrated,barraquand2019integrating}, that include both stage structure and interactions between species, might benefit even more from merging data types once we account for the possible randomness in time of their components. 


\subsubsection*{Acknowledgements}
FB was supported by Labex COTE (ANR-10-LABX-45) and OG by the French National Research Agency (grant ANR-16-CE02-0007). We thank Bret Elderd for insightful suggestions that helped improve the paper, as well as two referees and the associate editor for constructive comments. FB also thanks Wojciech Uszko for exchanges regarding functional responses. 

\clearpage

\part*{Appendices}
\renewcommand{\theequation}{A\arabic{equation}}
\renewcommand{\thesection}{A\arabic{section}}
\renewcommand{\thefigure}{A\arabic{figure}}

\setcounter{equation}{0}
\setcounter{section}{0}
\setcounter{figure}{0}

\section{Stability analysis of the deterministic model and stochastic implications}\label{sec:stability}

The model, as specified in log-scale by   eqs.~\eqref{eq:prey_discreteLeslieMay_logscale}--\eqref{eq:predator_discreteLeslieMay_logscale}, can be analysed by computing the Jacobian for the (non-trivial) fixed equilibrium point. For this, we need first the find the fixed point, which is defined by 

\begin{align}\label{eq:fixed-point}
    N & = N e^{\left(r - \frac{CP}{D+N}\right)} \frac{1}{1+\gamma N}\\
    P & = P \frac{e^s}{1+q\frac{P}{N}}
\end{align}

The second equation gives $1+q\frac{P}{N}=e^s$ so that $P=\frac{N}{q}(e^s-1)$. Insert this in eq.~\eqref{eq:fixed-point}, and we find

\begin{equation}
    1+\gamma N = e^{\left(r - \frac{AN}{D+N}\right)}
\end{equation}
with $A = \frac{C(e^s-1)}{q}$. Unfortunately, this is a transcendental equation so there is no closed form solution. But we can say a few things about the equilibrium by the study of the function 
\begin{equation}
f(x) =  e^{\left(r - \frac{Ax}{D+x}\right)} - (1+\gamma x)
\end{equation}
for $x \in [0,+\infty)$. The function $f(x)$ is zero at the fixed point. We have $f(0) = e^r-1$ and $\lim_{x\rightarrow +\infty} {f(x)} = \lim_{x\rightarrow +\infty} {- (1+\gamma x) } = - \infty$, thus the function goes through zero at least once for all positive $r$ values (there is at least one fixed point). The derivative of $f$ exists and is defined by 
\begin{equation}
    f'(x) = - \frac{Ax}{(D+x)^2} e^{\left(r-\frac{Ax}{D+x}\right)} - \gamma <0
\end{equation}
so for $A>0$, $D>0$ (which follows from the definition of parameters), there is a monotonic decrease of the function $f$ from $f(0)$ to $-\infty$. The function therefore passes only \emph{once} through zero, which means that there is a unique strictly positive equilibrium point $(N^*,P^*)$. This equilibrium can be found numerically by iterative convergence, we have done this using the R package `rootSolve'. 
\subsection{Derivation of the Jacobian}
The equations~\eqref{eq:prey_discreteLeslieMay_logscale}--\eqref{eq:predator_discreteLeslieMay_logscale} can be rewritten as

\begin{align}
    p_{t+1} & = g_1(n_t,p_t) \\
    n_{t+1} & = g_2(n_t,p_t) 
\end{align}
with $n=\ln(N)$, $p=\ln(P)$ and 
\begin{align}
    g_1(n,p) & = n +r- \frac{Ce^p}{D+e^n}-\ln(1+\gamma e^n)\\
    g_2(n,p) & = p +s-\ln(1+q e^p/e^n)
\end{align}

The Jacobian of the log-transformed system is by definition 
\begin{align}
    \mathbf{J} = 
    \begin{pmatrix}
    \frac{\partial g_1}{\partial n} & \frac{\partial g_1}{\partial p} \\
    \frac{\partial g_2}{\partial n} & \frac{\partial g_2}{\partial p} \\
    \end{pmatrix}
\end{align}

We introduce modified functions for convenience
\begin{align}
    \Tilde{g}_1(n,p) & = r- \frac{Ce^p}{D+e^n}-\ln(1+\gamma e^n)\\
    \Tilde{g}_2(n,p) & = s-\ln(1+q e^p/e^n)
\end{align}
which leads to a Jacobian 
\begin{align}
    \mathbf{J} = 
    \begin{pmatrix}
    1 + \frac{\partial \Tilde{g}_1}{\partial N} \frac{\partial N}{\partial n}& \frac{\partial \Tilde{g}_1}{\partial P} \frac{\partial P}{\partial p} \\
    \frac{\partial \Tilde{g}_2}{\partial N} \frac{\partial N}{\partial n} & 1+ \frac{\partial \Tilde{g}_2}{\partial P} \frac{\partial P}{\partial p}\\
    \end{pmatrix}
\end{align}

with $\frac{\partial N}{\partial n} = e^n = N$ and $\frac{\partial P}{\partial p} = e^p =P$. Doing so, we obtain

\begin{align}
    \mathbf{J} = 
    \begin{pmatrix}
    1+N \left( \frac{CP}{(D+N)^2} - \frac{\gamma}{1+\gamma N}\right) & 
    \frac{CP}{D+N} \\
    \frac{qP/N}{1+qP/N} & 
    1 - \frac{qP/N}{1+qP/N}\\
    \end{pmatrix}
\end{align}

which can be rewritten using the relationships at equilibrium

\begin{align}
    \mathbf{J} = 
    \begin{pmatrix}
    1+N^* \left( \frac{CP^*}{(D+N^*)^2} - \frac{\gamma}{1+\gamma N^*}\right) & 
    \frac{CP^*}{D+N^*} \\
    \frac{e^s-1}{e^s} & 
    \frac{1}{e^s}\\
    \end{pmatrix}
\end{align}

The Jacobian is then computed using the numerically found $(N^*,P^*)$, in the case of the fixed point parameter set $(N^*,P^*)=(5.44,0.35)$ and in the case of the limit cycle parameter set $(2.10,0.14)$. 

Using our two parameter sets, we found the following eigenvalues for the Jacobian, reported in Table~\ref{tab:eigenvalues_Jacobian}. 

\begin{table}[H]
    \centering
    \begin{tabular}{c|ccc}
     Attractor & $\Re(\lambda_{\mathbf{J}})$ & $\Im(\lambda_{\mathbf{J}})$ & $|\lambda_{\mathbf{J}}|$ \\
     \hline
     Fixed point & 0.44 & 0.16 & 0.47\\
     Limit cycle & 0.85 & 0.53 & 1.004
    \end{tabular}
    \caption{Eigenvalues of the Jacobian for both parameter sets (Fixed point, $(C,D) = (2.5,1)$; limit cycle, $(C,D) = (15,0.25)$). The values in columns are the real part, the imaginary part, and the modulus of the eigenvalues. A modulus below 1 indicates convergence of the fixed point in absence of stochasticity, a modulus above 1 indicates here an escape of the trajectory towards the invariant loop.}
    \label{tab:eigenvalues_Jacobian}
\end{table}

\section{Frequentist functional response parameter estimates }\label{sec:optim}

\begin{figure}[h]
    \centering
    \includegraphics[width=15cm]{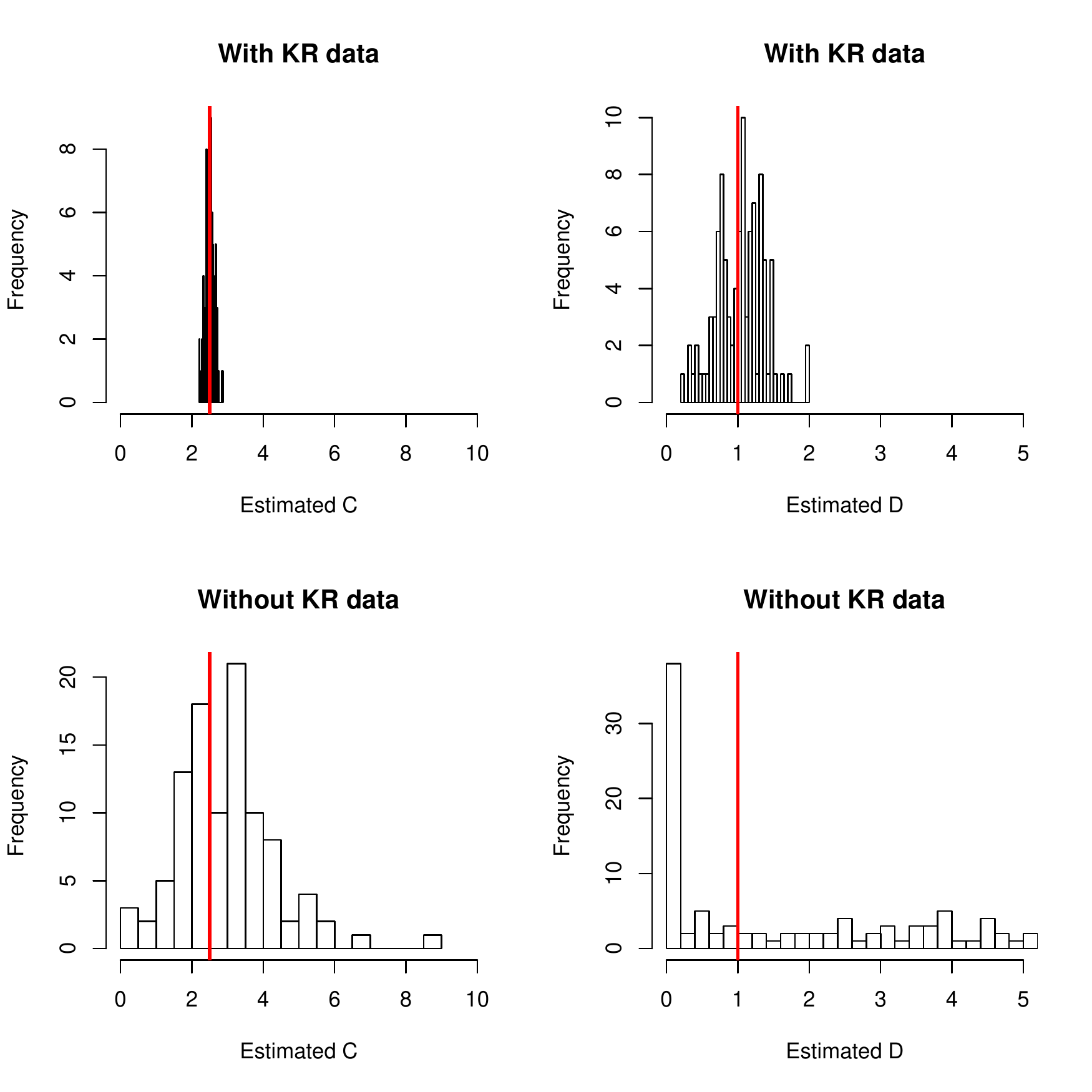}
    \caption{Histogram of estimated $C$ and $D$ functional response parameter values. True values as vertical red lines. Top row: with kill rate (KR) data, bottom: without kill rate data. Perturbed fixed point parameter set.}
    \label{fig:optimFP}
\end{figure}

\begin{figure}[h]
    \centering
    \includegraphics[width=15cm]{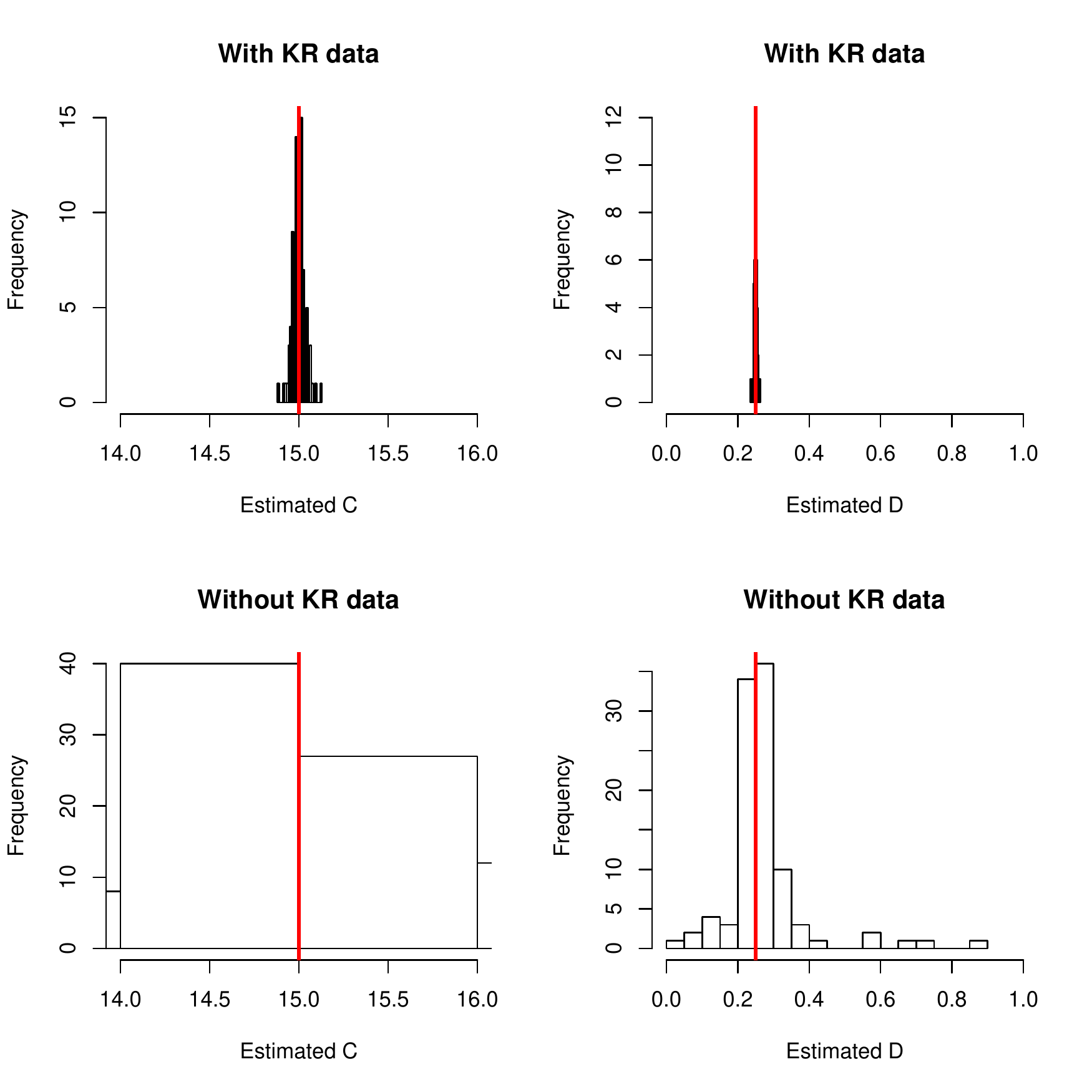}
    \caption{Histogram of estimated $C$ and $D$ functional response parameter values. True values as vertical red lines. Top row: with kill rate (KR) data, bottom: without kill rate data. Limit cycle parameter set.}
    \label{fig:optimLC}
\end{figure}

As shown in Fig.~\ref{fig:optimFP}, the bias and precision of estimators for $C$ and $D$ is very small with kill rate data and rather large without kill rate data, for the fixed point parameter set. Please keep in mind that the range of values without kill rate data largely reflects the starting values (although there is some central tendency in the distribution for $C$). Fig.~\ref{fig:optimLC} highlights that while unbiased estimators already exist without kill rate data for the limit cycle parameter set, adding kill rate data tremendously increases estimator precision. 

\section{Complementary results on the limit cycle parameter set}\label{sec:noisyLC-results}

\subsection{Fisher Information Matrix}

\begin{figure}[H]
    \centering
    \includegraphics[width=15cm]{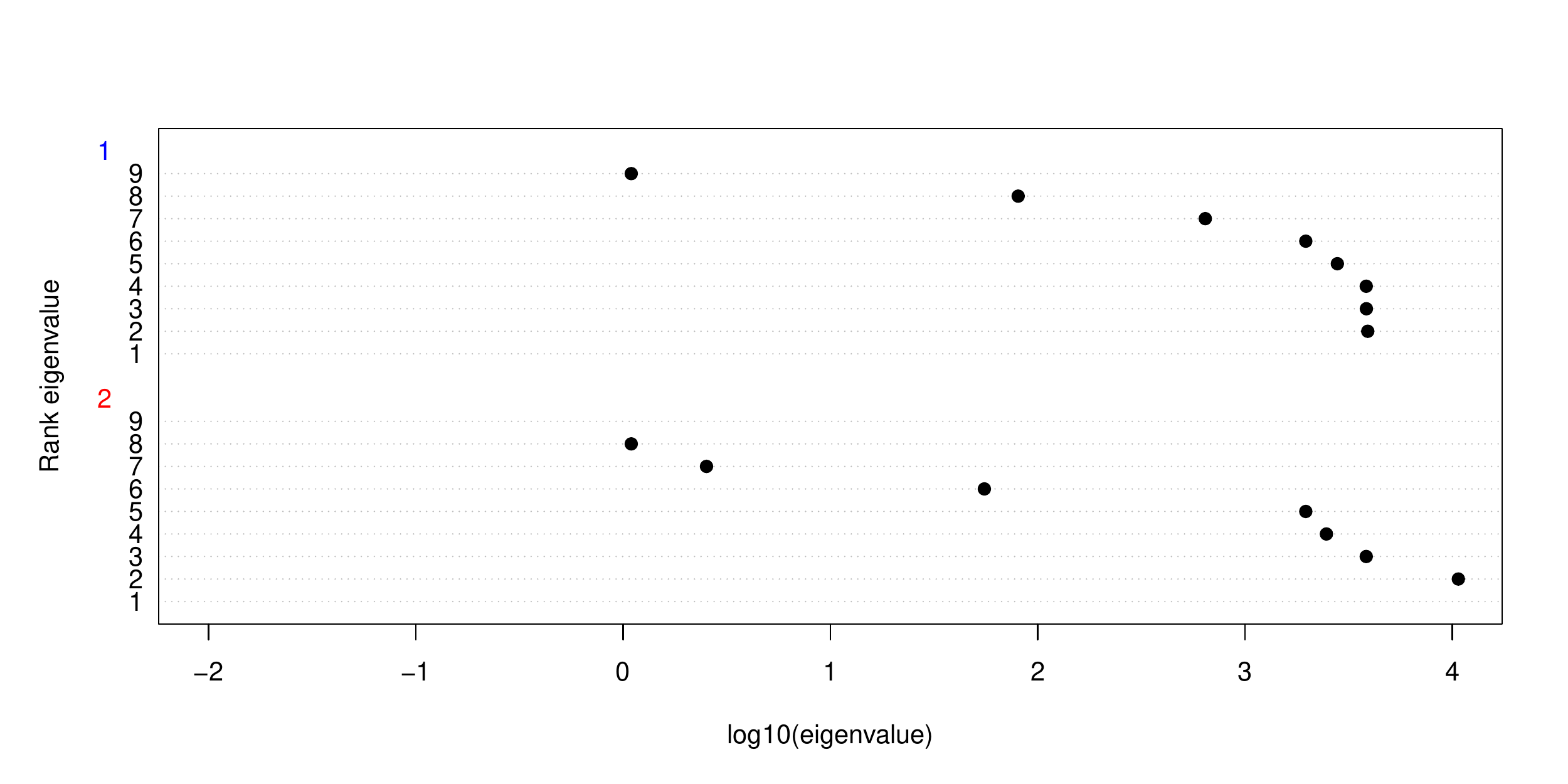}
    \caption{Distribution of eigenvalues of the FIM, noisy limit cycle case. First row, eigenvalues of the FIM for the model with kill rate data, second row, eigenvalues for the model without kill rate data.}
    \label{fig:FIM_eigenvalues_LC}
\end{figure}

\subsection{Posterior correlations}

\begin{figure}[H]
\begin{subfigure}[b]{0.45\textwidth}
    \centering
    \includegraphics[width=\textwidth]{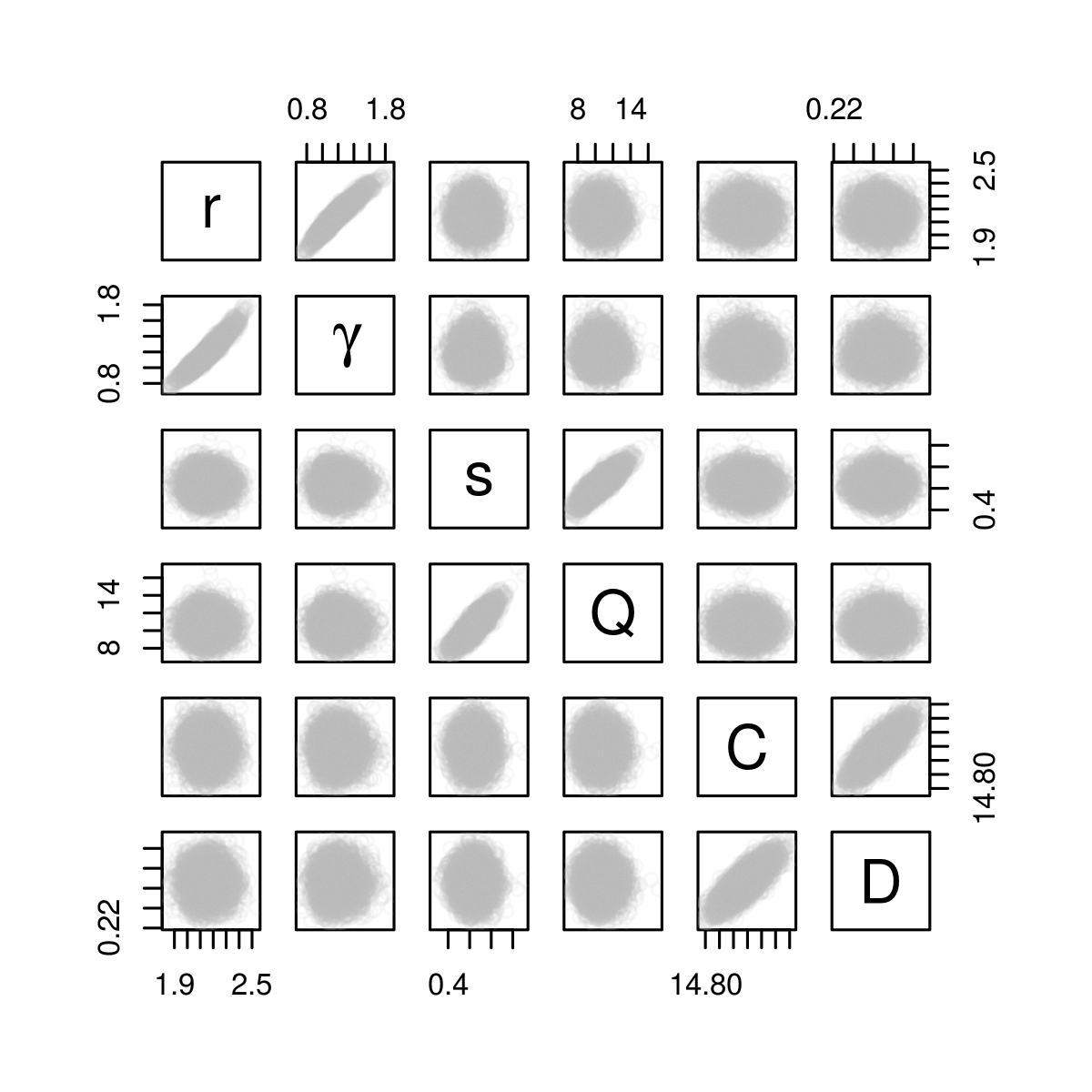}
    \caption{With kill rate data}
    \label{fig:correlation_posteriors_wKR_LC}
\end{subfigure}
~
\begin{subfigure}[b]{0.45\textwidth}
    \centering
    \includegraphics[width=\textwidth]{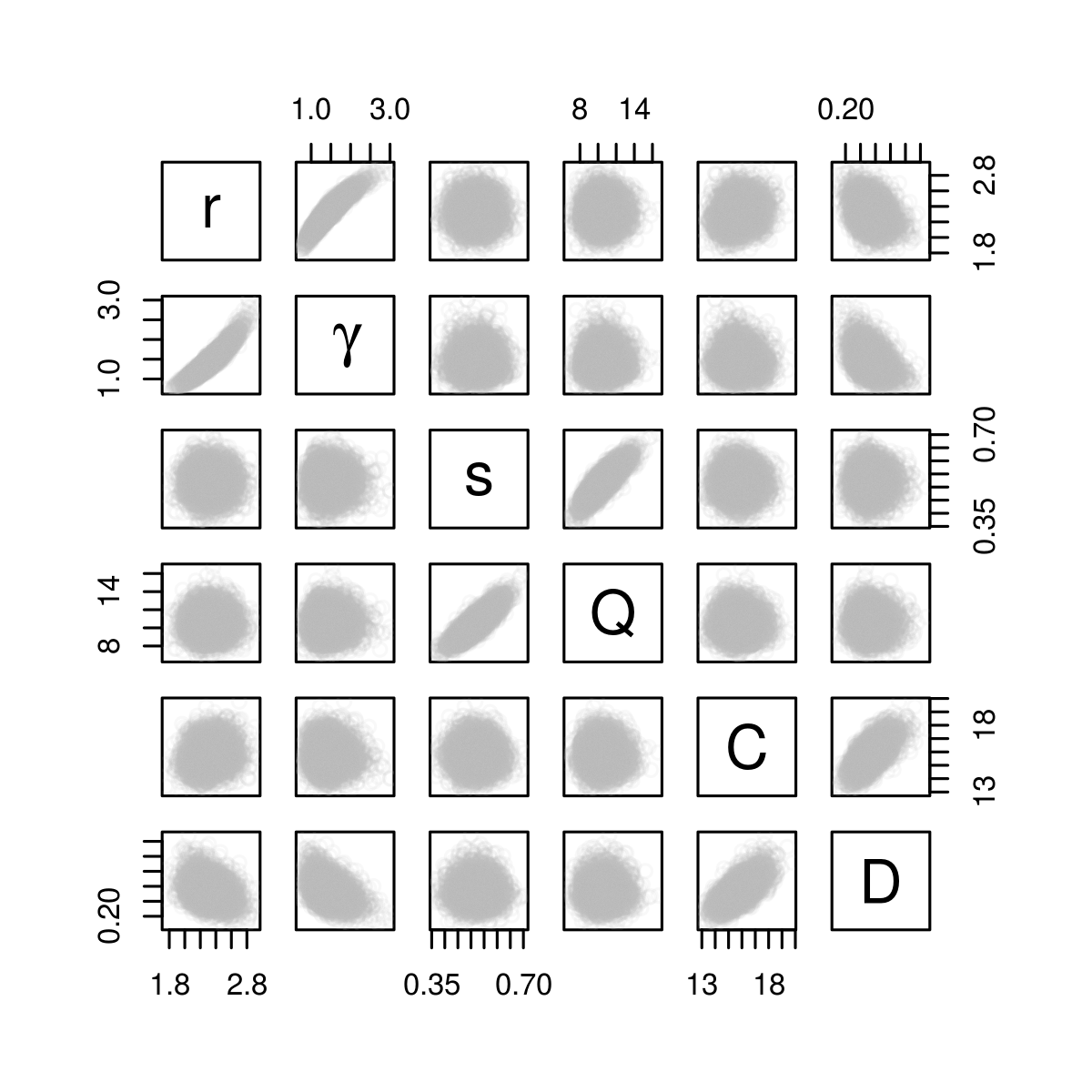}
    \caption{Without kill rate data}
    \label{fig:correlation_posteriors_woutKR_LC}
\end{subfigure}
 \caption{Plots of the MCMC samples per pair of parameters, for the noisy limit cycle parameter set.}
\label{fig:correlation_posteriors_LC}
\end{figure}

\subsection{Estimation of functional forms}

The functional response curve is plotted in Fig.~\ref{fig:estimation_curves_FR_LC} with vs without the correlations between parameters C and D. In the next Fig.~\ref{fig:estimation_curves_preyDD_LC}, results are shown for the prey density-dependence curve. These are similar to the main text results, except that the functional response can now be estimated without kill rate data for $T=100$. 

\begin{figure}[H]
    \centering
    \includegraphics[width=0.8\textwidth]{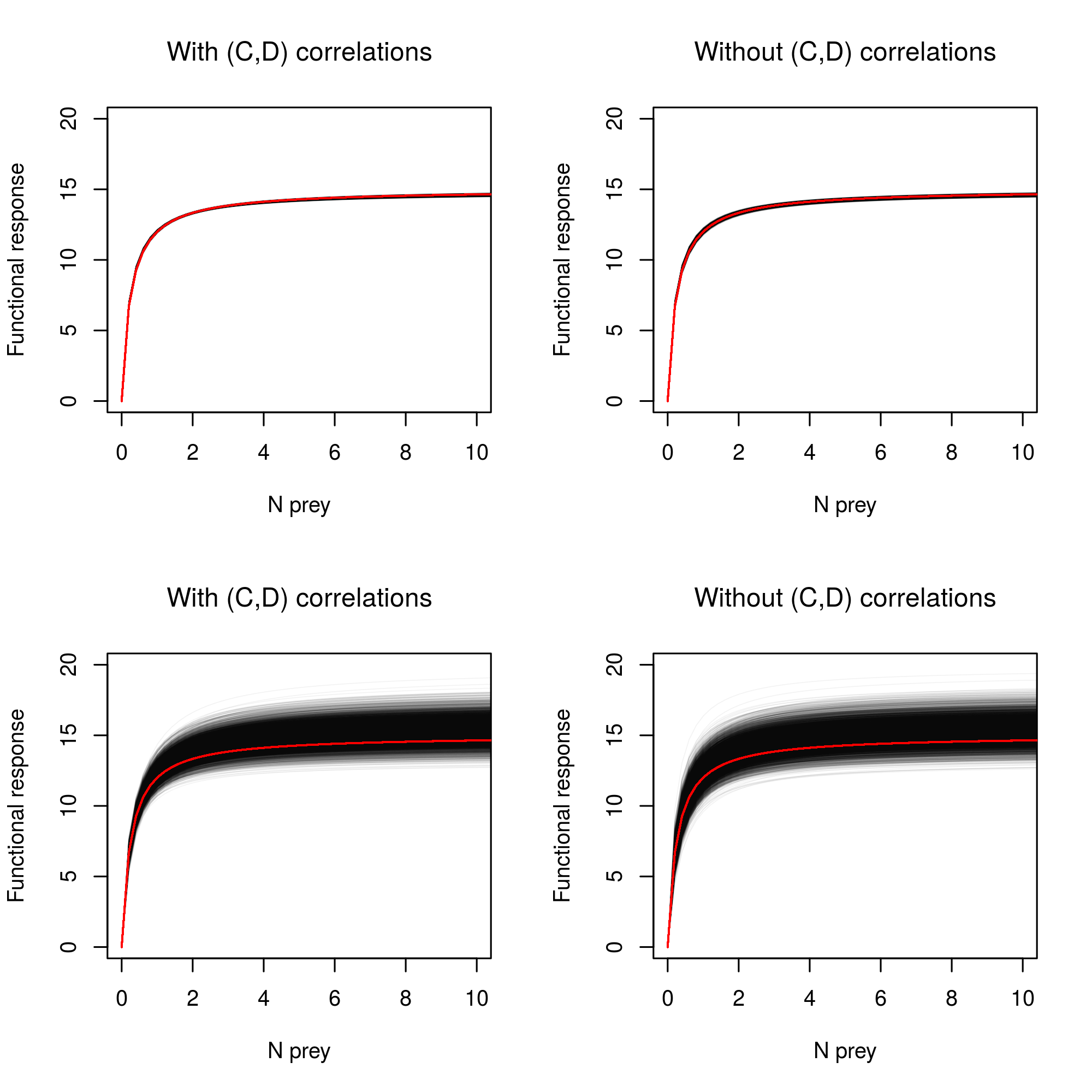}
    \caption{Average functional response with vs without correlation between parameters, with (top row) and without (bottom row) kill rate data, for the LC parameter set. The red line is the true, simulated average functional response.}
    \label{fig:estimation_curves_FR_LC}
\end{figure}

\begin{figure}[H]
    \centering
    \includegraphics[width=0.8\textwidth]{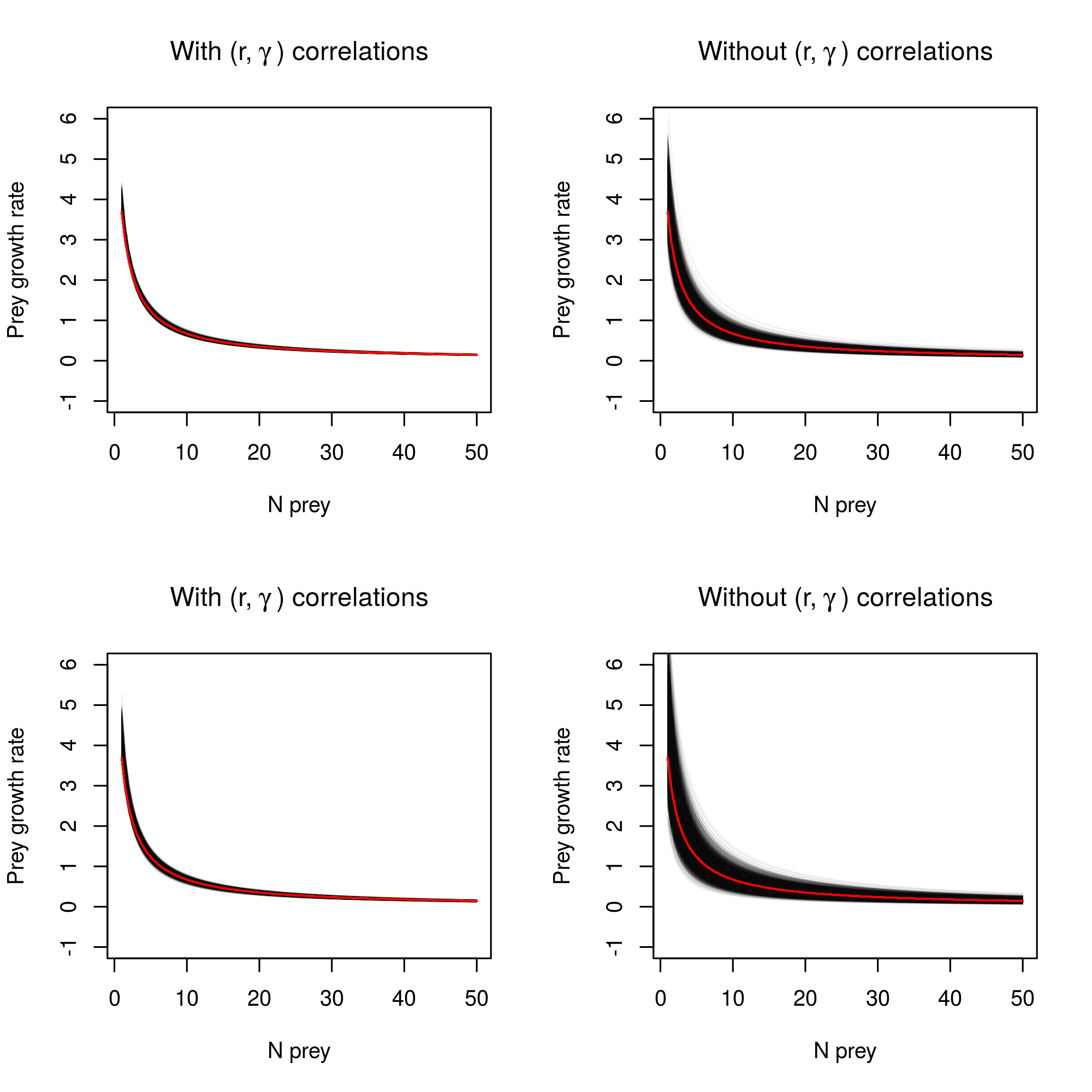}
    \caption{Prey growth rate-density curve with vs without correlation between parameters, with (top row) and without (bottom row) kill rate data, for the LC parameter set. The true, simulated curve is drawn in red.}
    \label{fig:estimation_curves_preyDD_LC}
\end{figure}

We present here plots for the prey density-dependence curve but similar results can be obtained for the predator density-dependence curve.

\subsection{Data availability scenarios}

The plots are similar to those of the main text, examining the posterior probability densities of $C$ and $D$, except that here we do this for the noisy limit cycle (LC) parameter set. 

\begin{figure}[H]
    \centering
    \includegraphics[width=0.9\textwidth]{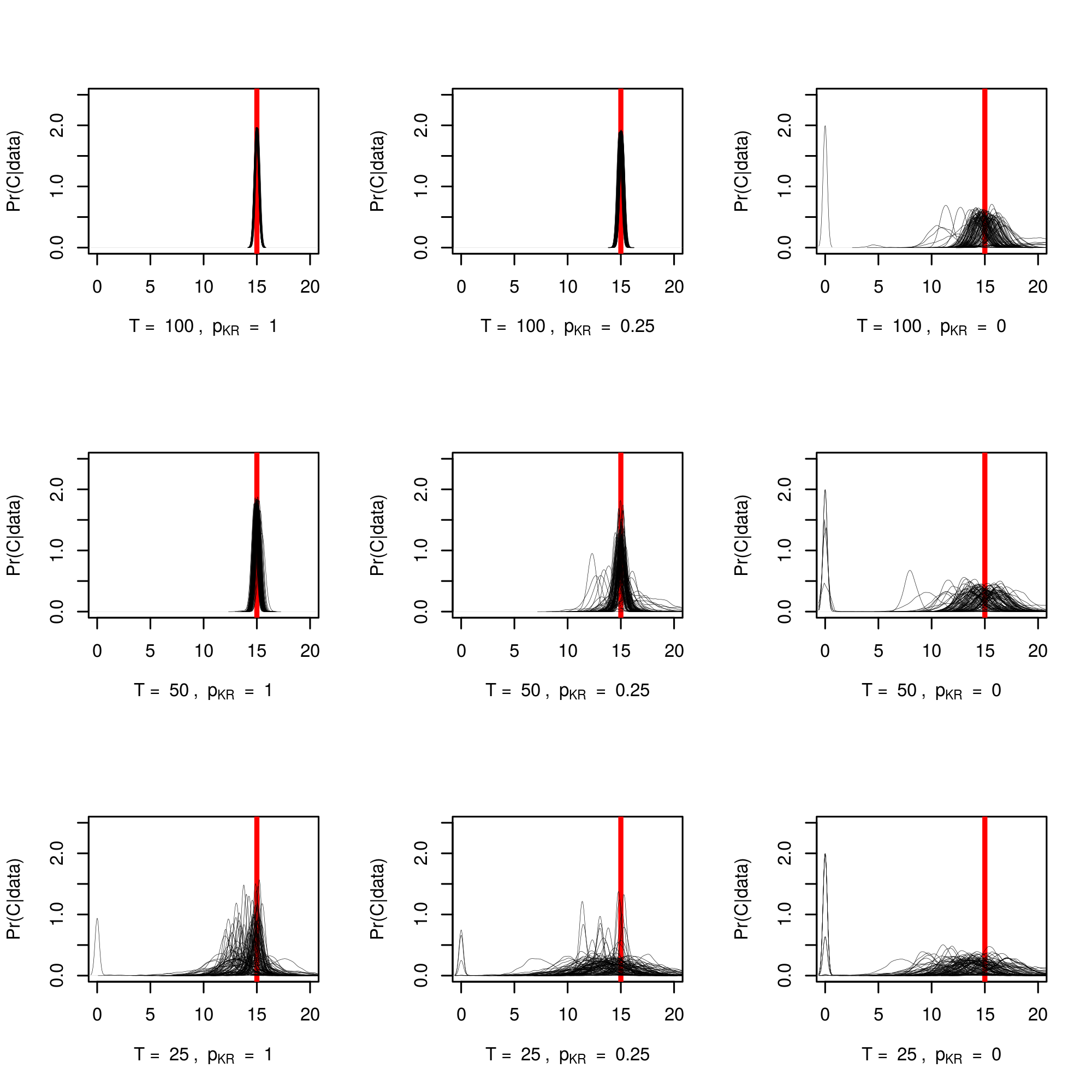}
    \caption{Posterior probability densities for the parameter $C$, for 100 simulations (each grey line is one posterior probability density) using the LC parameter set. The vertical red line materializes the true parameter value.}
    \label{fig:PosteriorsC_LC}
\end{figure}

\begin{figure}[H]
    \centering
    \includegraphics[width=0.9\textwidth]{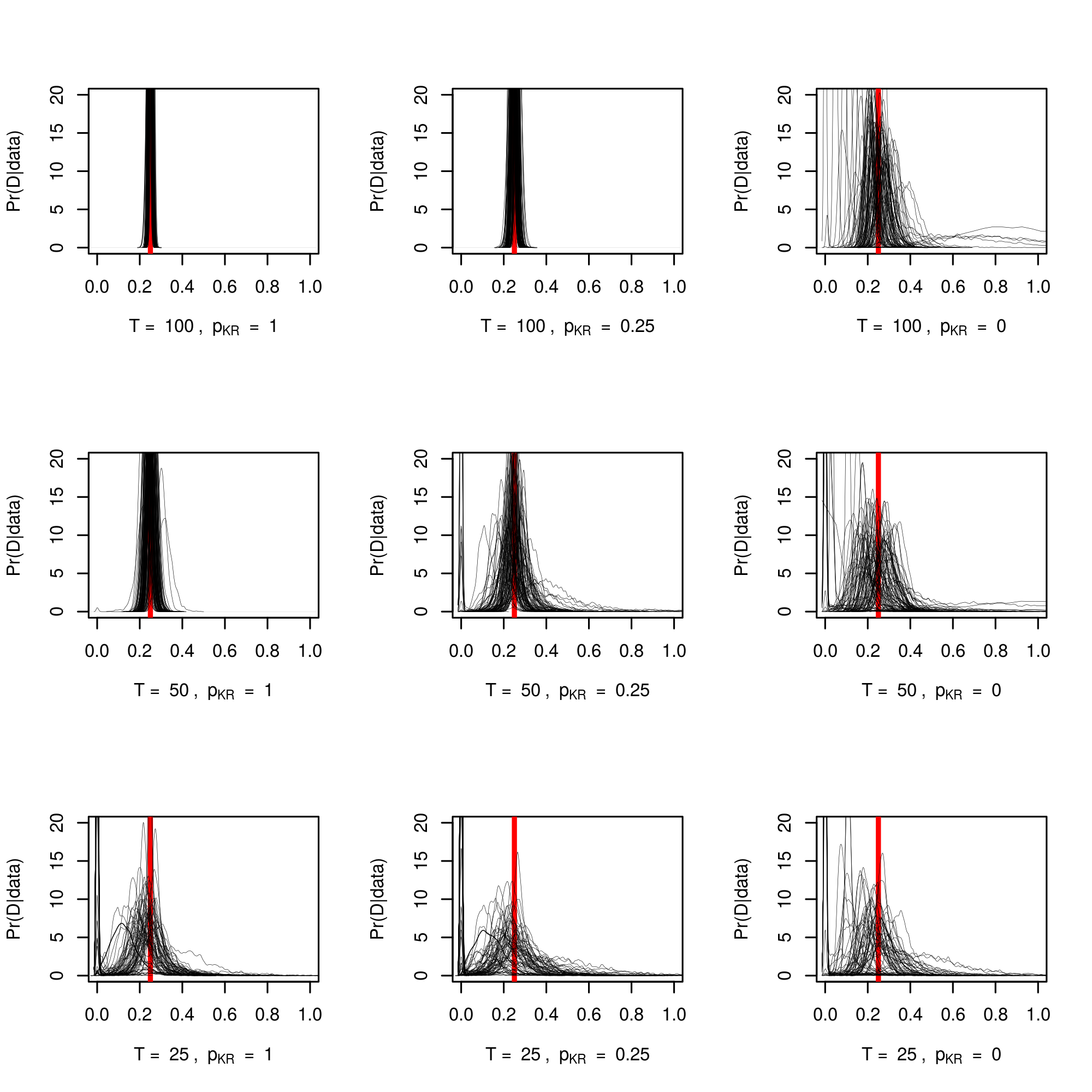}
    \caption{Posterior probability densities for the parameter $D$, for 100 simulations (each grey line is one posterior probability density) using the LC parameter set. The vertical red line materializes the true parameter value.}
    \label{fig:PosteriorsD_LC}
\end{figure}

\section{Complementary results on the reparameterized model}
\label{sec:reparam-results}

\begin{figure}[h]
    \centering
    \includegraphics[width=\textwidth]{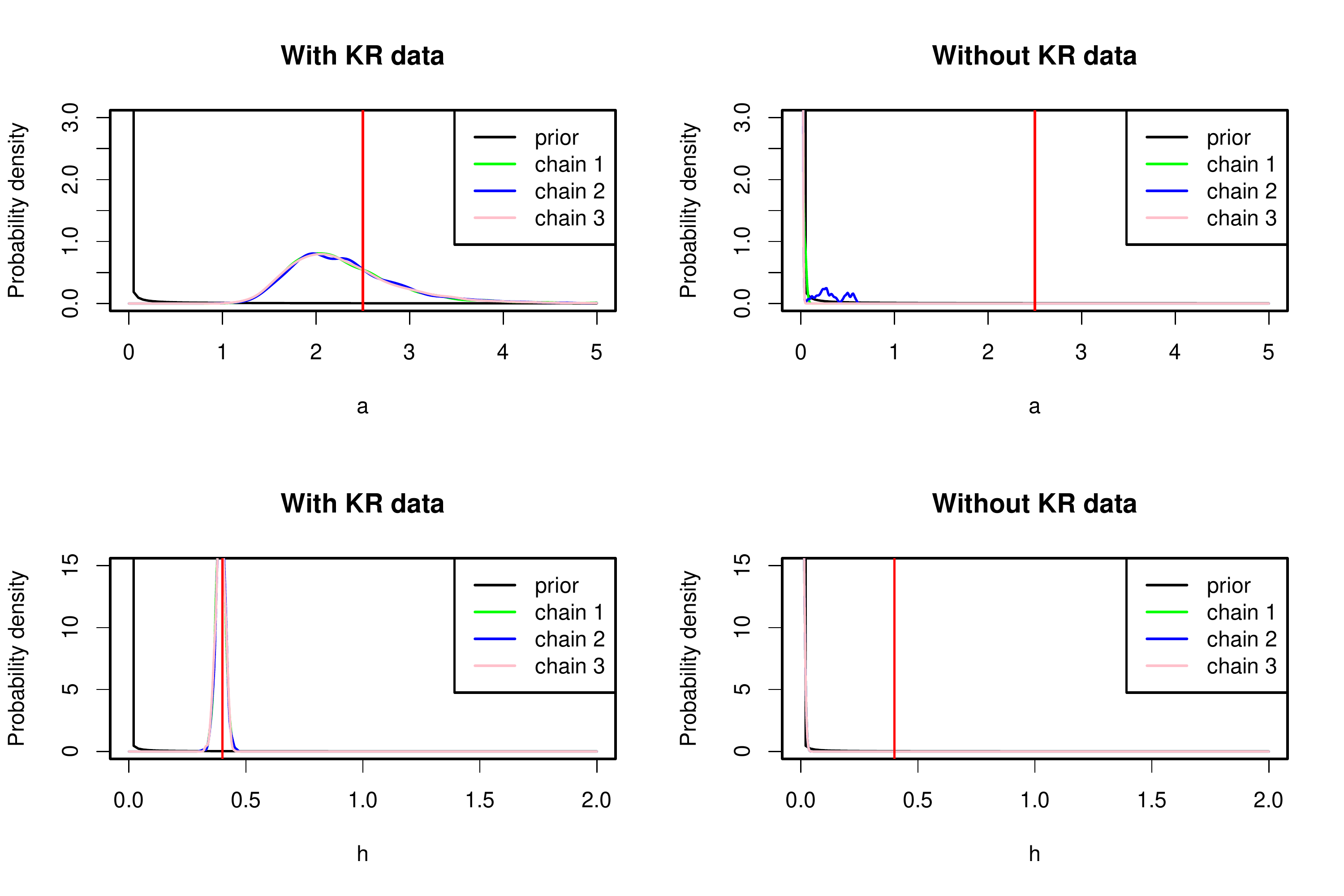}
    \caption{Perturbed fixed point dataset. Prior-posterior overlap for the reparameterized predator-prey model.}
    \label{fig:PPO_FP_reparam}
\end{figure}

Here we report the results of the Bayesian estimation of the reparameterized model of eqs.~\eqref{eq:prey_discreteLeslieMay_reparam}--\eqref{eq:predator_discreteLeslieMay_reparam} of the main text. Fig.~\ref{fig:PPO_FP_reparam} illustrates that $a$ and $h$ suffer from the same identification issues as $C$ and $D$ without kill rate data, for the fixed point parameter set (identical parameters and T = 100 simulation as used in the main text). 

\begin{figure}[h]
    \centering
    \includegraphics[width=0.8\textwidth]{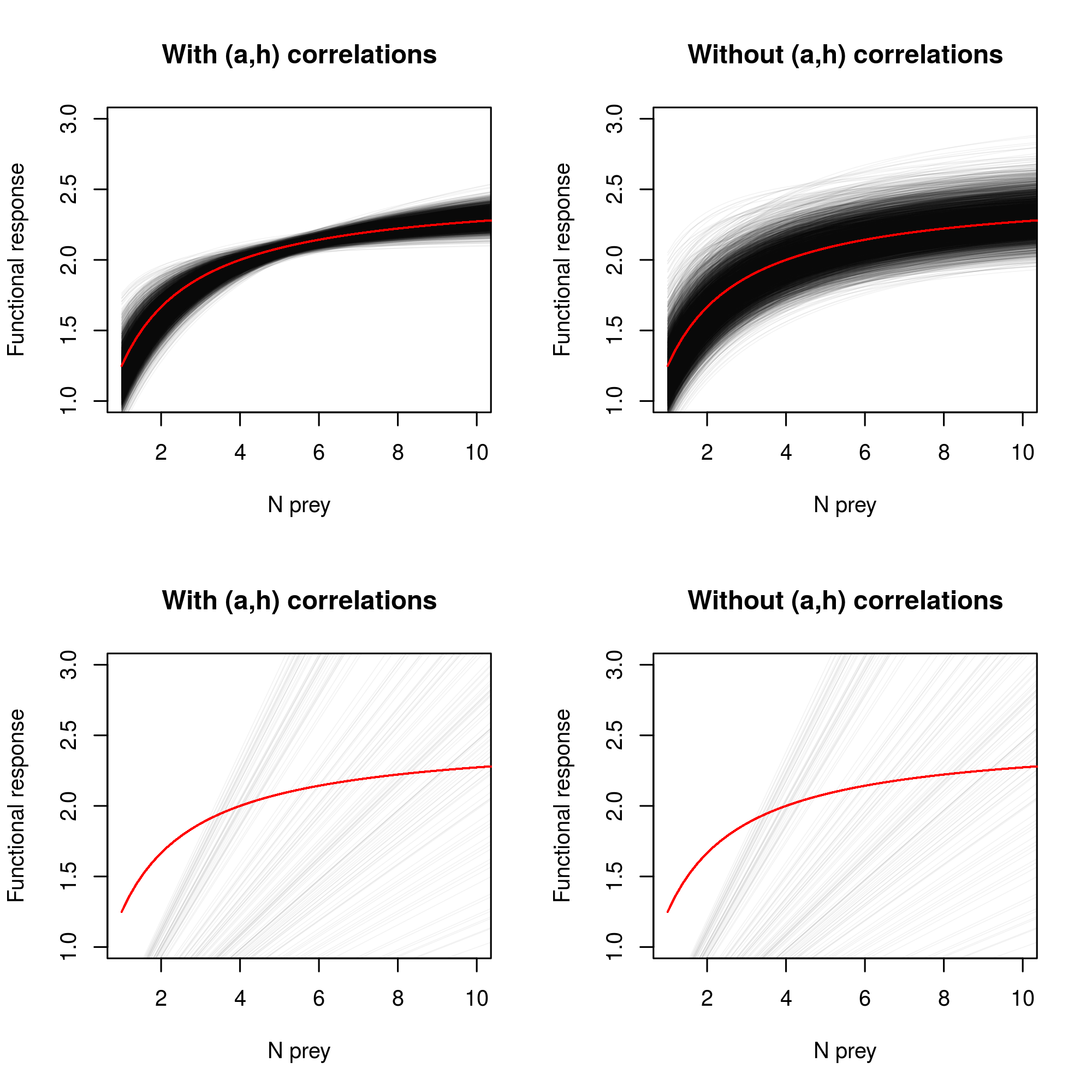}
    \caption{Average functional response with vs without correlation between parameters, with (top row) and without (bottom row) kill rate data. Perturbed fixed point dataset and re-parameterized model. The red line is the true, simulated average functional response.}
    \label{fig:estimation_curves_FR_FP_reparam}
\end{figure}

\begin{figure}[h]
    \centering
    \includegraphics[width=0.8\textwidth]{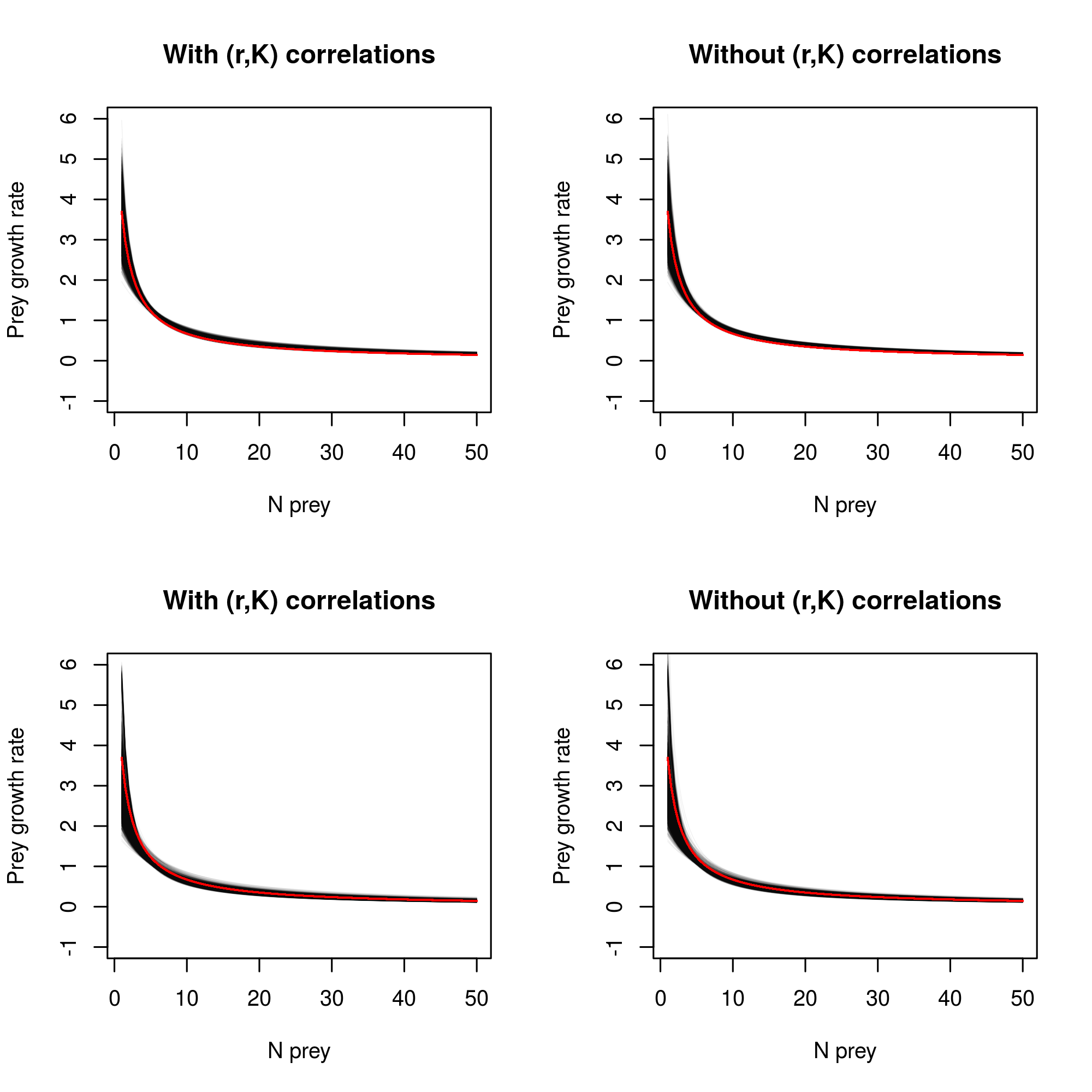}
    \caption{Prey growth rate-density curve with vs without correlation between parameters, with (top row) and without (bottom row) kill rate data. Perturbed fixed point dataset and re-parameterized model. The true, simulated curve is drawn in red.}
    \label{fig:estimation_curves_preydd_FP_reparam}
\end{figure}

Although $(a,h)$ are slightly less correlated than $(C,D)$, the average functional response curve as a whole is not better estimated (Fig.~\ref{fig:estimation_curves_FR_FP_reparam}).
We can see that the prey growth rate -- density curve is, on the other hand, estimated without notable correlations: independent permutations of the temporal order of parameters on $(r,K)$ chains does not affect the width of this curve anymore (Fig.~\ref{fig:estimation_curves_preydd_FP_reparam}). This is due to the use of a carrying capacity $K$ rather than a density-dependence coefficient $\gamma$, which removes posterior correlations (see also main text for the likelihood-based results). However, as we note in the main text, the $(r,K)$ parameterization does not yield an intrinsically more precise prey growth rate -- density curve than the $(r,\gamma)$ parameterization, when correlations between parameters in the posteriors are allowed. 
 
\clearpage

\bibliography{predpreyKRdata}
\bibliographystyle{amnat} 

\clearpage

\part*{Supplementary Material}

for \textit{Barraquand F. \& Gimenez O. (2021). Fitting stochastic predator–prey models using both population density and kill rate data. Theoretical Population Biology, doi:10.1016/j.tpb.2021.01.003}

\setcounter{equation}{0}
\renewcommand{\theequation}{B\arabic{equation}}

\setcounter{figure}{0}
\renewcommand{\thefigure}{B\arabic{figure}}

\setcounter{table}{0}
\renewcommand{\thetable}{B\arabic{table}}
\renewcommand{\thesection}{B\arabic{section}}

\setcounter{section}{0}

\section{Estimation and identifiability of a Rosenzweig-MacArthur model}


Due to the general congruence between Bayesian estimation through MCMC and ML estimation demonstrated in the main text, we present here only a Bayesian analysis of the discrete-time RMA model. 

The stochastic model is defined as 
\begin{align}
N_{t+1} & = N_{t}\frac{e^{r+\epsilon_{1t}}}{1+\gamma N_{t}}\exp\left(-g(N_{t},\epsilon_{3t})\frac{P_{t}}{N_{t}}\right),\,\epsilon_{1t}\sim\mathcal{N}(0,\sigma_{1}^{2})\label{eq:prey_discreteRMA}
\end{align}

\begin{align}
P_{t+1} & = P_{t} e^{\varepsilon g(N_{t},\epsilon_{3t}) - \mu},\,\epsilon_{2t}\sim\mathcal{N}(0,\sigma_{2}^{2})\label{eq:predator_discreteRMA}
\end{align}

which translates into 

\begin{align}
N_{t+1} & = N_{t}\frac{e^{r+\epsilon_{1t}}}{1+\gamma N_{t}}\exp\left(-\frac{C P_t}{D+N_t} + \frac{P_{t}}{N_{t}} \epsilon_{3t}\right),\,\epsilon_{1t}\sim\mathcal{N}(0,\sigma_{1}^{2})\label{eq:prey_discreteRMAexplicit}
\end{align}

\begin{align}
P_{t+1} & = P_{t} e^{\varepsilon \left(\frac{C N_t}{D+N_t} + \epsilon_{3t}\right) - \mu},\,\epsilon_{2t}\sim\mathcal{N}(0,\sigma_{2}^{2})\label{eq:predator_discreteRMAexplicit}
\end{align}


We have used the following parameter notations and parameter values 

\begin{table}[h]
    \centering
    \begin{tabular}{c|c|ccc}
     Name & Meaning & FP & QC & LC\\
     \hline
     $r$ & Prey intrinsic growth rate  & 2 & 2 & 1.8\\
     $\gamma$ & Density-dependence coefficient & 1 & 1 &1\\
     $\mu$ & Predator mortality & 0.2 & 0.2 & 0.7\\
     $\varepsilon$ & Conversion efficiency & 0.1 & 0.1 & 0.1\\
     $C$ & Max kill rate & 2.5 & 2.5 & 10\\
     $D$ & Half-saturation constant & 1 & 0.5 & 0.6\\
     $\sigma_1^2$ & Prey growth rate noise variance & 0.05 & 0.05 &0.05\\
     $\sigma_2^2$ & Pred. growth rate noise variance & 0.05 & 0.05 &0.05\\
     $\sigma_3^2$ & Functional response noise variance & 0.05 & 0.05 &0.05\\
    \end{tabular}
    \caption{Notation, meaning and values of the parameters for three parameter sets: FP, fixed point (in the deterministic model); QC, quasi-cycles, fixed point with strongly oscillatory dynamics in the presence of stochasticity and LC, noisy limit cycle.}
    \label{tab:RMA_parameters}
\end{table}

We added a parameter set leading to quasi-cycles (i.e., an excited stable focus, \citealp{nisbet1982modelling}, with strongly oscillatory decay to equilibrium in the absence of noise) that is an interesting intermediate case between a noisy limit cycle and a perturbed fixed point (the latter having near-monotonic decay to equilibrium in the absence of noise). 

\begin{figure}[h]
    \centering
    \includegraphics[width=16cm]{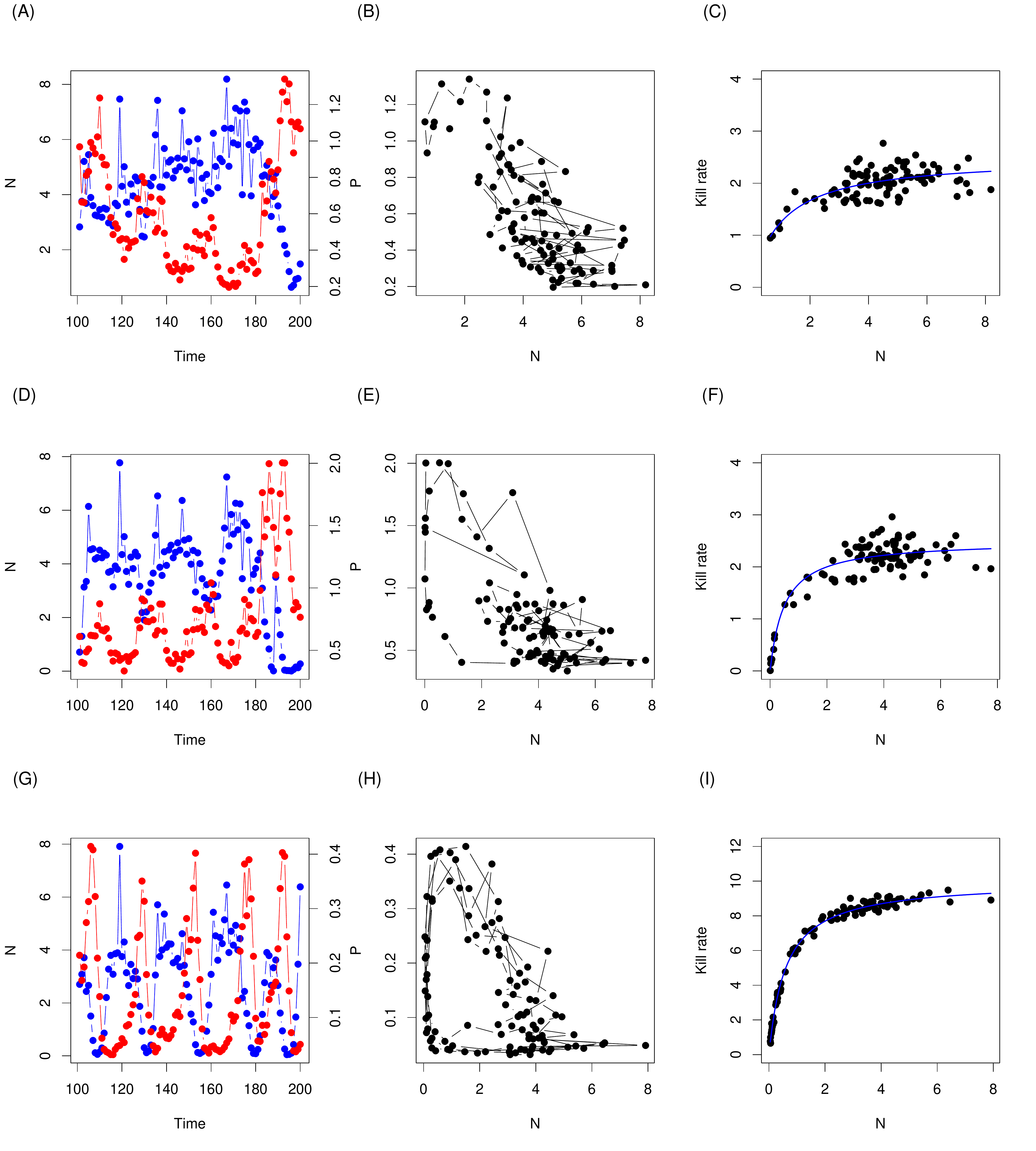}
    \caption{System dynamics for 100 timesteps (1 simulation), after transients (100 timesteps). In panel (A), we show densities of prey $N(t)$ in blue and predator in red $P(t)$, for the perturbed fixed point (FP) case, (B) corresponding trajectories in phase plane and (C) functional response: kill rate of individual predator as a function of prey density. In the second and third rows, identical panels (D-E-F) for the quasi-cycles (QC) parameter set and (G-H-I) for the noisy limit cycle (LC).}
    \label{fig:system_dynamics_RMA}
\end{figure}

Due to the presence of important transients in this model, we have removed the first 100 timesteps and done the estimation on the following 100. The correlations between pairs of parameters belonging to the same function were similar to that shown in the main text (omitted). 

\begin{figure}[h]
    \centering
    \includegraphics[width=\textwidth]{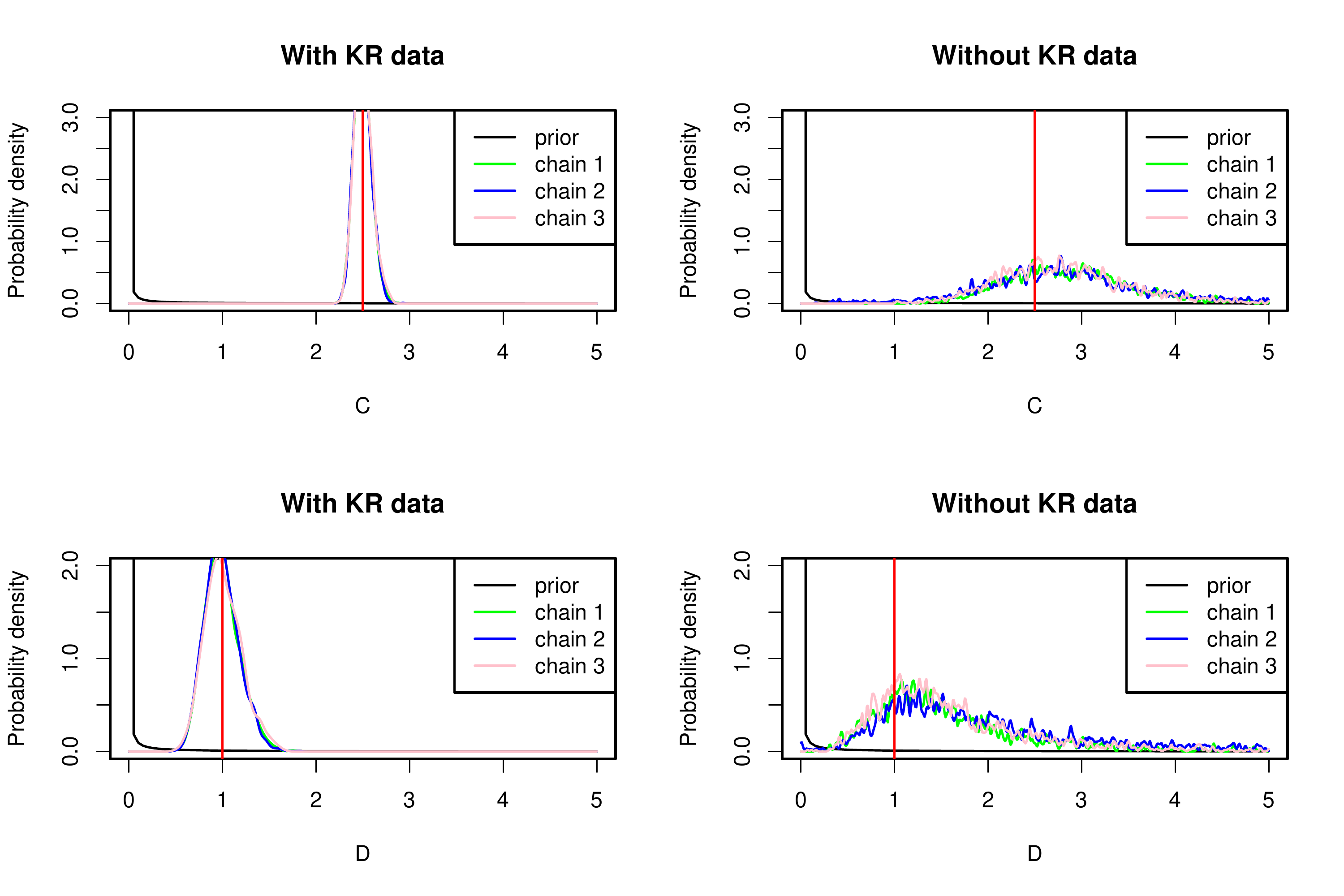}
    \caption{Prior-posterior overlap with kill rate (KR) data (left) and without kill rate data (right). RMA model, FP parameter set. Top row, parameter $C$; bottom row, parameter $D$. True parameter values are red vertical lines.}
    \label{fig:PPO_RMA_FP}
\end{figure}

\begin{figure}[h]
    \centering
    \includegraphics[width=\textwidth]{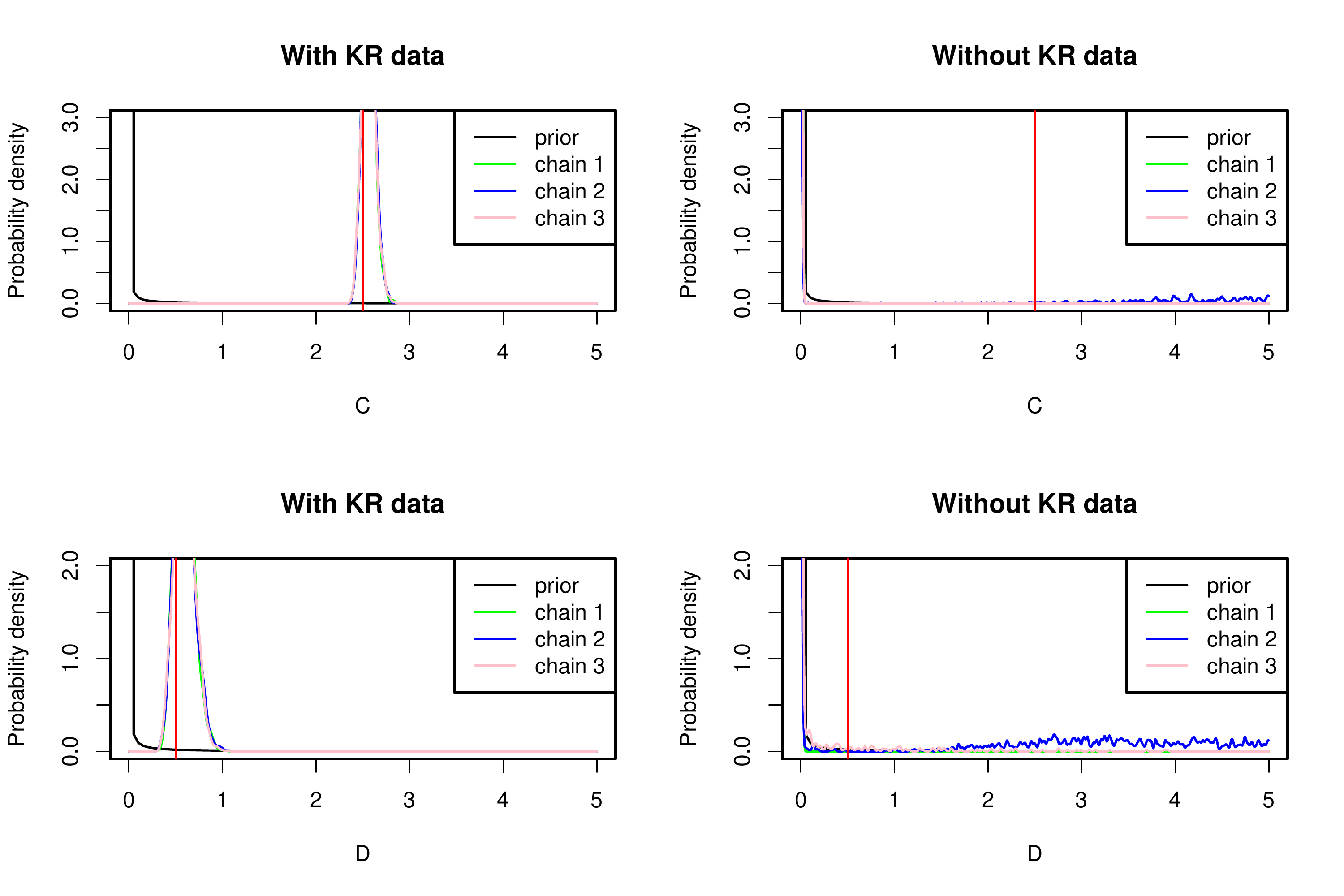}
    \caption{Prior-posterior overlap with kill rate data (left) and without kill rate data (right). RMA model, QC parameter set. Top row, parameter $C$; bottom row, parameter $D$. True parameter values are red vertical lines.}
    \label{fig:PPO_RMA_QC}
\end{figure}

\begin{figure}[h]
    \centering
    \includegraphics[width=\textwidth]{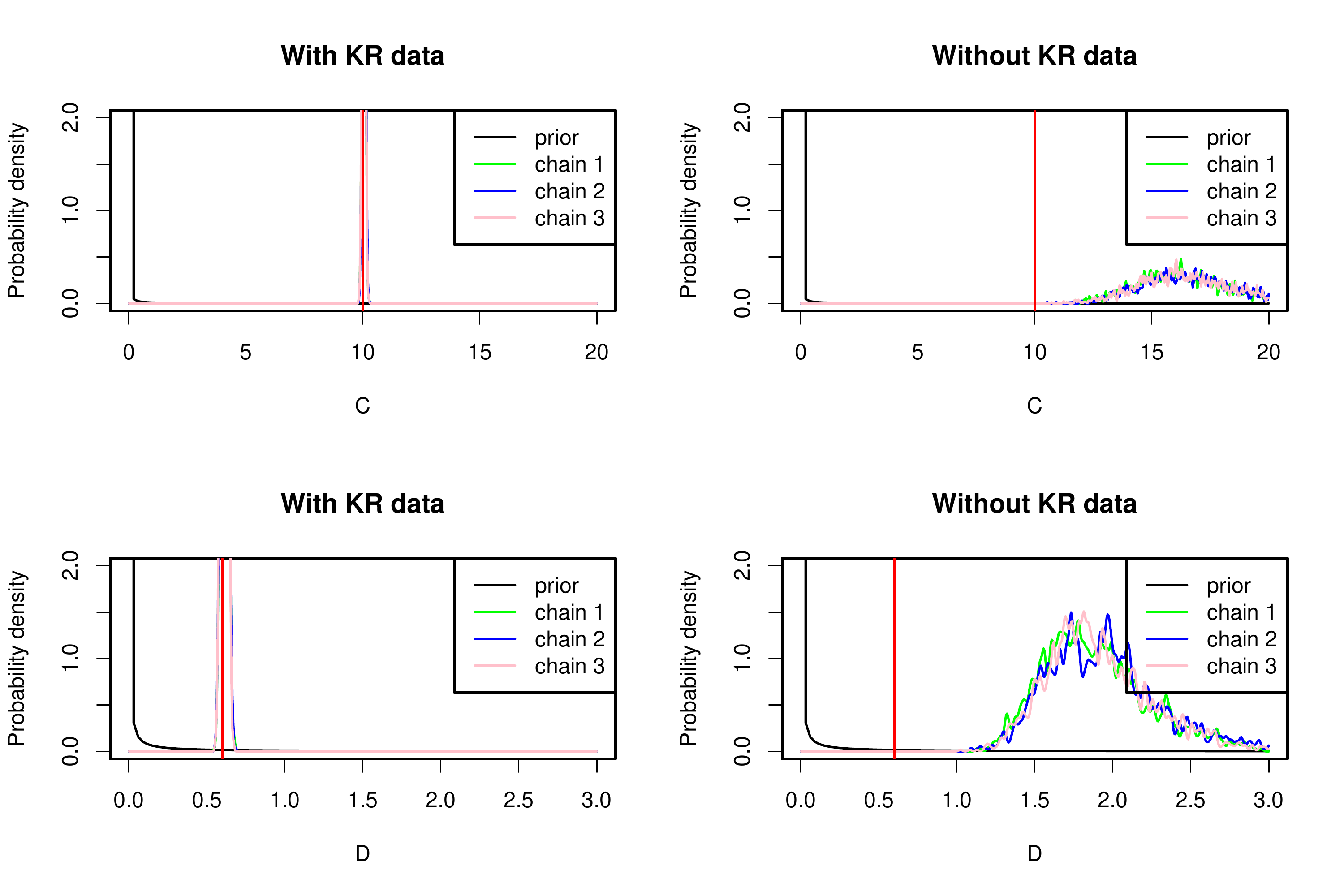}
    \caption{Prior-posterior overlap with kill rate data (left) and without kill rate data (right). RMA model, LC parameter set. Top row, parameter $C$; bottom row, parameter $D$. True parameter values are red vertical lines.}
    \label{fig:PPO_RMA_LC}
\end{figure}

The posterior-prior overlap curves are reported in the followings Figs.~\ref{fig:PPO_RMA_FP}--\ref{fig:PPO_RMA_LC}. For the perturbed fixed point parameter set, the estimation of parameters $C$ and $D$ is barely feasible without kill rate data while it is very good with kill rate data (Fig.~\ref{fig:PPO_RMA_FP}). The findings are similar for the QC parameter set (Fig.~\ref{fig:PPO_RMA_QC}), even though surprisingly --- since we are closer to cyclic dynamics --- the identifiability seems even worse in this case (but we caution this is only one simulation). The limit cycle parameter set leads to a less good identification of parameters $C$ and $D$ in absence of kill rate data than for the model of the main text. Indeed, the right order of magnitude is found for $C$ but $D$ is a bit misleading (Fig.~\ref{fig:PPO_RMA_LC}), and in absence of kill rate data, the posterior densities are a bit spiky even though the chains converge in the sense of $\hat{R}<1.1$. 

Therefore, our results with the RMA model confirm that a much more precise and less biased estimation of parameters is possible once the data on kill rates are added to the model. 

\clearpage

\section{Simulations and model fitting with very stochastic functional responses}

In this appendix, we model cases where the functional response exhibit large, seemingly stochastic temporal variation (i.e., large process noise). This is done by introducting temporal variation in the parameters $C$ and $D$ of the functional response $\frac{CN}{D+N}$. To data simulated with temporally variable $C_t$ and $D_t$, we then fit:
\begin{itemize}
    \item A model with a Gaussian noise on the kill rate (identical to the model fitted in the main text)
    \item A model without kill rate data
    \item The simulated model
\end{itemize}
Therefore, this supplement also serves as an evaluation of the robustness of the results presented in the main text, since we fit the model with a Gaussian functional response to data produced by a more complex and noisy simulation model. 

\subsection{Temporal variation in $C$}

\begin{figure}[H]
    \centering
    \includegraphics[width=0.7\textwidth]{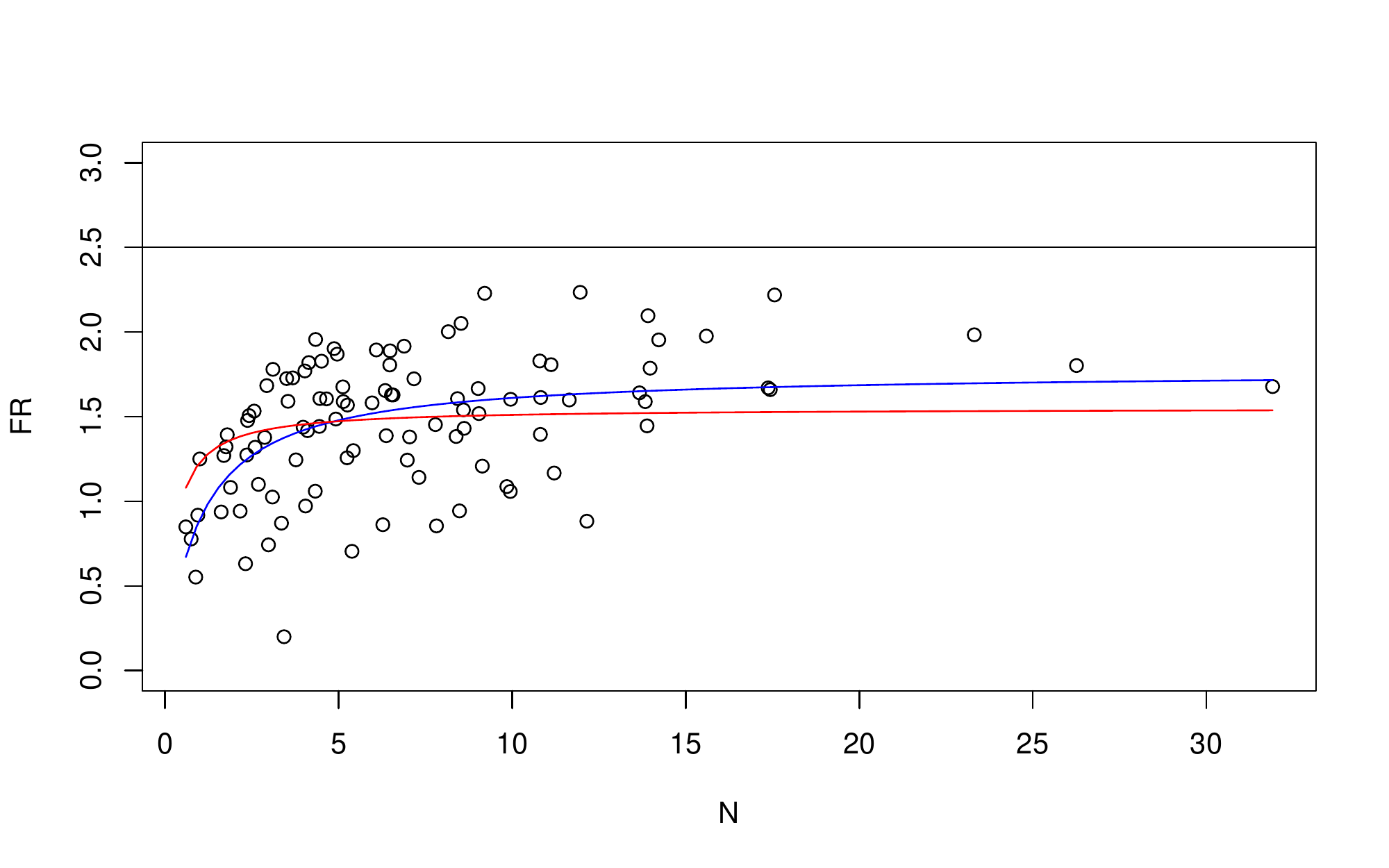}
    \caption{Functional response with temporally variable $C$. The max functional response value, $C_{\text{max}}=2.5$, is shown as a horizontal black line. The blue line is a least square fit of the functional response assuming a Gaussian noise on the kill rate (instead of a Beta noise on $C$); the red line the same fit in JAGS, with mildly informative priors, which performs typically a bit better.}
    \label{fig:funcrespCvariable}
\end{figure}

To make for a realistic point cloud in the functional response, we assume that there is a `hard limit' (say, $C_{\text{max}}$) to the kill rate with some variation below that limit. In this formulation of the model, parameters are equal to that of the fixed point (FP) parameter set considered previously, except that the functional response is now 
\begin{equation}
     G_t = \frac{C_{\text{max}}(1-B_t)N_t}{D+N_t}, \; B_t \sim \text{Beta}(\alpha,\beta)
    \label{eq:temporally_variableC}
\end{equation}
The Beta distribution $\text{Beta}(\alpha,\beta)$ allows to make for lifelike-looking functional response point clouds (Fig.~\ref{fig:funcrespCvariable}). We have used the FP parameter set of the main text, with added parameters $\alpha=2$ and $\beta =5$. To have commensurate noise levels on all components (and therefore test the general robustness of our results to increased levels of randomness), we chose a noise on growth rates with variance $\sigma^2 = 0.5$. 

\subsubsection{Model with a Gaussian functional response, with kill rate data}

JAGS estimates ($C_{\text{Gaussian}},D_{\text{Gaussian}}$) of a Gaussian functional response fitted to the simulated functional response data (without the densities data) with $C_t$ temporally variable yield $(C_{\text{Gaussian}},D_{\text{Gaussian}})=(1.55,0.26)$. These are therefore the values that we can hope to estimate when fitting the full model to data. Given this, the estimates of the functional response in Table~\ref{tab:simCvariable_fittedGaussianFR} are satisfactory, even though $D$ has some difficulties to converge (these discrepancies are to be expected to some degree, since the fitted model is not fully equal to the simulated model). 

\begin{table}[ht]
\centering
\begin{tabular}{rrrrrrrrrr}
  \hline
Parameter & mean & sd & 2.5\% & 25\% & 50\% & 75\% & 97.5\% & Rhat & n.eff \\ 
  \hline
$C$ & 1.50 & 0.08 & 1.37 & 1.44 & 1.50 & 1.56 & 1.65 & 1.01 & 230.00 \\ 
  $D$ & 0.16 & 0.14 & 0.00 & 0.00 & 0.16 & 0.26 & 0.45 & 1.17 & 41.00 \\ 
  $Q$ & 10.70 & 3.07 & 5.65 & 8.53 & 10.37 & 12.51 & 17.77 & 1.00 & 4200.00 \\ 
  $\gamma$ & 2.06 & 1.22 & 0.49 & 1.05 & 1.72 & 2.92 & 4.71 & 1.01 & 360.00 \\ 
  $s$ & 0.44 & 0.11 & 0.24 & 0.37 & 0.44 & 0.51 & 0.67 & 1.00 & 2100.00 \\ 
  $r$ & 2.50 & 0.56 & 1.47 & 2.06 & 2.49 & 2.97 & 3.43 & 1.01 & 350.00 \\ 
  $\sigma^2_P$ & 0.41 & 0.06 & 0.31 & 0.37 & 0.41 & 0.45 & 0.54 & 1.00 & 4900.00 \\ 
  $\sigma^2_V$ & 0.59 & 0.09 & 0.44 & 0.52 & 0.58 & 0.64 & 0.78 & 1.00 & 6000.00 \\ 
  $\tau_{\text{FR}}$ & 6.34 & 0.93 & 4.65 & 5.68 & 6.29 & 6.96 & 8.25 & 1.00 & 3000.00 \\ 
   \hline
\end{tabular}
    \caption{Estimated parameters for the model with Gaussian functional response.}
    \label{tab:simCvariable_fittedGaussianFR}
\end{table}

We therefore confirm here that:
\begin{itemize}
    \item it is possible to provide a reasonable first approximation of complex functional response with a simple Gaussian noise
    \item the main text results are robust to higher noise levels. 
\end{itemize}

\subsubsection{Model with a deterministic functional response, without additional kill rate data}

In this case, we confirm logically the main text results, the functional response is not identifiable and the chains do not converge for parameters $C$ and $D$ which are informed only by the priors. 

\begin{table}[ht]
\centering
\begin{tabular}{rrrrrrrrrr}
  \hline
Parameter & mean & sd & 2.5\% & 25\% & 50\% & 75\% & 97.5\% & Rhat & n.eff \\ 
  \hline
$C$ & 0.00 & 0.00 & 0.00 & 0.00 & 0.00 & 0.00 & 0.00 & 1.28 & 16.00 \\ 
  $D$ & 0.01 & 0.06 & 0.00 & 0.00 & 0.00 & 0.00 & 0.00 & 2.92 & 4.00 \\ 
  $Q$ & 10.76 & 3.12 & 5.68 & 8.52 & 10.46 & 12.63 & 17.87 & 1.00 & 6000.00 \\ 
  $\gamma$ & 1.24 & 0.94 & 0.30 & 0.60 & 0.91 & 1.57 & 3.93 & 1.00 & 680.00 \\ 
  $s$ & 0.45 & 0.11 & 0.23 & 0.37 & 0.44 & 0.52 & 0.67 & 1.00 & 3100.00 \\ 
  $r$ & 1.94 & 0.56 & 1.06 & 1.52 & 1.86 & 2.31 & 3.16 & 1.00 & 750.00 \\ 
  $\sigma_P^2$ & 0.41 & 0.06 & 0.31 & 0.37 & 0.41 & 0.45 & 0.55 & 1.00 & 5700.00 \\ 
  $\sigma_V^2$ & 0.55 & 0.08 & 0.41 & 0.49 & 0.55 & 0.60 & 0.74 & 1.00 & 2100.00 \\ 
   \hline
\end{tabular}
    \caption{Estimated parameters for the model with deterministic functional response, without kill rate data.}
    \label{tab:simCvariable_fittedDetFR}
\end{table}

\subsubsection{Model with a Beta-distributed max kill rate $C$, with kill rate data}

Here, we fit the same model that we simulated, with extra temporal variation on $C$ (eq.~\eqref{eq:temporally_variableC}). The model is now quite well estimated, although it should be noted that some chains can have difficulties to converge. 

\begin{table}[ht]
\centering
\begin{tabular}{rrrrrrrrrr}
  \hline
Parameter & mean & sd & 2.5\% & 25\% & 50\% & 75\% & 97.5\% & Rhat & n.eff \\ 
  \hline
$C$ & 2.37 & 0.11 & 2.17 & 2.28 & 2.36 & 2.44 & 2.61 & 1.00 & 1100.00 \\ 
  $D$ & 0.48 & 0.09 & 0.31 & 0.42 & 0.47 & 0.53 & 0.66 & 1.00 & 1200.00 \\ 
  $Q$ & 10.71 & 3.07 & 5.55 & 8.53 & 10.44 & 12.52 & 17.54 & 1.00 & 6000.00 \\ 
  $a$ & 1.70 & 0.42 & 1.01 & 1.40 & 1.65 & 1.96 & 2.65 & 1.00 & 3600.00 \\ 
  $b$ & 3.70 & 0.66 & 2.54 & 3.23 & 3.66 & 4.12 & 5.11 & 1.00 & 6000.00 \\ 
  $\gamma$ & 2.22 & 1.28 & 0.49 & 1.13 & 1.92 & 3.22 & 4.77 & 1.01 & 300.00 \\ 
  $s$ & 0.44 & 0.11 & 0.23 & 0.37 & 0.44 & 0.52 & 0.67 & 1.00 & 6000.00 \\ 
  $r$ & 2.56 & 0.57 & 1.46 & 2.12 & 2.58 & 3.05 & 3.45 & 1.01 & 310.00 \\ 
  $\sigma_P^2$ & 0.41 & 0.06 & 0.31 & 0.37 & 0.40 & 0.45 & 0.55 & 1.00 & 6000.00 \\ 
  $\sigma_V^2$ & 0.59 & 0.09 & 0.44 & 0.53 & 0.58 & 0.64 & 0.78 & 1.00 & 2600.00 \\ 
  $\sigma_{\text{FR}}$ & 0.09 & 0.01 & 0.07 & 0.09 & 0.09 & 0.10 & 0.10 & 1.00 & 1500.00 \\ 
   \hline
\end{tabular}
 \caption{Estimated parameters for the model with stochastic functional response and a temporally Beta-distributed C, with kill rate data. Here $\sigma_{\text{FR}}$ is a small SD term that is needed to estimate the model in JAGS with a proper mixing of the chains.}
    \label{tab:simCvariable_fittedBetaC_FR}
\end{table}

\subsection{Temporal variation in the half-saturation constant $D$}

Temporal variation in $D$ is added using a Gamma distribution for the half-saturation constant, with mean $m_D$ and standard deviation $s_D$ that are then converted into shape (sh) and rate (ra) parameters. We chose $m_D=2$ and $s_D=4$, all other parameters are equal to those of the fixed point simulations, save for the noise variances on the growth rates. To have commensurate noise on all components, we chose $\sigma^2 = 0.5$.

\begin{equation}
    G_t = \frac{C N_t}{D_t+N_t}, \; D_t \sim \text{Gamma(sh,ra)}
    \label{eq:temporally_variableD}
\end{equation}

with $\text{sh} = \frac{m_D^2}{s_D^2}$ and $\text{ra} = \frac{m_D}{s_D^2}$.

\begin{figure}[H]
    \centering
    \includegraphics[width=0.7\textwidth]{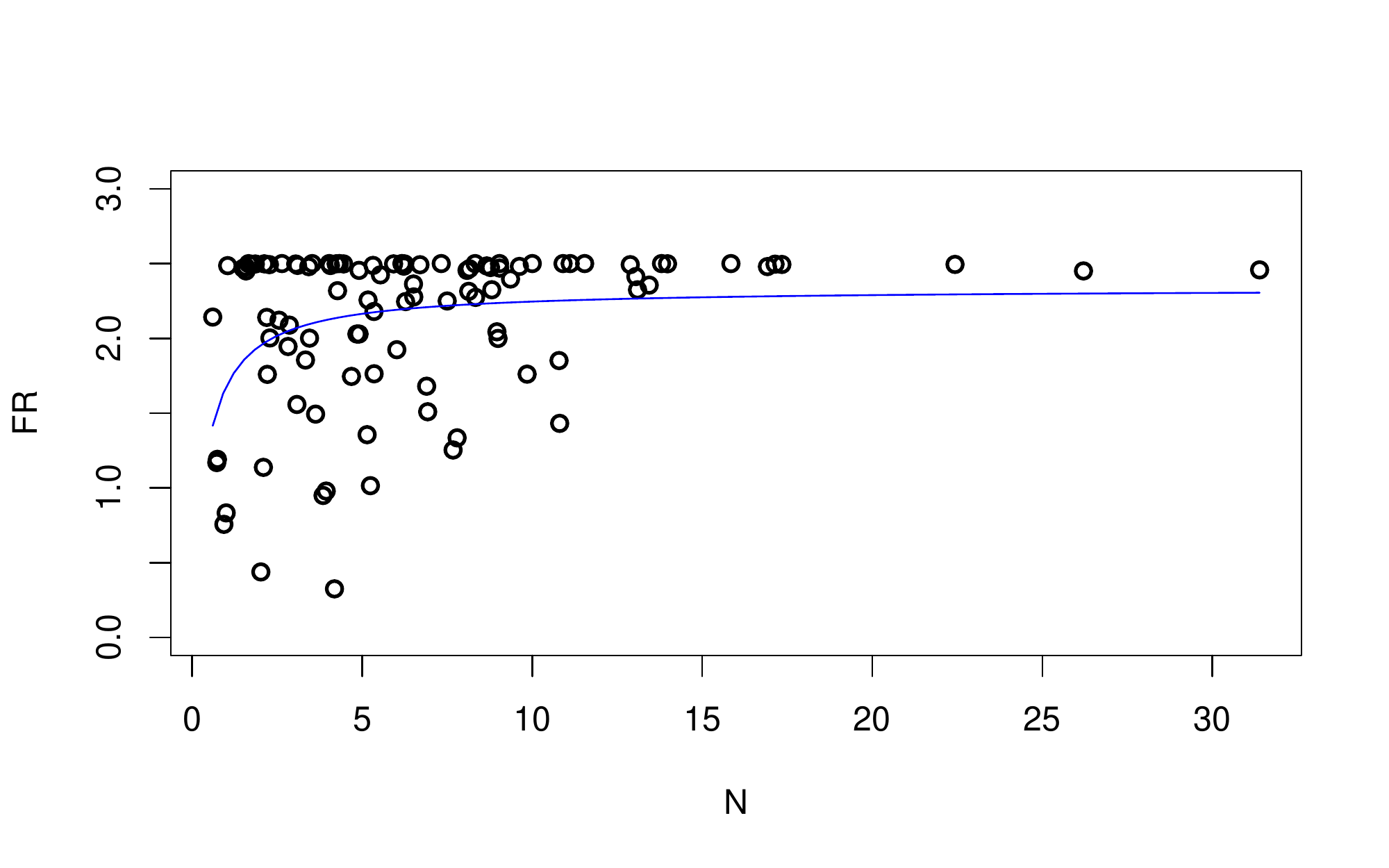}
    \caption{Functional response with temporally variable $D$. The blue line is a least-square-fit of the functional response assuming a Gaussian noise on the kill rate.}
    \label{fig:funcrespDvariable}
\end{figure}

\subsubsection{Model with a Gaussian functional response, with kill rate data}

Estimates of a least square fit of the Gaussian functional response based on the simulated functional response data with $D_t$ temporally variable yield $(C_{\text{LS}},D_{\text{LS}})=(2.33,0.39)$. Given this, the estimates of the functional response in Table~\ref{tab:simDvariable_fittedGaussianFR} are quite satisfactory, although $D$ is a little high. 

\begin{table}[ht]
\centering
\begin{tabular}{rrrrrrrrrr}
  \hline
Parameter & mean & sd & 2.5\% & 25\% & 50\% & 75\% & 97.5\% & Rhat & n.eff \\ 
  \hline
$C$ & 2.47 & 0.12 & 2.25 & 2.39 & 2.47 & 2.55 & 2.72 & 1.00 & 5800.00 \\ 
  $D$ & 1.05 & 0.21 & 0.69 & 0.90 & 1.04 & 1.18 & 1.51 & 1.00 & 3400.00 \\ 
  $Q$ & 10.53 & 3.03 & 5.55 & 8.39 & 10.23 & 12.31 & 17.35 & 1.00 & 6000.00 \\ 
  $\gamma$ & 1.96 & 1.18 & 0.48 & 1.01 & 1.67 & 2.72 & 4.67 & 1.01 & 300.00 \\ 
  $s$ & 0.44 & 0.11 & 0.22 & 0.36 & 0.44 & 0.51 & 0.67 & 1.00 & 6000.00 \\ 
  $r$ & 2.45 & 0.54 & 1.46 & 2.03 & 2.45 & 2.89 & 3.40 & 1.01 & 280.00 \\ 
  $\sigma^2_P$ & 0.41 & 0.06 & 0.31 & 0.37 & 0.41 & 0.45 & 0.55 & 1.00 & 6000.00 \\ 
  $\sigma^2_V$  & 0.59 & 0.09 & 0.45 & 0.53 & 0.58 & 0.64 & 0.78 & 1.00 & 6000.00 \\ 
  $\tau_{\text{FR}}$ & 2.55 & 0.41 & 1.84 & 2.27 & 2.53 & 2.81 & 3.42 & 1.00 & 6000.00 \\ 
   \hline
\end{tabular}
\caption{Estimated parameters for the model with Gaussian functional response.}
\label{tab:simDvariable_fittedGaussianFR}
\end{table}

\subsubsection{Model with a deterministic functional response, without additional kill rate data}

Here again, we confirm the main text results, the functional response parameters are not identifiable and dominated by the priors.

\begin{table}[H]
\centering
\begin{tabular}{rrrrrrrrrr}
  \hline
Parameter & mean & sd & 2.5\% & 25\% & 50\% & 75\% & 97.5\% & Rhat & n.eff \\ 
  \hline
$C$ & 0.00 & 0.00 & 0.00 & 0.00 & 0.00 & 0.00 & 0.00 & 1.32 & 15.00 \\ 
  $D$ & 0.00 & 0.00 & 0.00 & 0.00 & 0.00 & 0.00 & 0.00 & 1.67 & 6.00 \\ 
  $Q$ & 10.51 & 2.99 & 5.42 & 8.40 & 10.24 & 12.29 & 17.13 & 1.00 & 2300.00 \\ 
  $\gamma$ & 0.63 & 0.46 & 0.21 & 0.37 & 0.51 & 0.72 & 1.87 & 1.00 & 1200.00 \\ 
  $s$ & 0.44 & 0.11 & 0.23 & 0.36 & 0.43 & 0.51 & 0.65 & 1.00 & 2600.00 \\ 
  $r$ & 1.42 & 0.41 & 0.79 & 1.15 & 1.36 & 1.63 & 2.42 & 1.00 & 1400.00 \\ 
  $\sigma^2_P$ & 0.41 & 0.06 & 0.31 & 0.37 & 0.41 & 0.45 & 0.55 & 1.00 & 6000.00 \\ 
  $\sigma^2_V$ & 0.55 & 0.08 & 0.41 & 0.50 & 0.55 & 0.60 & 0.74 & 1.00 & 6000.00 \\ 
   \hline
\end{tabular}
\caption{Estimated parameters for the model with deterministic functional response and no kill rate data.}
\label{tab:simDvariable_fittedDeterFR}
\end{table}

\subsubsection{Model with a Gamma-distributed half-saturation $D$, with kill rate data}

Here, we fit the same model that we simulated, with extra temporal variation on $D$ (eq.~\eqref{eq:temporally_variableD}).

\begin{table}[ht]
\centering
\begin{tabular}{rrrrrrrrrr}
  \hline
Parameter & mean & sd & 2.5\% & 25\% & 50\% & 75\% & 97.5\% & Rhat & n.eff \\ 
  \hline
$C$ & 2.50 & 0.00 & 2.50 & 2.50 & 2.50 & 2.50 & 2.50 & 1.00 & 1.00 \\ 
  $Q$ & 10.65 & 3.04 & 5.53 & 8.48 & 10.36 & 12.50 & 17.59 & 1.00 & 2600.00 \\ 
  $\gamma$ & 1.91 & 1.18 & 0.47 & 0.99 & 1.56 & 2.58 & 4.70 & 1.00 & 520.00 \\ 
  $m_D$ & 2.07 & 0.30 & 1.57 & 1.86 & 2.04 & 2.25 & 2.74 & 1.00 & 1400.00 \\ 
  $s_D$ & 4.56 & 0.70 & 3.39 & 4.07 & 4.48 & 4.97 & 6.17 & 1.00 & 1400.00 \\
  $s$ & 0.44 & 0.11 & 0.24 & 0.37 & 0.44 & 0.51 & 0.66 & 1.00 & 4600.00 \\ 
  $r$ & 2.42 & 0.55 & 1.44 & 2.00 & 2.40 & 2.84 & 3.42 & 1.00 & 530.00 \\ 
  $\sigma^2_P$ & 0.41 & 0.06 & 0.31 & 0.37 & 0.41 & 0.45 & 0.55 & 1.00 & 6000.00 \\ 
  $\sigma^2_V$ & 0.59 & 0.09 & 0.45 & 0.53 & 0.58 & 0.64 & 0.79 & 1.00 & 3400.00 \\ 
  $\sigma_{\text{FR}}$ & 0.00 & 0.00 & 0.00 & 0.00 & 0.00 & 0.00 & 0.00 & 1.08 & 55.00 \\ 
   \hline
\end{tabular}
 \caption{Estimated parameters for the model with stochastic functional response and a temporally Gamma-distributed D, with kill rate data. Here $\sigma_{\text{FR}}$ is a small SD term that is needed to estimate the model in JAGS with a proper mixing of the chains.}
    \label{tab:simDvariable_fittedGammaD_FR}
\end{table}

This model is well estimated, with good convergence; for the parameters chosen here the Gaussian functional response and Gamma-distributed half-saturation $D$ yield fairly similar parameter values. The good performance of this Gamma-distributed $D$ model suggests that in many cases for which the functional response has a deterministic `hard' limit (e.g., the animal cannot possibly kill than $x$ items in a certain amount of time), so that $C$ is truly a constant, a model with $D=D_t$ temporally variable will be an interesting model to avoid the assumption of a normally distributed kill rate. This will be especially the case where variation in $D_t$ could be linked to another prey species, as in this case one expects from mechanistic multi-species functional response model that $D_t = D_0 + M_t$ where $M_t$ is the density of the additional species.

\end{document}